\documentclass[11pt,a4paper]{article}
\usepackage[utf8]{inputenc}
\usepackage{amsmath}
\usepackage{amsfonts}
\usepackage{amssymb}
\usepackage[left=2cm,right=2cm,top=2.5cm,bottom=2.5cm]{geometry}
\usepackage{cite}

\usepackage{slashed}
\usepackage{hyperref}
\usepackage{graphicx}
\usepackage{caption}
\usepackage{float}
\usepackage{subcaption}

\hypersetup{
	unicode=true,          
	pdftoolbar=true,        
	pdfmenubar=true,        
	pdffitwindow=false,     
	pdfstartview={FitH},    
	pdftitle={NSCFI},    
	pdfauthor={Author},     
	pdfsubject={Subject},   
	pdfcreator={},   
	pdfproducer={Producer}, 
	pdfkeywords={dark matter} {freeze-in} {non standard cosmology}, 
	pdfnewwindow=true,      
	colorlinks=true,       
	linkcolor=blue,          
	citecolor=red,        
	filecolor=black,      
	urlcolor=blue           
}

\usepackage{ifthen}
\usepackage[dvipsnames]{xcolor}
\usepackage{tikz}
\usepackage{mathtools}

\newcommand{\treh}{T_{\rm reh}}

\definecolor{ao}{rgb}{0.0, 0.0, 1.0}
\newenvironment{bl}[1]{{\color{ao}  #1}}

\newcommand{\dint}{  \displaystyle \int }
\newcommand{\ie}{{\em i.e.,} }
\newcommand{\eg}{{\em e.g.,} }
\newcommand{\GeV}{{\rm GeV}}
\newcommand{\TeV}{{\rm TeV}}
\newcommand{\MeV}{{\rm MeV}}
\newcommand{\keV}{{\rm keV}}
\newcommand{\rhs}{RHS }
\newcommand{\lhs}{LHS }

\newcommand{\rhoR}{ \rho_{R} }
\newcommand{\rhoPhi}{ \rho_{\Phi} }
\newcommand{\nDM}{ n_{\chi} }
\newcommand{\YDM}{ Y_{\chi} }
\newcommand{\NDM}{ N_{\chi} }

\newcommand{\GammaDM}{ \Gamma_{S \to \chi\chi} }
\newcommand{\GammaPhi}{ H_{\rm R}^{\rm (end)} }

\newcommand{\TEND}{ T_{\rm end} }

\newcommand{\TDI}{ T_{ {\rm D}_{1} } }
\newcommand{\TDII}{ T_{{\rm D}_{2} } } 

\newcommand{\EI}{ {\rm E}_1 } 
\newcommand{\EII}{ {\rm E}_2 }
\newcommand{\DI}{ {\rm D}_1 }
\newcommand{\DII}{ {\rm D}_2 }
\newcommand{\FI}{ {\rm FI} }

\newcommand{\Ti}{ T_{\rm ini}   }
\newcommand{\ai}{ a_{\rm ini}   }
\newcommand{\ini}{ {\rm ini}   }

\newcommand{\yDM}{ y_{\chi} }
\newcommand{\mSV}{ m_{S} }
\newcommand{\mST}{ m_{S,T} }
\newcommand{\mDM}{ m_{\chi} }
\newcommand{\TFI}{ T_{\rm FI} }
\newcommand{\Rmax}{ R_{\rm max} }
\newcommand{\relic}{ \Omega_{\rm DM} h^2 }

\newcommand{\Lint}{ \mathcal{L}_{\rm int} }

\newcommand{\vev}[1]{\langle #1 \rangle}
\newcommand{\Bvev}[1]{\Bigg\langle #1 \Bigg\rangle}
\newcommand{\bvev}[1]{\Big\langle #1 \Big\rangle}

\newcommand{\pDM}{  \vev{p_{\chi}} }
\newcommand{\pDMt}{  \vev{p_{\chi , 0}} }
\newcommand{\EDM}{  \vev{E_{\chi}} }
\newcommand{\uDM}{  u_{\chi} }

\newcommand{\pWDMt}{  \vev{p_{{\rm WDM}, 0}} }
\newcommand{\pWDMrel}{  \vev{p_{\rm WDM, rel} } }

\newcommand{\TWDMt}{  T_{{\rm WDM}, 0}} 
\newcommand{\TWDMrel}{  T_{\rm WDM, rel} } 

\newcommand{\vWDMt}{  \vev{v_{{\rm WDM}, 0}} }
\newcommand{\mWDM}{ m_{\rm WDM} }

\newcommand{\lrb}[1]{\left( #1 \right)}
\newcommand{\lrsb}[1]{\left[ #1 \right]}
\newcommand{\lrBigb}[1]{\Big( #1 \Big)}

\newcommand{\lrBiggb}[1]{\Bigg( #1 \Bigg)}
\newcommand{\lrBiggsb}[1]{\Bigg[ #1 \Bigg]}

\newcommand{\lrBiggcb}[1]{\Bigg\{ #1 \Bigg\}}

\newcounter{NumArgs}

\newcommand{\eqs}[1]{\setcounter{NumArgs}{0}\foreach\i in{#1}{\stepcounter{NumArgs}}%
	\ifthenelse{\equal{\theNumArgs}{1}}{eq.~(\ref{#1})}%
	{\ifthenelse{\equal{\theNumArgs}{2}}%
		{eqs.~\foreach\i[count=\q]in{#1}{\ifthenelse{\equal{\q}{\theNumArgs}}{and (\ref{\i})}{(\ref{\i})~}}}%
		{eqs.~\foreach\i[count=\q]in{#1}{\ifthenelse{\equal{\q}{\theNumArgs}}{and (\ref{\i})}{(\ref{\i}),~}}}}}

\newcommand{\Eqs}[1]{\setcounter{NumArgs}{0}\foreach\i in{#1}{\stepcounter{NumArgs}}%
	\ifthenelse{\equal{\theNumArgs}{1}}{Eq.~(\ref{#1})}%
	{\ifthenelse{\equal{\theNumArgs}{2}}%
		{Eqs.~\foreach\i[count=\q]in{#1}{\ifthenelse{\equal{\q}{\theNumArgs}}{and (\ref{\i})}{(\ref{\i})~}}}%
		{Eqs.~\foreach\i[count=\q]in{#1}{\ifthenelse{\equal{\q}{\theNumArgs}}{and (\ref{\i})}{(\ref{\i}),~}}}}}

\newcommand{\refs}[1]{\setcounter{NumArgs}{0}\foreach\i in{#1}{\stepcounter{NumArgs}}%
	\ifthenelse{\equal{\theNumArgs}{1}}{(\ref{#1})}%
	{\ifthenelse{\equal{\theNumArgs}{2}}%
		{\foreach\i[count=\q]in{#1}{\ifthenelse{\equal{\q}{\theNumArgs}}{and (\ref{\i})}{(\ref{\i})~}}}%
		{\foreach\i[count=\q]in{#1}{\ifthenelse{\equal{\q}{\theNumArgs}}{and (\ref{\i})}{(\ref{\i}),~}}}}}

\newcommand{\Figs}[1]{\setcounter{NumArgs}{0}\foreach\i in{#1}{\stepcounter{NumArgs}}%
	\ifthenelse{\equal{\theNumArgs}{1}}{Fig.~\ref{#1}}%
	{\ifthenelse{\equal{\theNumArgs}{2}}%
		{Figs.~\foreach\i[count=\q]in{#1}{\ifthenelse{\equal{\q}{\theNumArgs}}{and \ref{\i}}{\ref{\i}~}}}%
		{Figs.~\foreach\i[count=\q]in{#1}{\ifthenelse{\equal{\q}{\theNumArgs}}{and \ref{\i}}{\ref{\i},~}}}}}

\usepackage{authblk}

\usepackage[normalem]{ulem}

\title{Frozen-in fermionic singlet dark matter in non-standard cosmology with a decaying fluid}

\author[a,b]{Paola Arias}
\author[c]{Dimitrios Karamitros}
\author[b,c]{Leszek Roszkowski}

\affil[a]{Departamento de Fisica, Universidad de Santiago de Chile, Casilla 307, Santiago, Chile}
\affil[b]{Astrocent, Nicolaus Copernicus Astronomical Center of the Polish Academy of Sciences, ul. Rektorska 4, 00-614 Warsaw, Poland}
\affil[c]{National Centre for Nuclear Research, ul. Pasteura 7, 02-093 Warsaw, Poland}

\affil[ ]{  }
\affil[ ]{\textit{E-mail: }
	\href{mailto:paola.arias.r@usach.cl}{\bl{paola.arias.r@usach.cl}},
	\href{mailto:dimitrios.karamitros@ncbj.gov.pl}{\bl{dimitrios.karamitros@ncbj.gov.pl}},
	\href{mailto:leszek.roszkowski@ncbj.gov.pl}{\bl{leszek.roszkowski@ncbj.gov.pl} }}

\renewcommand{\theequation}{\arabic{section}.\arabic{equation}}

\begin{document}
	
	\maketitle
	\flushbottom
	
	\begin{abstract} 
		We perform a detailed study of dark matter production via freeze-in under the assumption that some fluid dominates the early Universe before depositing its energy to the plasma causing entropy injection.
		As a dark matter candidate we consider a fermionic singlet that is produced through its interactions with a scalar particle in the thermal plasma.
		The fluid alters the expansion rate of the Universe, as well as the scaling of the temperature, which  significantly affects the evolution of both the number density and the mean momentum of the dark matter particle.   
		We identify and discuss in detail the effects of the evolution of these quantities by considering several examples representing dark matter production at different stages of expansion and entropy injection. 
		We  find that, since the dark matter density is reduced when the entropy injection to the plasma continues after freeze-in, in order to reproduce its observational value an enhanced rate of dark matter production is required relative to standard cosmology. Furthermore, the impact of the assumed non-standard cosmological history on the dark matter mean momentum can result in either a relaxed or a tightened bound on the dark matter mass from large structure formation data. 
	\end{abstract}
	\flushbottom

	\section{Introduction}\label{sec:intro}
	\setcounter{equation}{0}
	While the existence of dark matter (DM) in the Universe is  well established (for recent review see, \eg~\cite{Massey:2010hh,Salucci2018DarkMI}), its nature remains unknown.  The most popular hypothesis assumes that DM is made up a thermally produced weakly-interacting massive particle (WIMP) that is stable, or extremely long lived (for recent reviews on WIMPs see, \eg~\cite{Arcadi:2017kky, Roszkowski:2017nbc}). One basic reason behind this is that particles of this class are ubiquitous in models of ``new physics" beyond the Standard Model (SM) that address other open questions of particle physics and cosmology. The second reason is that, assuming that the interactions of WIMPs with SM particles are of (electro)weak nature, even if strongly suppressed, then the correct relic abundance of DM can often be obtained by producing WIMPs via the mechanism of freeze-out, which is robust because it is usually rather efficient under very basic and well-justified assumptions of the standard cosmological model, in particular that of radiation dominance during a hot thermal period of  the expansion of the early Universe.
	
	However, while these two arguments remain attractive and natural, it is also well known that a wide range of alternatives for each of them exist. Firstly, WIMP interactions can be completely unrelated to the weak interactions of the SM, and may be very much weaker (or, actually, also stronger), even if it is produced thermally via freeze-out, thus in general disentangling DM WIMPs from the electroweak mass or interaction scale; see, \eg~\cite{Feng:2010tg}.  Secondly, DM does not have to be produced via freeze-out, as there exist other mechanisms of relic DM production that under some circumstances can reproduce the reproduce the correct density of DM, while the freeze-out cannot. 
	
	One long-known example of a general WIMP that is not thermally produced is a gravitino~\cite{Ellis:1984eq}, another is an axino~\cite{Covi:1999ty,Covi:2001nw}. Both are extremely weakly interacting, and therefore fall into the category of E-WIMPs~\cite{Choi:2005vq}, or equivalently FIMPs (from ``feebly''), or super-WIMPs; see, \eg~\cite{Baer:2014eja} for a review. The Universe was void of such relics  after it reheated -- as their primordial population was inflated away -- and they were subsequently reproduced via thermal scattering and decay processes involving SM particles. This mechanism -- later named the freeze-in mechanism~\cite{Hall:2009bx} in the context of renormalisable interactions but in fact it is more general and applies to FIMPs mentioned above and also to other cases~\cite{McDonald:2001vt}. A number of interesting models exhibiting extremely weak interaction strength with other particles (\eg the SM) has been studied in the literature, \eg~\cite{Blennow:2013jba,Drewes:2015eoa,Shakya:2015xnx,Dedes:2017shn, DeRomeri:2020wng,Seto:2020udg}; for a review, see, \eg \cite{Bernal:2017kxu}.

	The freeze-in mechanism is based on the assumption that FIMPs constituting DM are absent after reheating due to strongly suppressed interaction with the inflaton and SM particles, and are produced later through interactions taking place in the expanding plasma. Due to their assumed very feeble interactions, they cannot reach equilibrium with the plasma, \ie DM is produced during the early Universe with negligible back-reactions, which requires that FIMP renormalisable couplings are less than some $10^{-7}$~\cite{Bernal:2017kxu}, while, in order to obtain the correct DM relic density -- assuming standard cosmology -- FIMP Yukawa or gauge couplings need to be of the order of $10^{-11}$~\cite{Hall:2009bx}.

	Similarly to the case of thermal WIMPs, FIMP DM can be implemented in a variety of ``new physics" models, either involving renormalisable interactions or not. In the latter case high-temperature DM production near the reheating temperature $\treh$ after inflation dominates, as is the case with the gravitinos or axinos. The opposite is typically true in models with FIMPs exhibiting renormalisable interactions with the SM particles. For instance, when DM production involves a light mediator, the low-temperature production dominates over the high-temperature one and freeze-in is largely independent of $\treh$~\cite{Hall:2009bx}. 
	
	The process of freeze-in ends when the plasma cools down to some freeze-in temperature $\TFI$ which is the temperature at which the production rate effectively ends. Therefore, frozen-out and frozen-in DM scenarios are complementary to each other, as they usually apply at different ranges of interaction strength. 
	
	Like in the case of freeze-out, also DM production via freeze-in is sensitive to the assumed thermal history of the early Universe. The  number density of DM particles during the Universe's expansion is not given by their thermal phase-space distribution function; it depends on the process that produces them, the evolution of the plasma temperature, and the expansion rate of the Universe at all times. Furthermore, the DM relic abundance depends on both the DM and photon number densities. Therefore, the properties of the DM particle, like its mass and its interactions, do not determine uniquely the DM relic abundance, as it also depends on the assumed cosmological history. 
	
	In the standard cosmological scenario, once inflation ends (for a review see~\cite{Lyth:1998xn}), the Universe enters a period of radiation-dominated expansion. Constraints from measurements of the cosmic microwave background (CMB) and Big Bang Nucleosynthesis (BBN)~\cite{Kolb:206230,Peebles:1993}, suggest that the Universe was indeed dominated by radiation at temperatures around $\mathcal{O}(10)~\MeV$~\cite{Kawasaki:2000en, Hannestad:2004px, Ichikawa:2005vw, DeBernardis:2008zz}. However, the period between the end of inflation and the start of the process of nucleosynthesis is largely unconstrained, with the temperature range spanning up to twenty orders of magnitude.

	During that period, a wide range of non-standard cosmology (NSC) scenarios  are possible. Some early works invoked, \eg the presence of (slowly-decaying) heavy particles that dominate the energy density of the Universe~\cite{Vilenkin:1982wt,Coughlan:1983ci}, or  the dominance of a scalar field with various potentials~\cite{Wetterich:1987fm,Ratra:1987rm,Choi:1999xn,Gardner:2004in,Tsujikawa:2013fta}. For a review and their implications, see~\cite{Allahverdi:2020bys}.
	
	Since DM production usually happens after inflation and before BBN, a study of a NSC scenario is often accompanied by an attempt to explain the DM content of the Universe. Numerous analyses were performed for both freeze-out and freeze-in mechanisms for a variety of models~\cite{McDonald:1989jd,Moroi:1999zb,Giudice:2000ex,Acharya:2009zt,Co:2015pka,Drees:2017iod,DEramo:2017gpl,Hamdan:2017psw,Redmond:2017tja,DEramo:2017ecx,Hardy:2018bph,Bernal:2018kcw,Arias:2019uol,Allahverdi:2019jsc,Poulin:2019omz,Bernal:2019mhf,Cosme:2020mck,Bernal:2020bfj}. Typically, the parameter space becomes significantly altered, and often relaxed, compared to the standard cosmological model.
	In particular, in simple WIMP models (\eg the singlet DM model~\cite{Silveira:1985rk,McDonald:1993ex,Burgess:2000yq}) one is  able to easily evade the stringent direct DM-detection bounds~\cite{Aprile:2018dbl} by assuming some NSC scenario (\eg the singlet DM model in an NSC~\cite{Hardy:2018bph,Bernal:2018kcw}).   
	Furthermore, in the freeze-in scenario, the constraints on the interaction couplings between DM and other (\eg SM) particles can be relaxed  considerably, \eg by up to around $6$ orders of magnitude~\cite{Bernal:2018kcw} in some NSC scenarios.   
	
	In this article, we study the freeze-in production of DM in the context of some NSC scenarios.  We consider the case of a (slowly-decaying) fluid $\Phi$ that, after an initial (\ie after the end of inflation)  radiation-dominated period of the expansion of the Universe, dominates for a certain period of time.  Its domination ends when $\Phi$ decays away to plasma particles, increasing the entropy of radiation (\ie entropy injection). 
	We examine the evolution of DM production during the different periods of expansion and entropy injection. For the DM content, we consider a case where the DM particle $\chi$ interacts only with a scalar field $S$, with both assumed to be singlets of the SM symmetries. Unlike $\chi$, $S$ is assumed to remain in thermal equilibrium with the plasma. 

	Although this is arguably one of the simplest scenarios for an NSC, it turns out to be distinctive from other recently studied approaches  because the production of DM in the early Universe proceeds in different ways: via decays of $S$, forbidden decays due to thermal mass corrections  to $S$~\cite{Darme:2019wpd} and pair annihilations of $S$. The different production channels are open or effective at different times, allowing us to investigate different effects arising from competing channels.
	A detailed analysis of the DM production process exposes several effects (\eg how the pair-annihilation channel can dominate the relic abundance over the decays), with the most significant one being the general dilution of the DM number density by  entropy injection taking place between the time when DM production stops and the time when $\Phi$ decays away. (We define  the notion of diluted DM more precisely below \eqs{eq:dYdT_with_dNdT}). As in previous works, we find that the required interaction strength between  $\chi$ and $S$ can be much larger in NSC scenarios as long as DM production is followed by a period of entropy injection.

	Furthermore, we trace the evolution of the DM mean momentum during the domination and decay of $\Phi$. This allows us to uncover all effects the DM momentum experiences due to the presence of $\Phi$. By calculating the DM mean momentum at the present time, we are also able to re-examine in this NSC a bound on the frozen-in DM mass from large-scale structure formation (LSSF)~\cite{Irsic:2017ixq,Banik:2019smi}. We find that the momentum of diluted DM experiences a higher redshift, slightly relaxing the constraint on its mass, as  previously noted~\cite{Gelmini:2008sh, Evans:2019jcs}. However, we also show that there is another (finely tuned) case where the opposite effect takes place, namely the DM momentum today is slightly enhanced, leading to the tightening of the aforementioned constraint. 
	
	The article is organised as follows: 
	in Section~\ref{sec:model} we start by introducing the NSC scenario and we define the DM model, together with the corresponding Boltzmann equations. 
	To have a better insight into the impact of the different periods of the NSC scenario, in Section~\ref{sec:RhoPhi}, we introduce the points that correspond to a change in the behavior of energy densities or the overall expansion. Moreover, we show, and describe in detail in Appendix~\ref{app:approx}, analytical behavior of the energy densities of $\Phi$ and radiation  as a function of the scale factor $a$ in the  NSC scenario considered, and show its impact of on the entropy as well as the expansion rate of the Universe.    
	Section~\ref{sec:DM_production} is devoted to study the production of DM via freeze-in. First, we introduce the BE assuming only the decay production channel and identify the effect due to the presence of the $\Phi$ field. Next, we show their evolution during all epochs, and investigate their impact on the overall DM production. Closing this Section we discuss how the inclusion of the pair annihilation  production channel can affect the relic abundance.
	In Section~\ref{sec:DM_momentum} we derive a constraint on the mass of the DM particle by recasting the bound on warm dark matter from LSSF. To do so we introduce an equation (derived in Appendix~\ref{app:BE_Emean}) that describes the mean momentum of the DM particles. Furthermore,  we  point-out some additional NSC effects  show numerical examples during different periods, and trace  the DM momentum evolution. Finally, we show examples of the DM momentum today, and discuss the impact of the NSC scenario on the LSSF  constraint on the DM mass. 
	In Section~\ref{sec:benchmarks} we study some representative  points of the model's parameter space.  In particular, we delineate and discuss the shape of the allowed parameter space for both light and heavy DM and point out the impact of the LSSF bound, as well as the thermalisation constraints.
	We perform a detailed scan of the parameter space in Section~\ref{sec:ParameterSpace} and show that, in a wide range of NSC scenarios one is able to generate diluted a DM density -- relative to the standard case -- and point-out the impact of this scenario on both the relic abundance of DM as well as the LSSF constraint. 
	We summarize our findings in Section~\ref{sec:sum}. 
	

	\section{Assumptions and a dark matter model} \label{sec:model}
	\setcounter{equation}{0}
	We start by introducing the fluid $\Phi$, with an equation of state given by
	\begin{equation}
		p_{\Phi}=  w \, \rhoPhi \;.
		\label{eq:eos}
	\end{equation}   
	In order to study this scenario in a model independent way, we assume that $w$ is a constant.  We define
	\begin{equation}
		c \equiv 3(1+w) 
		\label{eq:c_def}
	\end{equation}
	which will prove to be a more convenient parameterization. Furthermore, we assume that initially (\ie after the end of inflation) the Universe was dominated by radiation, followed by a period of $\Phi$ dominance until its decay that increased the entropy of the plasma. We, therefore, assume $c<4$ and some energy transfer rate ($\Gamma_\Phi = {\rm const.}$) from $\Phi$ to the plasma.
	Moreover, in order not to alter too much the outcome of nucleosynthesis, $\Gamma_\Phi$ has to be such that $\Phi$ has decayed away some time before the plasma reaches $T \sim \mathcal{O}(10)~\MeV$~\cite{Kawasaki:2000en, Hannestad:2004px, Ichikawa:2005vw, DeBernardis:2008zz}. Also, we consider only the cases with $c>0$, since otherwise $\Phi$ would have decayed very slowly and it would have been difficult for it to decay away efficiently. 
	
	Since the mean life-time, $\tau_{\rm end}$, of the fluid is of the order $\tau_{\rm end} \sim \Gamma_\Phi^{-1}$, we expect most decays of $\Phi$ to have happened before $t \approx \tau_{\rm end} $, \ie when $\Gamma_\Phi \approx H|_{t=\tau_{\rm end}}$, where $H$ is the Hubble parameter. Time  is not a convenient variable here, and usually one expresses it in terms of the temperature of the plasma. In a radiation-dominated Universe this is straightforward, as $H_{\rm R} \sim T^2$, where $H_{\rm R}$ denotes the Hubble parameter  assuming radiation-dominated expansion. In an NSC setting, the Hubble parameter can deviate from $H_{\rm R}$ significantly. 
	However, it is still convenient to define $\Gamma_\Phi$ as the standard cosmological value of the Hubble  parameter at $\TEND$,
	\begin{equation}
		\Gamma_\Phi  \equiv \GammaPhi  
		\label{eq:TEND}
	\end{equation}
	where $\GammaPhi =  H_{\rm R} (\TEND)$ and $\TEND$ denotes the temperature that would correspond to $ H_{\rm R}|_{t=\tau_{\rm end} }$ in a radiation-dominated Universe.
	Therefore, we may choose $\TEND$ as a free (and more intuitive) parameter which determines $\Gamma_\Phi$. Notice that $\TEND$ and the actual temperature at which $\Phi$ has decayed away can be different, depending on how much $H$ deviates from $H_{\rm R}$. However, as we discuss later, these two temperatures are of the same order (at least for $c \gtrsim 0.5$), and we should expect  $\TEND \gtrsim \mathcal{O}(10~\MeV)$ in order to preserve the outcome of nucleosynthesis.

	In order to study the effect of a decaying fluid on DM production, we assume that the DM particle is a Majorana fermion $\chi$ that interacts with a real scalar $S$ via an interaction term (in Weyl notation~\cite{Dreiner:2008tw})
	\begin{eqnarray}
		\Lint =  -\dfrac{\yDM}{2}  \lrb{\chi \, \chi  + \chi^{\dagger} \, \chi^{\dagger}} \, S \; .
		\label{eq:Lint}
	\end{eqnarray}
	Any interaction between $\chi$ and the SM is assumed to be suppressed, \eg due to some $\mathcal{Z}_2$ symmetry. 
	We also assume that $S$ remains in equilibrium with the plasma via some interactions with the Higgs boson without having to specify them (but see the discussion below). This keeps the system as model independent as possible, but does not allow us to calculate the departure from equilibrium for $S$. 
	
	Generally, $S$  decouples from the plasma when the temperature falls well below its mass, provided that the Higgs boson is still in equilibrium with the rest of plasma. On the other hand, if $S$ is light enough, the Higgs boson decouples first. In this case, $S$ can either decouple along with the Higgs boson, or remain in equilibrium due to interactions with SM fermions (\eg ref.~\cite{Fradette:2017sdd}). Therefore, for consistency, we restrict our discussion to a case of a rather heavy $S$ ($\mSV \geq 50 ~\GeV$), where the details of the decoupling of $S$ are irrelevant to the DM production, \ie the freeze-in production of $\chi$ is expected to occur while $S$ is still in equilibrium. 
	Furthermore, we   also assume  that the $S$-Higgs boson (and possible $S$ self-interactions) result in negligible contributions to the DM production rate. This, in particular, means that possible production via channels such as $SH \to \chi \chi$ is assumed to be suppressed.  
	
	\subsubsection*{Interaction of the scalar $S$ and the Higgs doublet $H$}
	We have assumed above that $S$ remains in thermal equilibrium with the plasma via interactions with the Higgs boson. The interaction can take several forms. The simplest possibility is a quadratic interaction term 
	\begin{equation}
		\mathcal{L}_{HS}= -\dfrac{\lambda_{HS}}{2} \ S^2 \ H^{\dagger}  H \;.
		\label{eq:LHS}
	\end{equation}
	An experimental limit on the branching ratio for the invisible decay mode $h \to SS$ is constrained to be below $20\%$ for $\mSV<m_H/2$~\cite{Sirunyan:2018owy}, which translates to $\lambda_{HS} \lesssim 10^{-2}$. This is consistent with our earlier assumption that $\lambda_{HS}$ is suppressed, in order for the production of $\chi$ to be dominated by the Yukawa interaction of \eqs{eq:Lint}, while keeping $S$ in equilibrium at early times which in turn requires $\lambda_{HS} \gtrsim 10^{-5}$.~\footnote{The equilibrium condition is $H < \vev{\Gamma_{ SS \leftrightarrow HH}}$. In a radiation dominated Universe this gives $T / M_p \lesssim \lambda_{HS}^2$. Therefore, even for  $\lambda_{HS} \sim 10^{-5}$, $S$ can be kept in equilibrium for $T \lesssim 10^{9}~\GeV$, which is larger than any freeze-in temperature we consider in our analysis.} 
	Therefore, we will assume $10^{-5} \lesssim \lambda_{HS} \lesssim 10^{-2}$, which is both phenomenologically viable and allows $S$ to be in equilibrium at high temperature.
	
	If there remains any population of $S$ that has not decayed to $\chi$ pairs, \eg because it becomes kinematically forbidden, can be assumed to decay to some other light particle in a dark sector, such that their relic abundance is suppressed by their mass ratio. Note that we can assume that these particles are well below the $\keV$ scale if they comprise less than $10\%$ of the total DM relic abundance~\cite{DEramo:2020gpr}.
	In the above scenario $S$ does not develop a vacuum expectation value. One can also consider an interaction term of the form
	\begin{equation}
		\mathcal{L}_{SHH}= A \ S H^{\dagger}H \;,
		\label{eq:VA}	
	\end{equation}
	which introduces a perturbation in the potential (with $A\ll \mSV, m_H$), such that $\vev{S} \sim A$, which induces a mixing with the Higgs boson. In this case, a suppressed $A$ (\eg  $A \sim 10^{-9}~\GeV$) is sufficient for $S$ to decay to SM particles without violating any phenomenological bounds including BBN~\cite{Fradette:2017sdd} due to the negligible mixing angle between $S$ and the Higgs (\eg $\theta \sim 10^{-11}$ for $\mSV \sim 100~\GeV$).

	\subsection*{Boltzmann Equations}
	The Boltzmann equations (BEs) that  describe the system under study are given by~\footnote{In order to solve the system of the BEs, we employ {\tt NaBBODES}~\cite{NaBBODES} and {\tt BB\_VEGAS}~\cite{BB_VEGAS}. We  also perform some of the calculations using {\tt scipy}~\cite{scipy}, and find a good agreement between the two methods.}
	\begin{subequations}
		\begin{align}
			\dfrac{d s}{dt}&=-3 \, H \, s + \dfrac{\GammaPhi}{T} \, \rho_{\Phi} - \dfrac{  \dot{Q}_{DM} }{T} \;,  \label{eq:BE_R} \\
			\dfrac{d \nDM}{dt}&=-3 \, H \, \nDM + 2 \mST  \, \GammaDM \,  n_S^{(-1)}  + C_{22}(T) \label{eq:BE_DM}  \;,\\
			\dfrac{d \rho_{\Phi}}{dt}&=-c  \,  H  \,  \rho_{\Phi} -\GammaPhi \, \rho_{\Phi}  \label{eq:BE_Phi} 
		\end{align}\label{eq:BEs}%
	\end{subequations}
	where $s$ denotes the entropy density of the plasma, $\nDM$ the DM number density,  and $\rhoPhi$ the energy density of $\Phi$. Moreover, $\GammaPhi$ determines the energy transfer rate from $\Phi$ to the plasma -- see above -- $\GammaDM$ denotes the decay rate of $S \to \chi \chi$, and $C_{22}$ the contribution of $SS \to \chi \chi$ to the  DM production rate.  The last term in \eqs{eq:BE_R} describes the energy transfer from the plasma to the DM, due to decays and pair annihilations of $S$, which is assumed to be negligible  as long as $\chi$ remains far away from equilibrium, which is one of the key assumptions behind the freeze-in relic production.
	
	Assuming for simplicity that the decay products of $\Phi$ thermalize instantly,~\footnote{In reality, thermalization lasts for some time and \eqs{eq:BE_R} needs to be modified~\cite{Harigaya:2014waa,Mukaida:2015ria,Harigaya:2019tzu} but such a detailed analysis is beyond the scope of this work.} the entropy density of radiation is given by the familiar formula
	\begin{equation}
		s=\dfrac{2 \pi^2}{45} h_{\rm eff} T^3 
		\label{eq:s_def}
	\end{equation}
	with $h_{\rm eff} $ denoting the so-called relativistic internal degrees of freedom.~\footnote{In our work we use the data provided by~\cite{Saikawa:2020swg}, but we should note that there are differences in the literature (\eg~\cite{Drees:2015exa}), especially close to the QCD phase transition.} 
	
	Without loss of generality, we assume that $S$ develops a thermal mass of the form 
	\begin{equation}
		\mST^2 = \mSV^2 +  \alpha^2 T^2 
		\label{eq:msT}
	\end{equation}
	where $\mSV$ stands for the mass of the scalar in the vacuum and $\alpha$ is some (small) constant. This results in a temperature-dependent decay width
	\begin{equation}
		\GammaDM = \dfrac{\yDM^2 }{16 \pi} \ \mST  \lrsb{ 1 - \lrb{ \dfrac{2 \mDM}{ \mST } }^2 }^{3/2} \;.
		\label{eq:GammaDM}
	\end{equation}
	Furthermore, the moment $n_S^{(-1)}$ is defined as
	\begin{equation}
		n_S^{(-1)} = \int \dfrac{d^3 \vec p}{(2 \pi)^3} \dfrac{1}{E}\dfrac{1}{e^{E/T} - 1}  = \dfrac{T^2}{2\pi^2}  \int_{x}^{\infty} dw \dfrac{\sqrt{w^2 - x^2}}{e^{w} - 1}  \; , 
		\label{eq:n1s}
	\end{equation}
	where $x = \mST/T$.

	In addition, following the procedure delineated in refs.~\cite{Lebedev:2019ton,DeRomeri:2020wng}, the contribution from $S S \to \chi \chi$ to the DM production rate can be written as
	\begin{equation}
		C_{22} = \dfrac{ T^4 }{4\pi^4} \dint_{w_{\rm min}}^{\infty} dw \dint_{1}^{\infty} dt \ \lrb{ \hat s \, \sigma v_{\rm rel} }
		\ \dfrac{w^2}{e^{2wt} - 1} \
		\log{ \dfrac {\sinh{\frac{1}{2} \lrb{ wt + \sqrt{ w^2 + x^2  } \sqrt{t^2 - 1}  }  } }{  \sinh{\frac{1}{2} \lrb{ wt - \sqrt{ w^2 + x^2  } \sqrt{t^2 - 1}  }  } }      } \; , 
		\label{eq:C22}
	\end{equation}
	with 
	$w = \hat{s}/4T$, where $\hat s$ is the center-of-mass energy squared, $w_{\rm min} = Max( \mSV, \mDM )/T$, $x = \mST/T$, and 
	\begin{align}
		\lrb{ \hat s \, \sigma v_{\rm rel} } =& \dfrac{\yDM^4}{\pi}  \sqrt{ 1- \dfrac{4\mDM^2}{\hat s}  } 
		\lrBiggsb{ -1- \dfrac{1}{2} \dfrac{ \lrb{  \mST^2 - 4\mDM^2 }^2  }{  \mST^4 + \hat s \; \mDM^2  -4 \mST^2 \mDM^2 }   	\label{eq:csx} \\
			+ & \dfrac{ \hat{s}^2 - 4 \hat s \lrb{  \mST^2 - 4\mDM^2 } -16 \mDM^2 \mST^2 -32\mDM^4 +6 \mST^4 }
			{ \lrb{ \hat s -2 \mST^2 } \sqrt{ \lrb{ \hat s -4 \mST^2 } \lrb{ \hat s -4 \mDM^2 } } }
			\tanh^{-1} \dfrac{\sqrt{\lrb{ \hat s -4 \mST^2 } \lrb{ \hat s -4 \mDM^2 }}}{ \hat s -2 \mST^2 }
		} \;. \nonumber
	\end{align}

	As already mentioned, the assumption that $\chi$ is a frozen-in relic implies that back reactions (\eg $\chi \chi \to S S$), as well as the energy transfer from $S$ to DM, are negligible. Therefore, \eqs{eq:BE_DM} becomes decoupled from the other equations in~(\ref{eq:BEs}) and the plasma-$\Phi$ system can be treated as a background in which DM evolves. In the following Sections we make use of this observation to simplify the analysis.

	\section{The radiation-$\Phi$ system}\label{sec:RhoPhi}
\begin{figure}[t!]
		\centering	\includegraphics[width=0.8\textwidth]{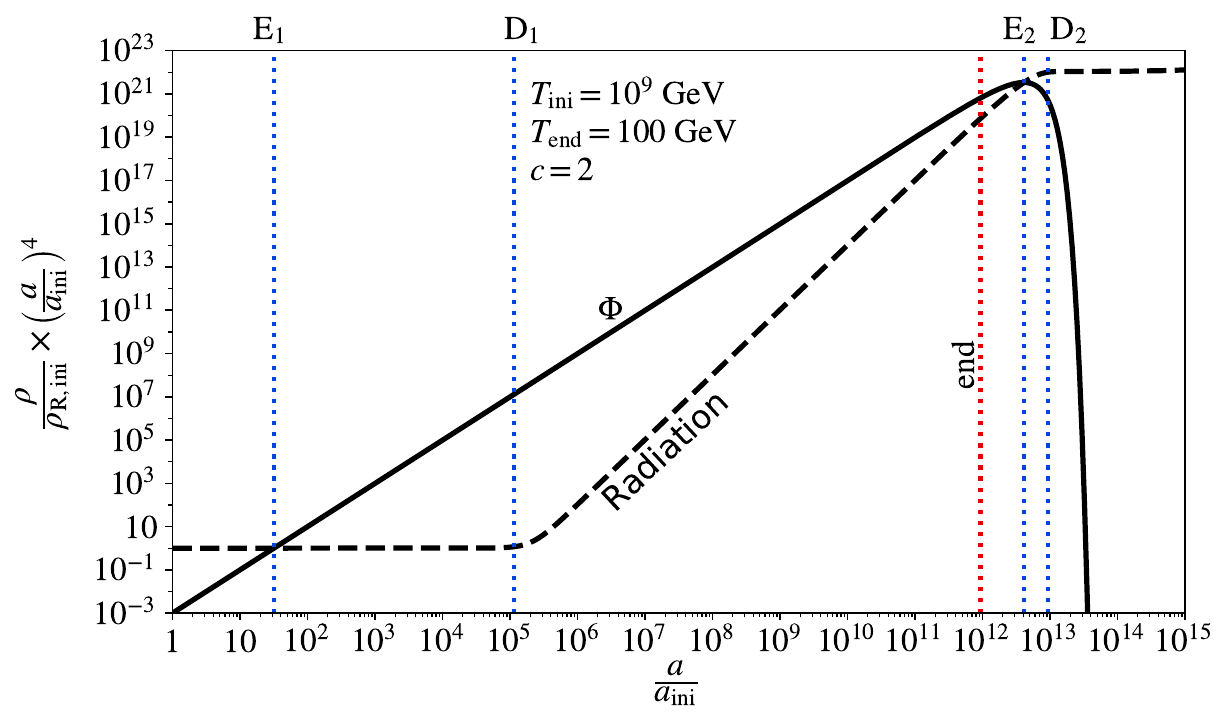}
		\caption{The evolution of the comoving energy densities of radiation and $\Phi$ for $\TEND = 100 ~\GeV$, $c=2$, $\Ti = 10^9 ~\GeV$ and $r=10^{-3}$. The dotted blue vertical lines correspond to the different
			points of reference discussed in the text. Note that the temperature at point $\DII$ (denoted $\TDII$) is expected to be
			lower than $\TEND$ since the Hubble parameter deviates from its value in the standard cosmological scenario. }
		\label{fig:rhoR_rhoPhi}
	\end{figure}

	\setcounter{equation}{0}
	In this Section we examine quantitatively the evolution of the energy densities of radiation and $\Phi$. To begin with, we show in \Figs{fig:rhoR_rhoPhi} a numerical solution for a particular choice of parameters. 
	The figure represents a typical behavior of the system. We observe that starting (at $T=\Ti$) with $\rhoPhi \ll \rhoR$, and denoting with $a$ the scale factor, the comoving energy density $\rhoR \times a^4$ remains constant while  $\rhoPhi \times a^4$ increases (since $c<4$). 
	At some point (denoted $\EI$) these two quantities become equal, but the comoving energy density of the plasma remains constant until it starts to increase. This is the point where a contribution from the decays of $\Phi$ starts affecting the evolution (denoted $\DI$) of $\rhoR$. 
	From this point on, the comoving energy density of both components increase until the decays of $\Phi$ start affecting the evolution of $\rhoPhi$ as well, and $\rhoPhi \times a^4$ begins to slow down. 
	As $\Phi$ continues to decay, $\rhoR \times a^4$ and $\rhoPhi\times a^4$ become equal again at point $\EII$. 
	Then, for a small period of time the comoving energy density of radiation continues to increase until it stops at point $\DII$ when $\Phi$ has effectively decayed away. 
	Beyond this point, $\rhoR \sim a^{-4}$ while $\Phi$ continues to decay exponentially. 
	While a detailed description of an approximate solution to the system of \eqs{eq:BE_R,eq:BE_Phi} is given in Appendix~\ref{app:approx}, here we summarize the analytical approximations in order to quantify the main features of the evolution of the system as described above.
	
	%
	
	\subsection*{Evolution of the energy densities}
	
	Initially, at $T=\Ti$ (corresponding to $a=\ai$), the Universe is dominated by radiation. In addition, the ratio of the energy densities of the $\Phi$-field and radiation is
	\begin{equation}
		r \equiv \dfrac{\rho_{\Phi, {\rm ini} }}{\rho_{R , {\rm ini}}} \ll 1 \; ,
		\label{eq:r_def}
	\end{equation}
	with  the decays of $\Phi$  being subdominant compared to the dilution terms of \eqs{eq:BE_R,eq:BE_Phi} (\ie $\GammaPhi \ll   H(\Ti)$). This means that the energy (and entropy) of the plasma is initially constant (unless there are annihilations happening in the plasma, which happens for $T \lesssim 10~\GeV$), and the energy densities evolve as free-falling fluids, \ie   
	\begin{align}
		\rhoR  \sim a^{-4} \; , \quad \rhoPhi  \sim a^{-c} \;.
		\label{eq:init_RD}
	\end{align} 
	It is worth pointing out that, since $c<4$ an initial condition ($\Ti, r$) is equivalent to another one with ($\Ti^{\prime} ,r^{\prime}$), as long as $r, r^{\prime} \ll 1$ and $\Ti, \Ti^{\prime} \gg \TDII$.
	From \eqs{eq:r_def,eq:init_RD}, we find that this equivalence is expressed as 
	\begin{equation}
		\dfrac{r}{S(\Ti)^{(c-4)/3}} = \dfrac{r^{\prime}}{S(\Ti^{\prime})^{(c-4)/3}} \, ,
		\label{eq:init_curve}
	\end{equation}
	where $S(T)$ is the comoving entropy density of the plasma at temperature $T$. Therefore, without much loss of generality, we can fix $r = 10^{-3}$ for the rest of the analysis.
	
	As both components of the system undergo a free-fall (with $\Phi$ more slowly than radiation), their energy densities become equal at
	\begin{equation}
		a_{\EI} = r^{\frac{1}{c-4}} \ai \; ,
		\label{eq:rel_aEI}
	\end{equation}
	where we note again that $r<1$ implies $c<4$. 
	
	At some time the entropy injection to the plasma due to $\Phi$ decays becomes  large enough to start affecting the evolution of the energy density $\rhoR$. The definition of $\DI$ is somewhat ambiguous since $\Phi$ continues to decay as long as $\rhoPhi \neq 0$. For definiteness, we define $\DI$ as the point where the energy injection rate surpasses $10\%$ of the dilution rate of $\rhoR$; \ie when  $(\GammaPhi/ H) ( \rhoPhi/\rhoR) \Big|_{a = a_{\DI}} = 4/10$, which can be determined numerically.
	
	After significant energy injection begins around $\DI$, the evolution of $\rhoR$ starts to deviate from its free-falling behavior and can be expressed approximately as
	\begin{equation}
		\rhoR \approx \rho_{\EI}   \lrsb{ \lrb{\dfrac{a_{\EI} }{a}}^{4 }  + \dfrac{2 \GammaPhi }{(8-c) H_{\EI}} 
			\lrb{\dfrac{a_{\EI}}{a}}^{ c/2 } -\dfrac{1}{c} \lrb{ \dfrac{\GammaPhi }{2 H_{\EI} } }^2     } \; ,
		\label{eq:Approx-rhoR_Phi-domination}
	\end{equation}
	where the subscript $\EI$ denotes the corresponding quantity obtained at the time of the first radiation-$\Phi$ equality. We note that the energy injection to the plasma causes its comoving energy density to increase, as the dominant (second) term of \eqs{eq:Approx-rhoR_Phi-domination} scales as $a^{ 4 - c/2 }$. Moreover, since $c<4$ the energy density of the plasma continues to fall. 
	
	The evolution of the energy density  of $\Phi$ during the same period can be approximated as
	\begin{equation}
		\rhoPhi  \approx \rho_{\EI} \lrsb{  \lrb{\dfrac{a_{\EI}}{a}}^{ c} - \dfrac{2\GammaPhi}{c \; H_{\EI}}  \lrb{\dfrac{a_{\EI}}{a}}^{c/2} + \lrb{\dfrac{\GammaPhi}{c \; H_{\EI}}}^2  }  \; .
		\label{eq:Approx-rhoPhi_Phi-domination}
	\end{equation}
	As expected, the decay of $\Phi$ tends to cause its energy density to decrease faster than before. At some point the second term of \eqs{eq:Approx-rhoPhi_Phi-domination} becomes important and $\rhoPhi$ starts to decrease much faster, leading to the a second equality point $\EII$ where $\rhoR = \rhoPhi$. After this point the Universe is again dominated by the energy density of the plasma. However, for some time after $\EII$ the energy of the plasma continues to increase until $\Phi$ has basically decayed away at $\DII$. 
	During this period, the approximate form of the energy densities is 
	\begin{align}
		\rhoR &\approx \rho_{\EII} \; \lrb{ \dfrac{a_{\EII}}{a} }^{4}
		\lrsb{ 1 -  \dfrac{\GammaPhi}{ \GammaPhi + H_{\EII}(c-6) } 
			\lrb{ \lrb{\dfrac{a_{\EII}}{a}}^{ c - 6 + \frac{ \GammaPhi}{ H_{\EII} } } -1} } \label{eq:Approx-rhoR_Final_RD} \; , \\
		\rhoPhi &\approx  \rho_{\EII}  \; \lrb{\dfrac{a_{\EII}}{a}}^{c +\frac{\GammaPhi}{2H_{\EII}} \lrsb{1- \lrb{ \frac{a}{a_{\EII} } }^2  }  }    \label{eq:Approx-rhoPhi_Final_RD} \;.
	\end{align}
	Similarly to the definition of $\DI$, we define $\DII$ as the point where the energy injection rate becomes less than $10 \%$ of the dilution rate of $\rhoR$.  We should point-out that $\TDII$ and $\TEND$ appear to be of the same order. 
	Although $\TDII$ depends on several parameters, numerically we find that  $\TDII \approx \TEND/3$ as long as $c \gtrsim 0.5$. This also implies that the BBN constraint can be translated roughly into $\TEND \gtrsim 30~\MeV$.

	Finally, for $T<\TDII$ the plasma expands freely (\ie $\rhoR \sim a^{-4}$), while  $\Phi$ continues to decay exponentially, with
	\begin{equation}
		\rhoPhi \approx \rho_{\Phi , \DII}  \lrb{\dfrac{a_{\DII}}{a}}^c \; e^{- \frac{1}{2}\frac{\GammaPhi}{H_{\DII}} \lrsb{ \lrb{ \frac{a}{a_{\DII}} }^{2} -1  }  }    \;. 
		\label{eq:Approx-rhoPhi-Free_RD} 
	\end{equation}
	%

	\subsection*{Effects on $H$ and Entropy Injection}
\begin{figure}[t!]
		\centering 
		\begin{subfigure}[b]{0.5\textwidth}
			\centering\includegraphics[width=1\textwidth ]{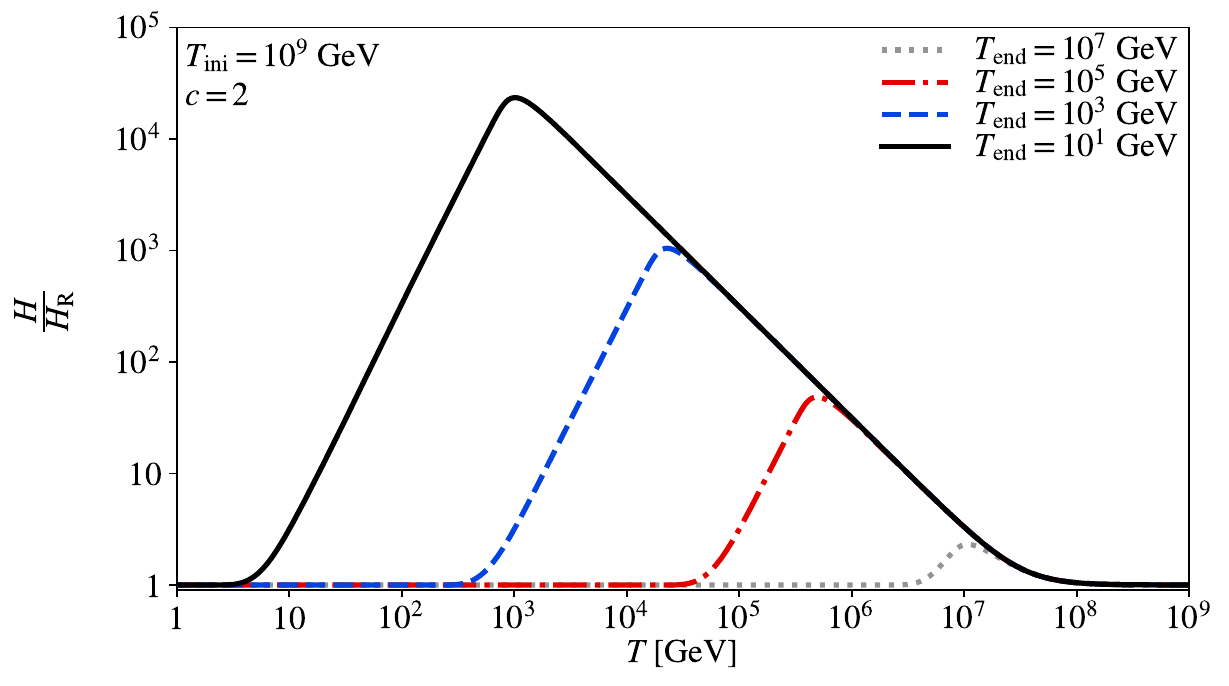}
			\caption{\label{fig:rel_H_TEND}}
		\end{subfigure}%
		\begin{subfigure}[b]{0.5\textwidth}
			\centering\includegraphics[width=1\textwidth]{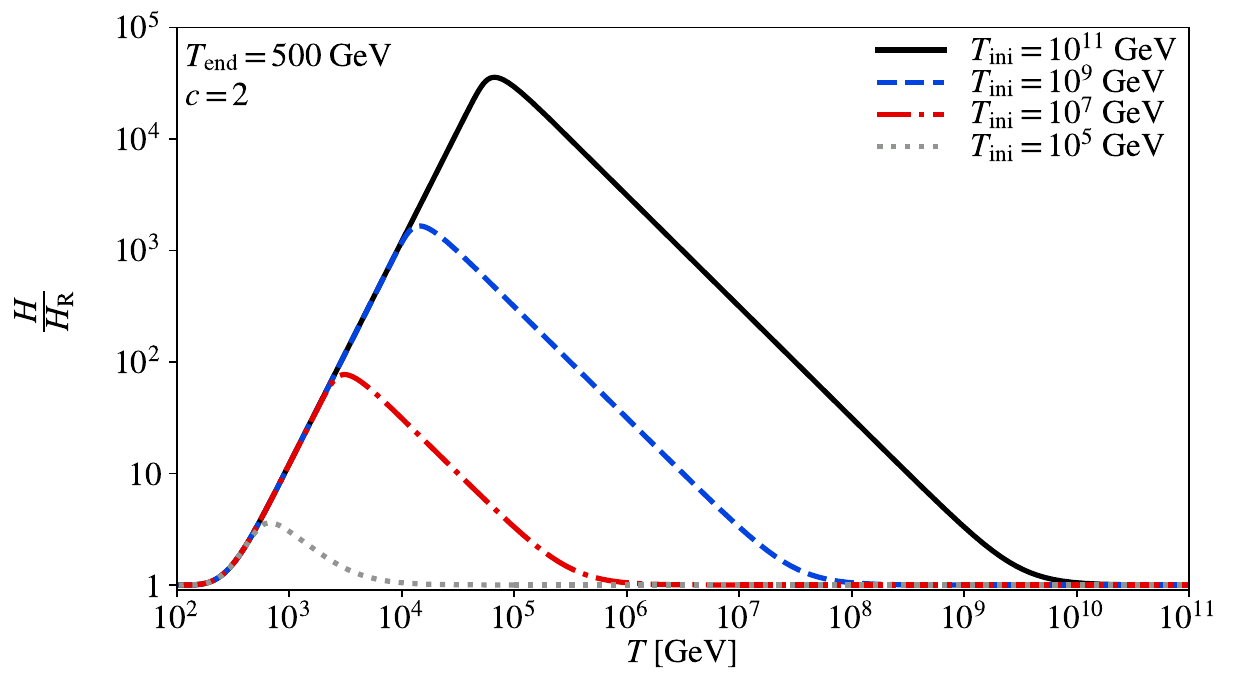}
			\caption{\label{fig:rel_H_Ti}}
		\end{subfigure}
		\begin{subfigure}[b]{0.5\textwidth}
			\centering\includegraphics[width=1\textwidth]{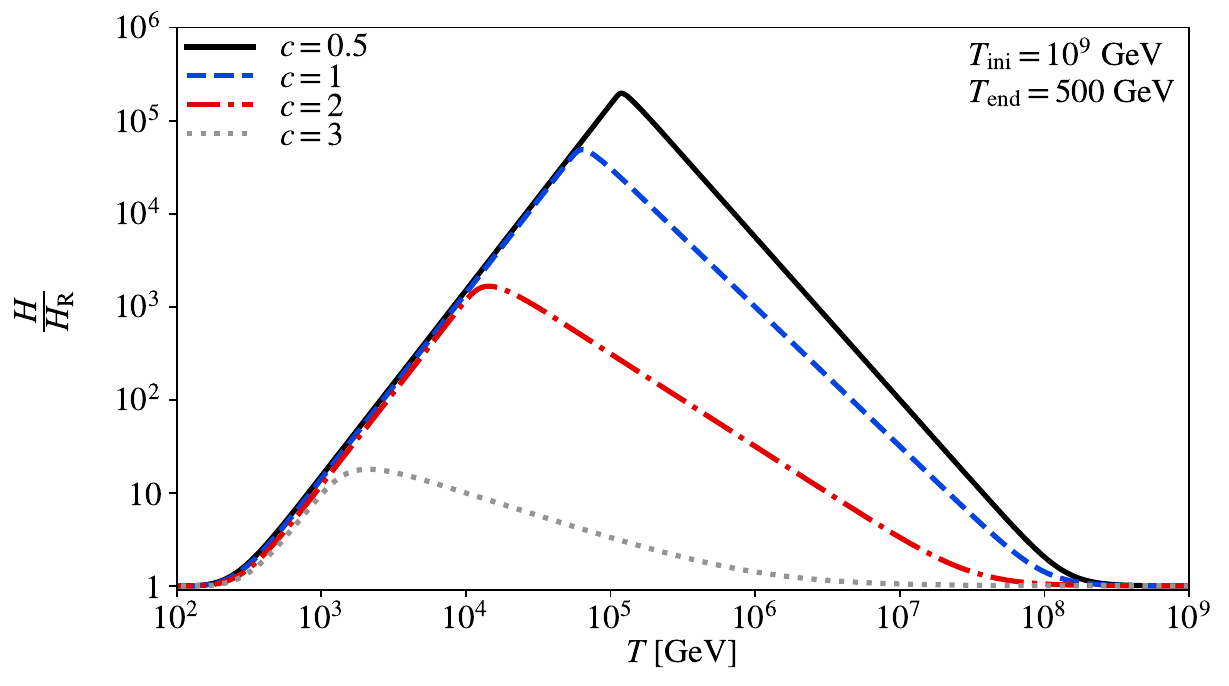}
			\caption{\label{fig:rel_H_c}}
		\end{subfigure}
		\caption{The evolution of $H/H_{\rm R}$ as a function of temperature for different values of $\TEND$ (a), $\Ti$ (b), and $c$ (c).
			The fixed values correspond to $\TEND = 500~\GeV$, $\Ti = 10^9 ~\GeV$, and $c=2$. We note that all the parameters can have similar effects since the Hubble parameter takes similar values at similar temperatures for different choices of their values.}
		\label{fig:rel_H}
	\end{figure}

	During the period of $\Phi$ dominance the Universe expands faster than it would during a radiation-dominated era at the same temperature. 
	Therefore, smaller values of $\TEND$ should result in a larger value of the Hubble parameter since the difference between $\rhoPhi$ and $\rhoR$ increases as the Universe expands and the temperature decreases. This effect can be seen in \Figs{fig:rel_H_TEND} where we show the ratio of the Hubble parameter over its corresponding value for a radiation-dominated expansion ($H_{\rm R}$) as a function of temperature for several choices of $\TEND$ and for $c=2$, $\Ti=10^9~\GeV$ and $r=10^{-3}$. We note that, for $T$  close to $\Ti$ the Universe's expansion rate approaches its radiation-dominated rate ($H/H_{\rm R} =1$), while for $T$ away from $\Ti$ $H$ deviates from $H_{\rm R}$ by up to a few orders of magnitude. Since $T_{\EII}$ is close to $\TDII$, $H$ and $H_{\rm R}$ become equal again when the temperature drops somewhat below $\TEND$, which is of the same order as $\TDII$.  
	
	Since a longer period of $\Phi$ domination results in a faster expansion rate of the Universe, taking a larger value of $\Ti$ (with all other parameters fixed) should have the same effect as selecting a lower value of $\TEND$. This can be seen in \Figs{fig:rel_H_Ti} which shows how the ratio $H/H_{\rm R}$ evolves with the temperature for different $\Ti$ and for $\TEND=500~\GeV$ and $c=2$. Since taking larger values of $\Ti$ has, by definition, the effect of pushing the initial values of the energy densities to higher temperatures, the ratio $H/H_{\rm R}$ approaches $1$ at higher temperatures, depending on $\Ti$. Once $\Phi$ starts to dominate,  $H$ increases relative to $H_{\rm R}$  until $\Phi$ has decayed away and the Universe becomes radiation-dominated again. It is also worth noting  that this happens at a temperature that is (almost) independent of  $\Ti$, \ie the initial condition has a negligible effect on $T_{\EII}$ (and also $\TDII$).  

\begin{figure}[t!]
		\centering 
		\begin{subfigure}[b]{0.5\textwidth}
			\centering\includegraphics[width=1\textwidth ]{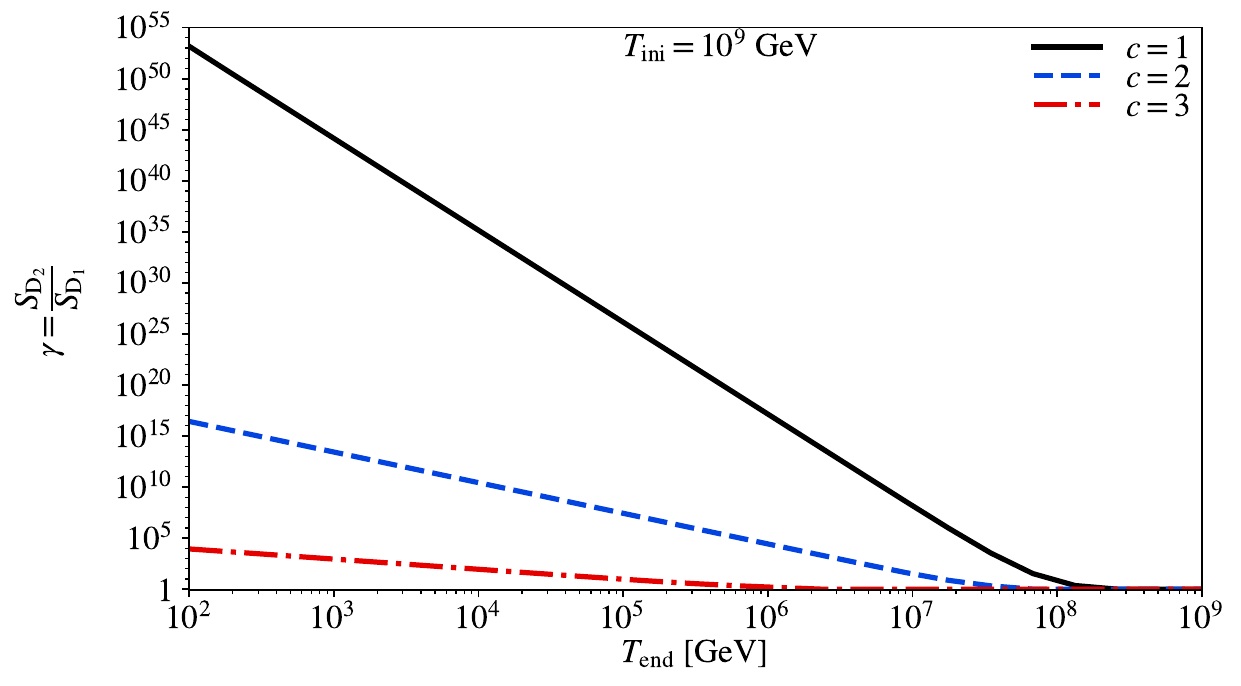}
			\caption{\label{fig:gamma_TEND}}
		\end{subfigure}%
		\begin{subfigure}[b]{0.5\textwidth}
			\centering\includegraphics[width=1\textwidth]{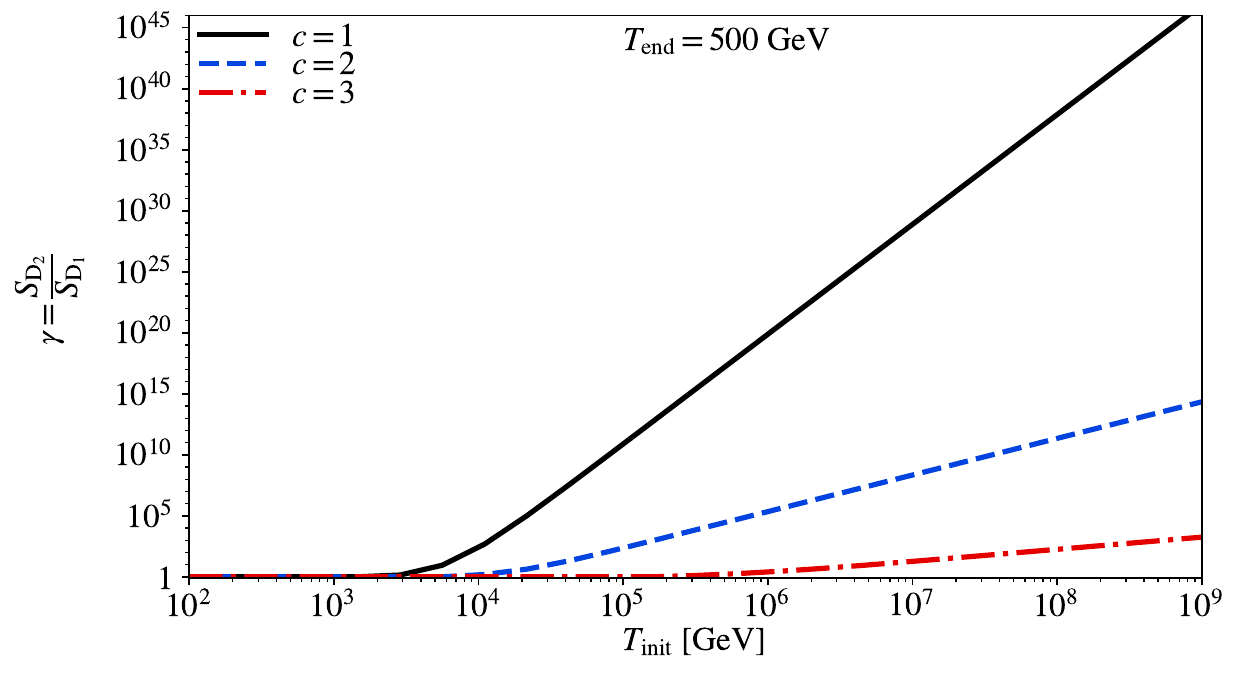}
			\caption{\label{fig:gamma_Ti}}
		\end{subfigure}
		\caption{The amount of entropy injection, $\gamma=\dfrac{S_{\DII}}{S_{\DI}}$, as a function of $\TEND$ (a)  and $\Ti$ (b), for $c=1,\,2,\,3$ (the fixed values are $\TEND = 500~\GeV$ and $\Ti = 10^9 ~\GeV$). Notice that the choice of parameters affect significantly $\gamma$. Furthermore, again there are different choices of the parameters that can result in a similar increase in entropy.}
		\label{fig:gamma}
	\end{figure}

	Moreover, small values of $c$ result in a $\rhoPhi$ that dominates over $\rhoR$ much faster and for a longer period of time compared to larger values of $c$. This is because, since we assume that $r \ll 1$, there are values of $c$ for which $\Phi$ does not dominate at all, \ie the  expansion of the Universe becomes faster as $c$ decreases. This effect is shown in \Figs{fig:rel_H_c} which shows the ratio $H/H_{\rm R}$ as a function of $T$ for several values of $c$ and for $\TEND = 500 ~\GeV$ and $\Ti = 10^9 ~\GeV$. Since for the values of $c$ that we consider here $\TEND$ and $\TDII$ (and also $T_{\EII}$) are close to each other, the expansion rate of the Universe starts being again dominated by radiation as $T$ drops below $\TEND$, \ie $H/H_{\rm R} \approx 1$ at $T \lesssim \TEND$. It is worth pointing out that the expansion rate is actually sensitive to $c$ since this parameter  quite strongly influences the range of temperatures for the period during which $\Phi$ dominates.
	A period of $\Phi$ dominance cannot be distinguished from a period of entropy injection to the plasma. This is clear because increasing the dominance of $\Phi$ has the effect of increasing the entropy of the plasma by a greater amount, as $\Phi$ has to decay in order for the Universe to return to a radiation-dominated expansion at least for $T \lesssim \mathcal{O}(10~\MeV)$. According to our discussion of the expansion rate above, taking higher values of $\Ti$ has the same effect as taking lower ones for $\TEND$ and $c$. In other words, the amount of entropy injection, defined as $\gamma \equiv S_{\DII} / S_{\DI}$ becomes larger as $\Ti$ increases or else as $\TEND$ (or $c$) decreases. 
	This behavior can be seen in \Figs{fig:gamma} where we show the dependence of $\gamma$ on $\TEND$ (\Figs{fig:gamma_TEND}) for  $\Ti = 10^{9} ~\GeV$, and on $\Ti$  (\Figs{fig:gamma_Ti}) for $\TEND = 500 ~\GeV$ and for different values of $c$. Notice that all the lines converge to $\gamma = 1$ as $\Ti$ and $\TEND$ come close to each other. On the other hand, the deviation from $\gamma=1$ depends strongly on $c$ which dictates how quickly (if at all) $\Phi$ takes over the expansion of the Universe over radiation. For example, the choice of $\TEND=10^7~\GeV$ and $c=3$ results in $\gamma = 1$ since $\Phi$ remains a subdominant component of the Universe, while taking $c=1$ at the same value of $\TEND$ gives $\gamma \sim 10^{15} $. In other words,  for lower values of $c$ the ratio of $\rhoPhi$ and $\rhoR$ increases more rapidly, until close to $\DII$, than for larger values of $c$.

	\section{Dark Matter production in NSC}\label{sec:DM_production}
	\setcounter{equation}{0}

	It has been shown above that the evolution of the energy densities of radiation and $\Phi$ exhibits a similar qualitative behavior as long as $\Phi$ dominates the expansion of the Universe for some period of time, fairly independently of the specific values of the parameters $\Ti$, $\TEND$, and $c$. However, we expect these parameters to have an important role in  DM production. In order to quantify the role played by each of the different NSC parameters in determining the DM yield, we start by writing down its evolution equation. The DM production is described by \eqs{eq:BE_DM}. Assuming that the DM number density is negligible (compared to its equilibrium value) during its production, this BE can be studied independently of the others.
	As above, we will study the DM evolution starting from $\Ti$. The initial condition for DM can be calculated following~\cite{Hall:2009bx}, or in the forbidden freeze-in case~\cite{Darme:2019wpd}, since at temperatures above $\Ti$ the Universe is assumed to be radiation-dominated.
	
	From the definition of the DM yield, $\YDM \equiv \nDM/s$, the BE for the DM evolution assumes its general form
	\begin{equation}
		\dfrac{d\log\YDM}{d \log \frac{\Ti}{T}} =   \dfrac{d \log\NDM}{d \log \frac{\Ti}{T}}
		\lrsb{ 1 -  \dfrac{d \log S}{d \log \frac{\Ti}{T} } \lrb {\dfrac{d \log \NDM}{d \log \frac{\Ti}{T}} }^{-1} }\;,
		\label{eq:dYdT_with_dNdT}
	\end{equation}
	with $S$ as before denoting the comoving entropy density of the plasma, and $\NDM$ the comoving number density of DM particles. 
	
	As has been shown in Section~\ref{sec:RhoPhi}, between $\EI$ and $\DII$ both the expansion rate of the Universe and the evolution of the plasma can behave significantly differently from what is expected in the standard cosmological scenario. The DM relic abundance is therefore expected to be strongly dependent on the period during which most of the DM particles are produced, which is for temperatures close to its freeze-in temperature. 
	When DM production stops then its yield is proportional to the comoving number of DM particles only when entropy is conserved, which is clear from \eqs{eq:dYdT_with_dNdT}. If, on the other hand, the entropy of the plasma increases after DM production has stopped then obviously the resulting DM yield decreases. Therefore we define the term ``diluted DM" as the situation when the freeze-in temperature is higher than -- or equal to-- the temperature at which entropy injection stops, \ie $\TFI \geq \TDII$.
	
	In the following, we will examine qualitatively the effects of the considered NSC scenario on the different parameters that will define the overall DM evolution and production.

	\subsection{Dark Matter production from $S\rightarrow \chi\chi$}{\label{sec:DM_production_12}}
	
	Production of DM particles is model dependent. In the case studied here, DM production proceeds via decays $S \to \chi \chi$ and pair annihilations $S S\to \chi \chi$. We expect the decays to be dominant in most of the parameter space. This is because the pair annihilation cross-section is proportional to $\yDM^4$, while the decay is proportional to $\yDM^2$. In addition, there are other suppression factors coming from the phase-space integration. We have verified numerically that the pair annihilation dominates only in the case of a large amount of dilution, for sizeable Yukawa coupling ($\yDM \gtrsim 10^{-6}$), and only where the decay proceeds in the forbidden freeze-in zone. Therefore, below we discuss in detail the DM production taking into account only the decay channel for which  \eqs{eq:dYdT_with_dNdT} takes the form        
	\begin{equation}
		s \ \dfrac{d\YDM}{d\log \frac{\Ti}{T}} = \delta_h \ \dfrac{2\GammaDM }{H_{\rm R}} \mST \ n_{s}^{(-1)}\ \dfrac{H_{\rm R}}{H} \ 
		\lrb{ 1-\dfrac{\GammaPhi }{ H } \dfrac{\rhoPhi}{3 \, s \, T} }^{-1} \ \lrb{1-\dfrac{\GammaPhi}{2\GammaDM}  \dfrac{\rhoPhi}{ \mST \ n_{s}^{(-1)} T} \ \YDM } \; ,
		\label{eq:dYdT}
	\end{equation}
	where $\delta_h = 1+ 1/3\, d\log h_{\rm eff} / d\log T$. 
	
	This equation is equivalent to \eqs{eq:BE_DM} with $C_{22} = 0$. However, it is more convenient for our discussion as shows clearly how $\YDM$ deviates from its standard cosmological evolution due to the appearance of the following three extra factors:
	\begin{align}
		&\mathcal{F}_1 =    \dfrac{H_{\rm R}}{H} \; , \nonumber \\
		&\mathcal{F}_2 = \lrb{ 1- \dfrac{\GammaPhi }{ H } \dfrac{\rhoPhi}{ 3\, s \, T} }^{-1} \; , 
		\label{eq:factors} \\
		&\mathcal{F}_3 = \lrb{ 1 -  \dfrac{d \log S}{d \log \frac{\Ti}{T} } \lrb {\dfrac{d \log \NDM}{d \log \frac{\Ti}{T}} }^{-1} } =\lrb{1-\dfrac{\GammaPhi}{2\GammaDM}  \dfrac{\rhoPhi}{ \mST \ n_{s}^{(-1)} T} \ \YDM } \;.
		\nonumber
	\end{align} 
	
	Although analytical treatment of \eqs{eq:dYdT} is very difficult, and in any case would not provide much information due to the complicated nature of the problem, we can still qualitatively see how DM production proceeds by examining how the different factors evolve over the different epochs. In the following we take a closer look on the expressions of $\mathcal F_1, \mathcal F_2$ and $\mathcal F_3$ and their behavior for different cosmological scenarios.
	
	\subsubsection*{Behavior of  $\mathcal{F}_1$}
	The first  factor, $\mathcal{F}_1$, represents the deviation of the expansion rate of the Universe from the radiation-dominated case.  Since the Hubble parameter increases during the period of $\Phi$ dominance  (between $\EI$ and $\EII$), see \Figs{fig:rel_H}, the Universe  expands faster, and the DM production rate is effectively reduced.   Therefore, independently of the values of the parameters,  $\mathcal{F}_1 \ll 1$ as long as $\Phi$ dominates the expansion of the Universe.
	
	\subsubsection*{Behavior of  $\mathcal{F}_2$}
\begin{figure}[t!]
		\centering\includegraphics[width=0.8\textwidth]{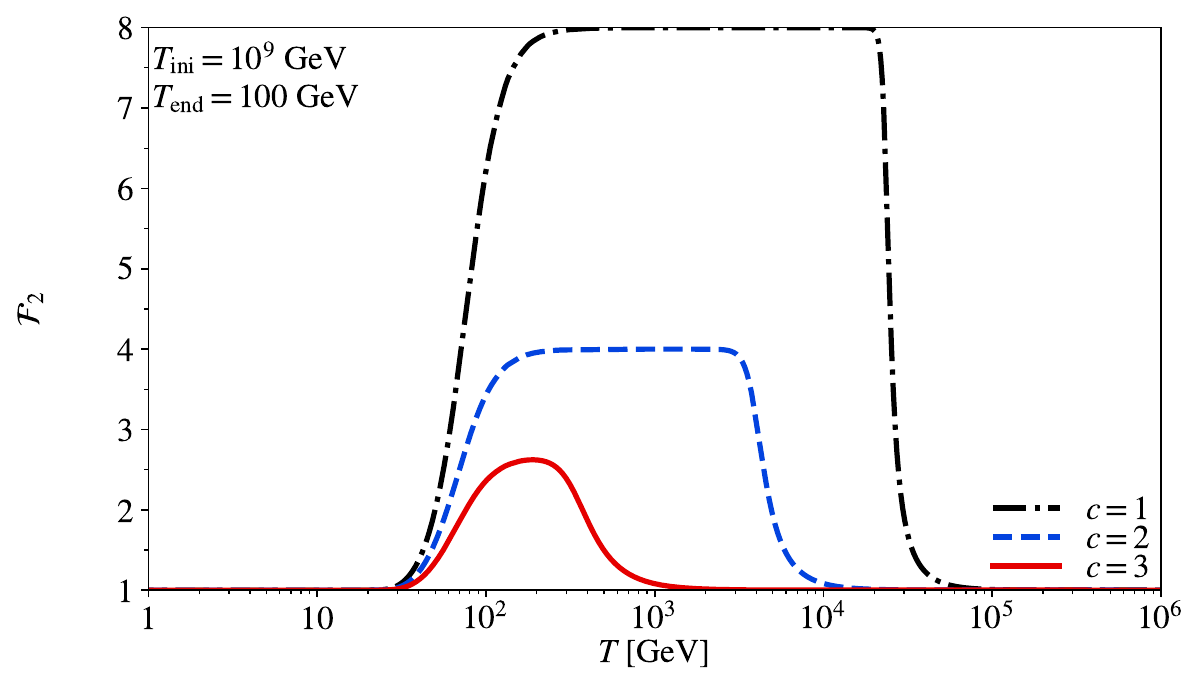}
		\caption{The dependence of $\mathcal{F}_2$ on temperature $T$ for the same parameter choice as in Fig.~\ref{fig:rhoR_rhoPhi}. In particular, this figure corresponds to $c=1,2,3$ where, according to our discussion, $\mathcal{F}_2 \lesssim 8, 4,  8/3 $, respectively.}
		\label{fig:F2}
	\end{figure}

	Energy injection to the plasma tends to increase the DM production rate since the factor $\mathcal{F}_2$ is expected to increase between $\DI$ and $\DII$. We can examine the behavior of this factor approximately in the case where the energy injection is active at high temperatures, with $ 4 \, \rhoR \approx 3 \, s \, T$. We further assume that this term is maximized while $\Phi$ is still dominant because its energy density decreases rapidly after $\EII$. Thus, taking the time derivative of $\mathcal{F}_2$ with $H \sim \sqrt{\rhoPhi} \gg \GammaPhi $ and $\rhoPhi \gg \rhoR$, we find that its maximum  is $8/c$, \ie
	\begin{equation}
		\lrb{1- \dfrac{\GammaPhi}{H}\dfrac{\rhoPhi}{3 \, s \,T}}^{-1} \lesssim \dfrac{8}{c} \; .
		\label{eq:2nd_term_MAX}
	\end{equation}
	Outside the period of energy injection $\mathcal{F}_2$  rapidly decreases to one. This can be seen in \Figs{fig:F2} where we show how this term reaches its maximum for different values of $c$. Notice that the approximation of \eqs{eq:2nd_term_MAX} is very accurate as long as  $ 4 \, \rhoR \approx 3 \, s \, T$ between $\DI$ and $\DII$. Beyond this approximation, the behavior is expected to be similar, however, a numerical treatment is necessary.

	\subsubsection*{Behavior of  $\mathcal{F}_3$}
	The factor $\mathcal{F}_3$ is  responsible for the dilution of DM if the production rate is small.~\footnote{This means that the factor  does not affect the number of DM particles produced but only $\nDM$ relative to the plasma.} This is clear if we consider the evolution of $\YDM$ after the DM production ends, \ie $\NDM = {\rm const}$, which in this case becomes
	\begin{equation}
		\dfrac{d\log \YDM}{d \log \frac{\Ti}{T}} =  -  \dfrac{d \log S}{d \log \frac{\Ti}{T} }  \;.
		\label{eq:dYdT_diluted}
	\end{equation}
	That is, away from the period of DM production, and since the entropy cannot decrease, $\YDM$ either decreases or remains  constant and proportional to $\NDM$. Thus, in  the presence of a decaying fluid, if the DM production stops before $\DII$ we expect $\YDM$ to decrease. This scenario corresponds to the case of diluted DM. If DM, however, is produced mostly after $\Phi$ has decayed away, we expect the entropy injection to only slightly affect the DM relic abundance, \ie the DM relic abundance is expected to be almost the same as in the case of purely radiation-dominated Universe.  Therefore, in general, we expect today's value of $\YDM$ to depend strongly on the relative values of  the freeze-in temperature $\TFI$ and $\TDII$. 

\begin{figure}[t!]
		\centering 
		\begin{subfigure}[b]{0.5\textwidth}
			\centering\includegraphics[width=1\textwidth ]{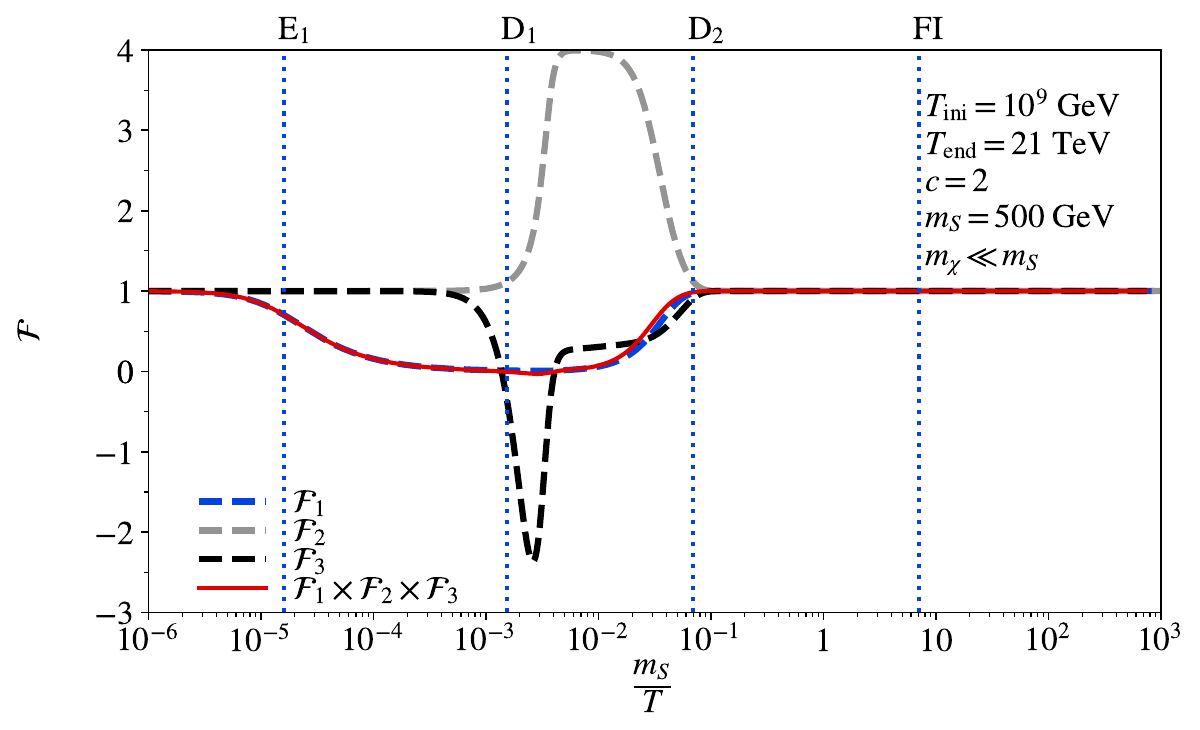}
			\caption{\label{fig:F3_noD_std-FI}}
		\end{subfigure}%
		\begin{subfigure}[b]{0.5\textwidth}
			\centering\includegraphics[width=1\textwidth]{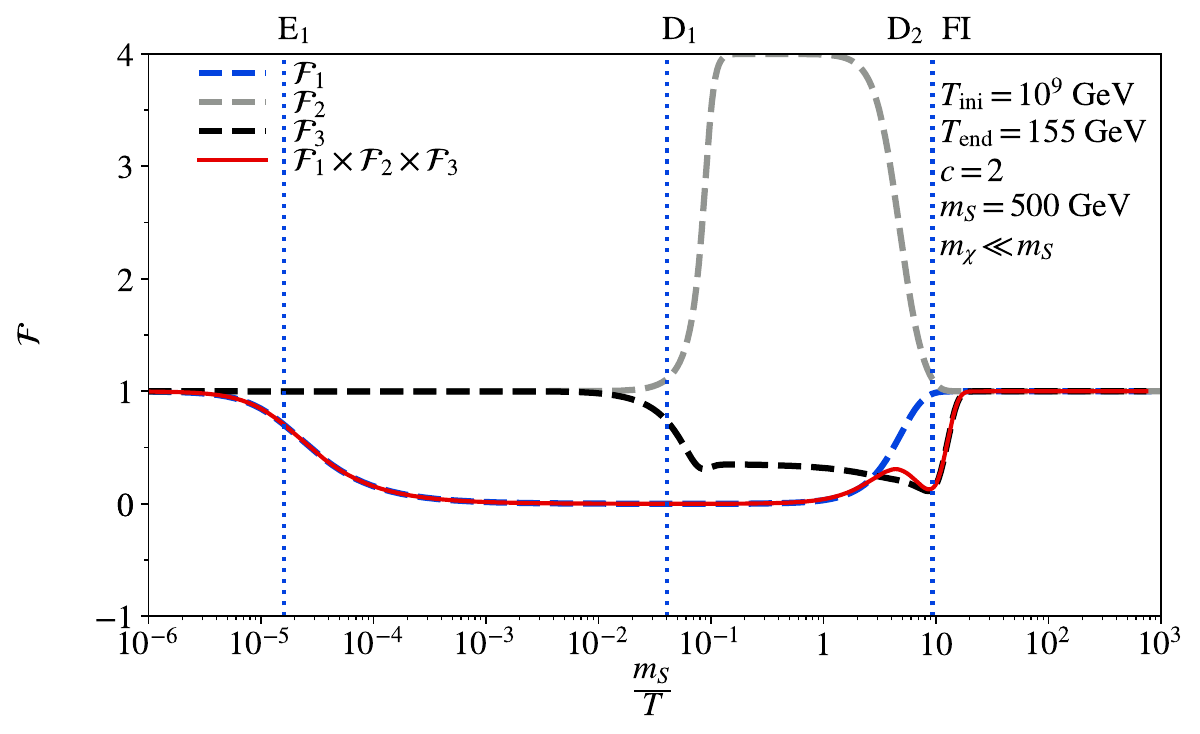}
			\caption{\label{fig:F3_mD_std-FI}}
		\end{subfigure}
		\begin{subfigure}[b]{0.5\textwidth}
			\centering\includegraphics[width=1\textwidth]{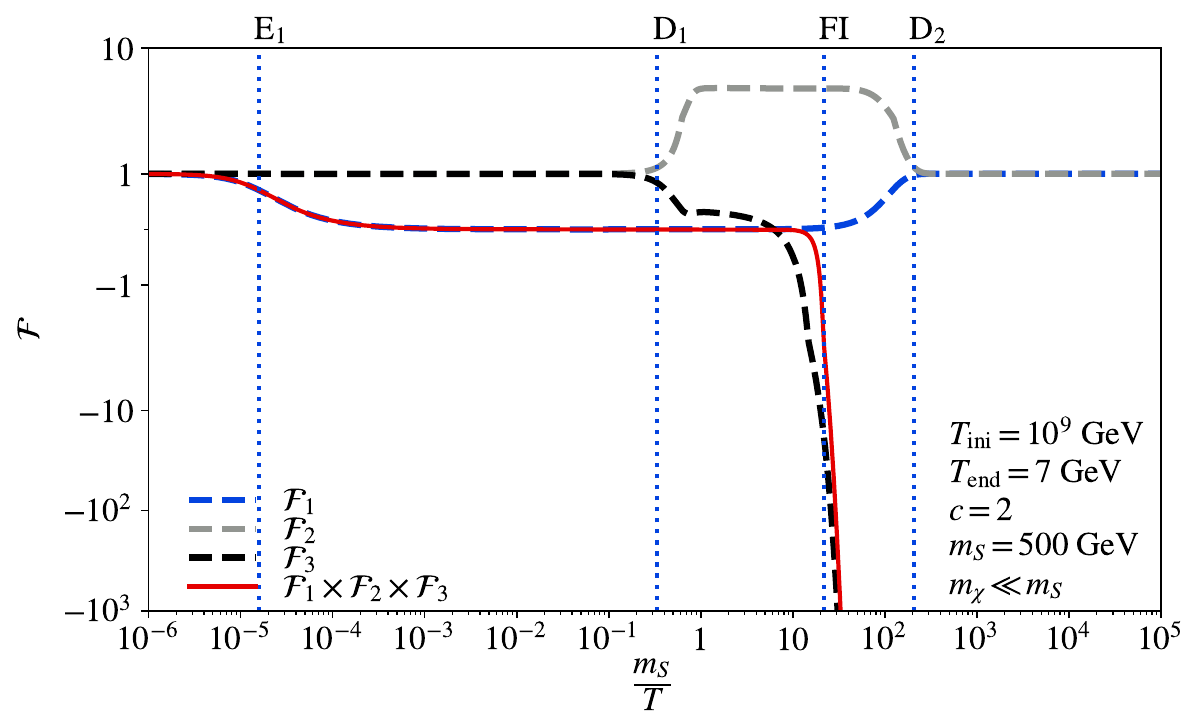}
			\caption{\label{fig:F3_D_std-FI}}
		\end{subfigure}
		\caption{The evolution of $\mathcal{F}_3$ (black line) as a function of the $\mSV/T$ for $\mSV=500~\GeV$, $\mSV \gg 2 \mDM$,  $\Ti =10^9 ~\GeV$, $c=2$. The three panels shown correspond to $\TEND = 21 ~\TeV$ (a), $\TEND =155~\GeV $ (b), $\TEND = 7~\GeV$ (c), with $\TFI = 10^{-2} \ \TDII $, $\TFI = \TDII$, and $\TFI = 10 \ \TDII $, respectively. For comparison, we also show $\mathcal{F}_{1,2}$ (blue dotted and red lines, respectively) along with the product $\mathcal{F}_{1} \times \mathcal{F}_{2} \times \mathcal{F}_{3}$ (gray line). The blue dotted vertical lines  correspond to the points $\EI$, $\DI$, and $\DII$, where the behavior of each term changes. The point $\EII$ is not shown  because there are no visible effects associated with it since it is very close to $\DII$.
			Notice that in all the cases the product of the three factors is smaller than $1$, implying that the relic abundance cannot increase above its standard cosmological history value. The behavior in the case of $\mSV \ll 2\mDM$ is qualitatively similar. }
		\label{fig:F3}
	\end{figure}

	The different kinds of expected behavior of $\mathcal{F}_3$  are presented (black line)  in \Figs{fig:F3}, along with $\mathcal{F}_1$ (blue line) and $\mathcal{F}_2$ (gray line), as well as their product (red line) for comparison. The parameter choice is: $\mSV=500~\GeV$, $\mSV \gg 2 \mDM$,  $\Ti =10^9 ~\GeV$, $c=2$. The three panels correspond to $\TEND = 21 ~\TeV$ (a), $\TEND = 155~\GeV $ (b), $\TEND = 7~\GeV$ (c), with $\TFI = 10^{-2} \ \TDII $, $\TFI = \TDII$, and $\TFI = 10 \ \TDII $, respectively.
	In \Figs{fig:F3_noD_std-FI}, we can see that the product of the three factors is close to zero between $\EI$ and $\DII$, and changes sign at $\DI$ from positive to negative, although it remains very close to zero.
	However, since freeze-in takes place after $\DII$, most of DM is produced when $\mathcal{F}_{1}\times\mathcal{F}_{2}\times\mathcal{F}_{3} = 1$, and we expect the presence of $\Phi$ and its decays to have no effect on $\YDM$ after the freeze-in. 
	In the case with $\TFI \approx \TDII$, shown in \Figs{fig:F3_mD_std-FI}, the effect of the dilution term marginally wins over the other two (close to the freeze-in) and we expect $\YDM$ to be slightly reduced after DM production terminates. 
	The case of $\TFI > \TDII$ when DM production is mostly suppressed is shown in \Figs{fig:F3_D_std-FI}. As explained previously, since the freeze-in happens before $\DII$ the DM production is inactive during some period of entropy injection. In this case the evolution of DM yield is described approximately by \eqs{eq:dYdT_diluted} where the number density of $\chi$ (with $\nDM \sim a^{-3}$) falls faster than the entropy of the plasma, since $s$ deviates from $s \sim a^{-3}$, resulting in an apparently smaller DM relic abundance.
	We expect a similar qualitative behavior also for DM production via classically forbidden decays, \ie when $\mSV \ll 2 \mDM$.

	We conclude this discussion by pointing out that we expect each case  to result in a decreased relic abundance (or $\YDM$) compared to the standard cosmological scenario, so long as $c<4$ and remains nonzero, because $\mathcal{F}_1 \ll 1$ while $\mathcal{F}_2 \lesssim 8/c$. 
	Only the case when $\TFI \approx \TDII$ could result in an increased $\YDM$, since the product of $\mathcal{F}_{1,2,3}$ is increased at $ T \approx \TDII$, but this case corresponds to a  negative contribution from $\mathcal{F}_3$ (see \Figs{fig:F3_mD_std-FI}). Therefore, we expect that the relic abundance in the absence of $\Phi$ and its decays to reach its maximum value interdependently of when  DM freezes-in.
	
	\subsubsection{Overall Dark Matter Production}

	It is clear that the factors $\mathcal{F}_{1,2,3}$ compete with each other in their contribution to the evolution of $\YDM$.  The first factor effectively decreases DM production rate, while $\mathcal{F}_2$ enhances the production of  DM due to the energy injection to the plasma, and tends to increase $\NDM$. The third factor shows by how much DM yield is reduced due to the increased entropy in the plasma.

	For  $T \gtrsim \TDI$ and $T \lesssim \TDII$, $\mathcal{F}_3$ can change sign, \ie $\YDM$ can exhibit local extrema.
	These extrema can be either minima and maxima depending on how $d\;\log \NDM / d\; \log \frac{\Ti}{T}$  changes compared to $ d\; \log S / d\; \log \frac{\Ti}{T}$. 
	For instance, consider a period when DM production is slow while  the entropy is increased. If this period is followed by one with an increased DM production rate then $\YDM$ can exhibit a local minimum. 
	\begin{figure}[t!]
		\centering 
		\begin{subfigure}[b]{0.5\textwidth}
			\centering\includegraphics[width=1\textwidth ]{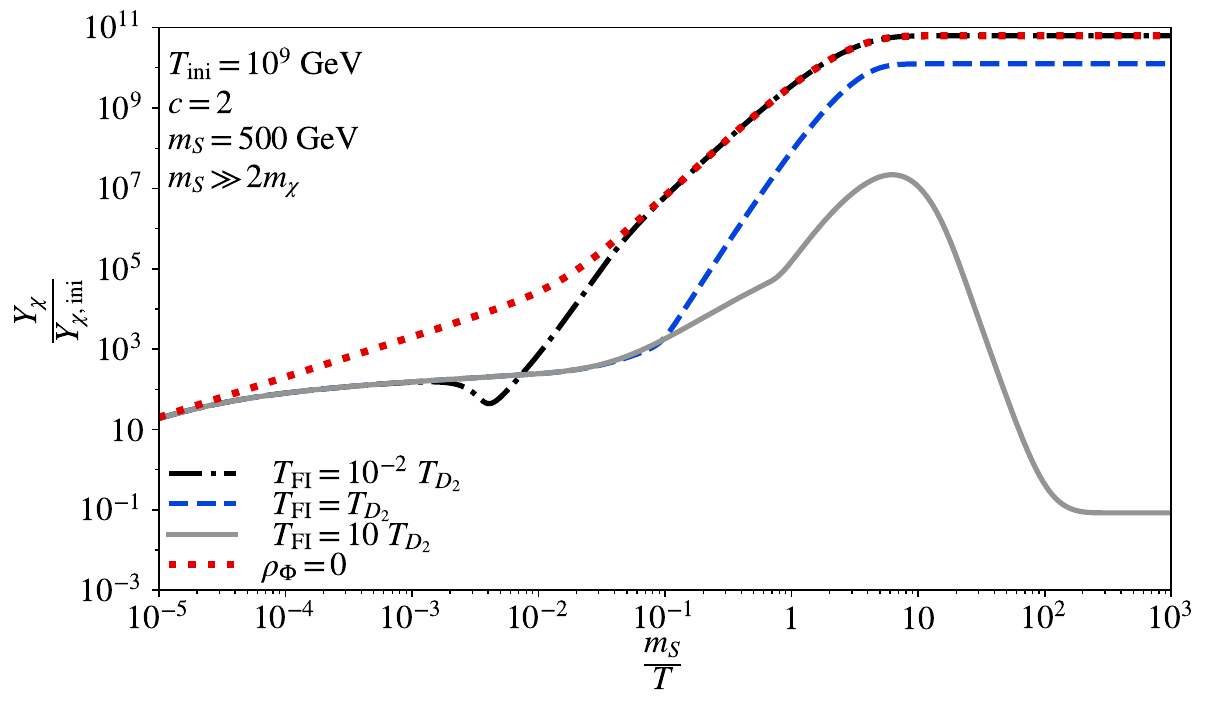}
			\caption{\label{fig:YDM_std-FI}}
		\end{subfigure}%
		\begin{subfigure}[b]{0.5\textwidth}
			\centering\includegraphics[width=1\textwidth]{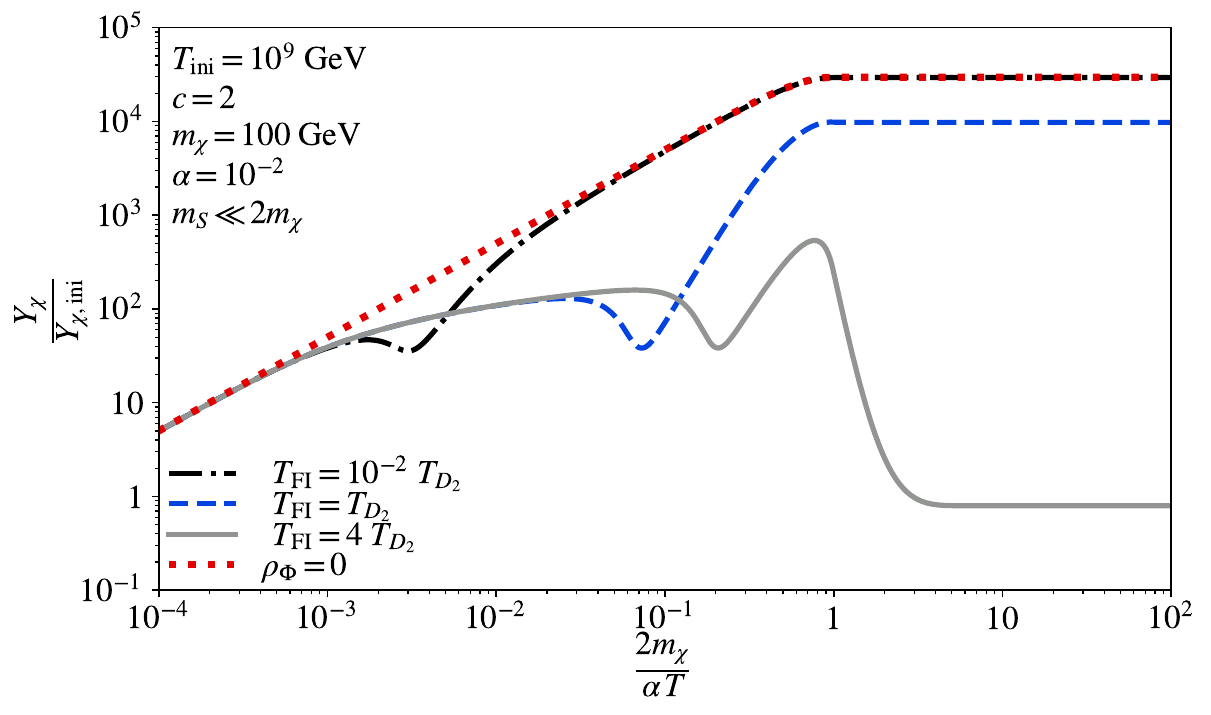}
			\caption{\label{fig:YDM_FFI}}
		\end{subfigure}
		\caption{The evolution of $\YDM$ for different production scenarios. The parameter choice regarding the radiation-$\Phi$ system is the same as in Fig.~\ref{fig:rhoR_rhoPhi}. The left figure shows $\YDM$ as a function of $\mSV/T$ for $\mSV = 500 ~\GeV$,  $\mDM \ll \mSV/2$, and
			$\TFI/T_{\DII} = 10^{-2}, 1, 10$ that correspond to black (with $\TEND = 21 ~\TeV$), blue (with $\TEND = 155~\GeV $), and gray (with $\TEND = 7~\GeV$) lines, respectively. Notice that the black, blue, and gray lines, also correspond to Figs.~\ref{fig:F3_noD_std-FI}, \ref{fig:F3_mD_std-FI}, and \ref{fig:F3_D_std-FI}, respectively.
			The right figure shows the forbidden freeze-in case  with $\mDM= 100 ~\GeV \gg \mSV/2 $, $\alpha=10^{-2}$, and
			$\TFI/T_{\DII} \approx 10^{-2}, 1, 4$ that correspond to black (with $\TEND = 8 \times 10^6 ~\GeV$), blue (with $\TEND = 70~\TeV $), and gray (with $\TEND = 15~\TeV$) lines, respectively.
			In both figures the red line shows the standard cosmological history, \ie $\rhoPhi = 0$. }
		\label{fig:YDM}
	\end{figure}
	The production of DM is determined by the combination of all the effects described above, and it would be helpful to show some numerical examples in order to see how $\YDM$ changes over time. 
	In \Figs{fig:YDM_std-FI} we show  $\YDM$ over its value at $T=\Ti$ as a function of $\mSV/T$ for $\mSV \gg 2 \mDM$ and for several choices of $\TFI/T_{\DII}$ that are obtained by varying $\TEND$ and the same choice of parameters as in \Figs{fig:F3}. The black curve corresponds to \Figs{fig:F3_noD_std-FI}, the blue one to \Figs{fig:F3_mD_std-FI}, and the gray one to \Figs{fig:F3_D_std-FI}.
	First, we observe that for all the choices of  $\TFI/T_{\DII}$, there is a period where $\YDM$ deviates from the standard cosmological scenario. This is because, during the period of the domination of $\Phi$, the expansion rate of the Universe changes,  \ie the DM production is slower since $\mathcal{F}_1$ decreases. 

	Although the evolution of $\YDM$ due to the presence of $\Phi$ deviates from the case of a standard cosmological history, in the end this deviation can be erased if freeze-in takes place after $\DII$ -- the black line approaches the red one on the  right side of the figure. 
	This is because most of DM particles are produced at low temperatures $T \approx \mSV$, since we assume that DM interacts via renormalizable operators~\cite{Hall:2009bx}.

	If, however, DM production ends before -- or close to -- $\DII$, then this is followed by  a period of DM dilution, \ie $\mathcal{F}_3$ becomes important. This is what we observe in the cases $\TFI = T_{\DII}$ (blue line) and  $\TFI = 10 T_{\DII} $ (gray line).  In fact, in the latter case the dilution is so effective that it leads to a strong reduction of the DM yield even relative to its initial value. As already mentioned, this is caused by a negative contribution of $\mathcal{F}_3$ for a period of  vanishing or slow DM production. Moreover, this is a case when $ \YDM $ can exhibit a maximum, which happens for $\TFI = 10 T_{\DII} $ since the \rhs of \eqs{eq:dYdT} at $\FI$ changes sign from positive to negative.
	It is worth noting that, once entropy injection begins at $\DI$ there can be a brief period of time when $\mathcal{F}_3$ becomes negative followed by a period when it is positive -- see also \Figs{fig:F3_D_std-FI} -- resulting in a local minimum of $\YDM$. This is what we can see for  $\TFI = 10^{-2} T_{\DII}$ (black line) around $\mSV/T \approx 10^{-2}$. 

	In \Figs{fig:YDM_FFI} we show the evolution of $\YDM$ as a function of $ 2\mDM /\alpha T$ for the forbidden freeze-in scenario~\cite{Darme:2019wpd}. In this case, since the production without a thermal mass correction is kinematically forbidden, the freeze-in temperature is  $T \approx 2\mDM /\alpha$. 
	We can see that, generally, in this case the evolution of $\YDM$ is similar to the previous case. In particular, $\YDM$ deviates significantly from the case of $\rhoPhi=0$ but in the end the DM yield becomes different only if $\TFI \gtrsim T_{\DII}$
	However, we note that now for all the different values of $\TFI/T_{\DII}$ $\YDM$ exhibits a minimum. This is because, in the forbidden freeze-in case DM production is in general quite inefficient, as already shown in~\cite{Darme:2019wpd}, that at early times $\mathcal{F}_3 < 0$ until the production rate of DM particles increases resulting in $\mathcal{F}_3 > 0$.

	In conclusion, we can see that DM production in NSC scenario that we consider is never more efficient than in the standard radiation-dominated case. In fact it can often be significantly suppressed, depending on the specific values of the parameters of the model. The biggest suppression takes place when DM production from freeze-in takes place before or during the period of $\Phi$ dominance.
	
	\subsection{Contributions from the pair annihilation channel}\label{sec:DM_production_22}

	We expect other DM production channels to exhibit a similar evolution of their DM yield. In the case at hand, the pair annihilation channel $SS \to \chi \chi$ needs to be included. Its contribution to DM production is given by \eqs{eq:C22}. 
	As mentioned at the beginning of Section~\ref{sec:DM_production_12}, we expect this channel to be generally subdominant since the coupling required to produce the observed relic abundance has to be small. However, for $\mSV \ll 2 \mDM$ the  pair annihilation channel may take over due to the dilution of DM  produced via forbidden decays.  The reason for this comes from the difference between the freeze-in temperatures of the two channels. Approximately, the temperature at which the decays stop is $\TFI^{(12)} \approx 2\mDM/\alpha$, while the one for the pair annihilation channel is $\TFI^{(22)} \approx \mDM$. Therefore, if $\alpha \ll 1$ the two temperatures can be significantly different and the DM population produced via $S \to \chi \chi$ can be diluted due to entropy injection to the plasma.	Clearly, the dilution of DM allows the $2 \to 2$ production to take over because the freeze-in mechanism requires small couplings in order to avoid both overclosing the Universe and thermalization. The actual region where this channel dominates, however, needs to be determined numerically.  
	\begin{figure}[t!]
		\centering 
		\begin{subfigure}[b]{0.5\textwidth}
			\centering\includegraphics[width=1\textwidth]{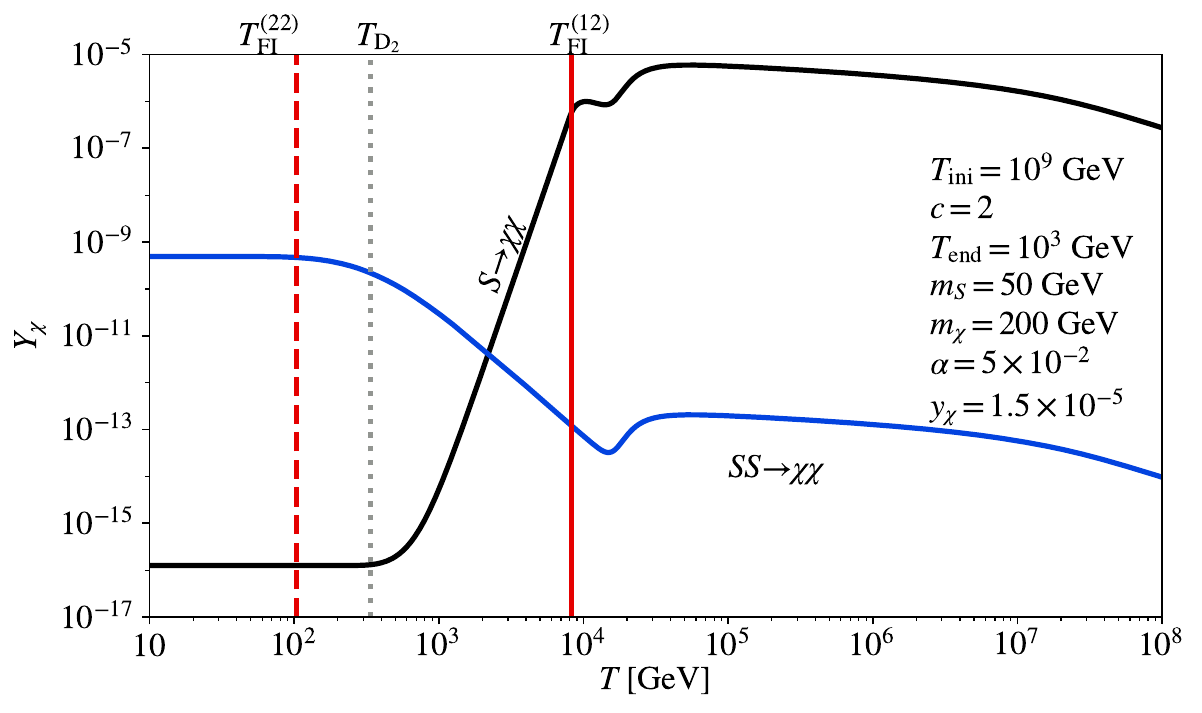}
			\caption{\label{fig:12_vs_22_medium_TEND}}
		\end{subfigure}%
		\begin{subfigure}[b]{0.5\textwidth}
			\centering\includegraphics[width=1\textwidth]{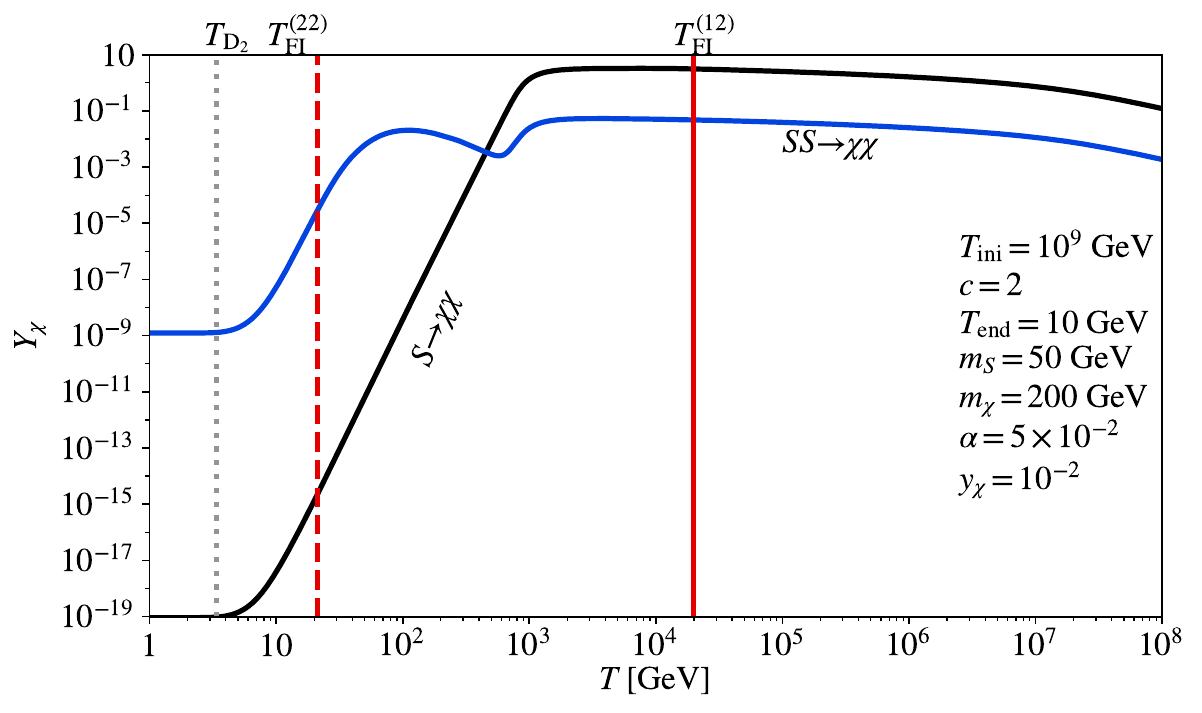}
			\caption{\label{fig:12_vs_22_low_TEND}}
		\end{subfigure}
		\caption{The evolution of the contribution from the two production channels, $S \to \chi \chi$ and $S S\to \chi \chi$, to the DM yield as a function of temperature for $\Ti = 10^9~\GeV$, $c=2$, $\mSV=50~\GeV$, $\mDM =200~\GeV$ and $\alpha = 5 \times 10^{-2}$. The selected values of $\TEND$ and $\yDM$ correspond to two typical ways that the pair annihilation channel can dominate over the one due to decays. In case (a) for $\TEND=10^3~\GeV$ and $\DII$ takes place after the kinematically forbidden decays stop at $T=\TFI^{(12)}$ while the pair-annihilation channel remains active until $T=\TFI^{(22)}$. This results in a DM population from $S$ decays whose  number density becomes rapidly diluted until it falls below the one for the population produced via $S S \to \chi \chi$. The value of the Yukawa coupling, $\yDM = 1.5\times 10^{-5}$, is chosen such as to approximately produce  the observed DM relic abundance. Case  (b) corresponds to  $\TEND = 10~\GeV$ which results in a $\TDII$ lower than both $\TFI^{(12)}$ and $\TFI^{(22)}$. In this case, both contributions correspond to diluted DM, but taking $\yDM = 10^{-2}$, that gives $\relic \sim 0.1$, is sufficient for the pair annihilation channel to dominate.
		}
		\label{fig:12_vs_22}
	\end{figure}

	In \Figs{fig:12_vs_22} we compare the two production channels. 
	In \Figs{fig:12_vs_22_medium_TEND} we show a case where  $\TFI^{(22)} < \TDII < \TFI^{(12)}$, corresponding to a dominant pair annihilation channel due to the dilution of the population produced via the decays. 
	A second possibility for a similar effect is shown in \Figs{fig:12_vs_22_low_TEND}, where the entropy injection stops after both DM production channels have become inefficient. In this case, although both DM populations are diluted, the coupling needed to reproduce  $\relic \sim 0.1$ is such that the $2 \to 2$ channel dominates.
	Finally, we note that the shape of the $\YDM$ curve that corresponds to the pair annihilation channel exhibits a similar behavior to the curves shown in  \Figs{fig:YDM}, which is expected because the evolution of both contributions is given by \eqs{eq:dYdT_with_dNdT}.

	\section{Dark Matter mean momentum and bound from LSSF}
	\label{sec:DM_momentum}
	\setcounter{equation}{0}
	A role played by DM in the formation of large structures in the Universe is a very strong argument for its existence. In particular, in order for these structures to have formed, DM had to be sufficiently slow in order for the initial density perturbations to grow. 
	The free streaming of warm, semi-relativistic relics, with masses in the $\keV$ range, impacts structure growth on scales that are probed by a Lyman-$\alpha$ forest of distant quasars ($0.5-100~h^{-1} ~{\rm Mpc}$).  By combining the  data of the XQ-100 and HIRES/MIKE sets, a bound of $\mWDM>5.3~\keV$ at $2 \sigma$ was obtained in ref.~\cite{Irsic:2017ixq} for a fermionic frozen-out DM relativistic at its decoupling (\ie WDM case). More recently a tighter bound $\mWDM>6.3~\keV$ was derived by using stellar stream observation~\cite{Banik:2019smi}. We will apply it in our work.
	
	\subsection{Recasting the LSSF bound}\label{sec:LSSF_recast} 

	We need to recast the above bound as a bound on the mass of the WDM particle in order to apply it in the DM production scenario considered here. As pointed out in~\cite{Huo:2019bjf}, matching  today's DM velocity with the WDM scenario can yield an accurate constraint for an alternative DM scenario.~\footnote{In ref.~\cite{Kamada:2019kpe} the matching  quantity is the ``warmness" $\sigma \sim \vev{p_{\chi}^2}$. However, our aim is to apply \eqs{eq:dpDMdt}, which does not take into account the variance of the momentum.}  
	This is done by using the definition of the average WDM velocity today $\vWDMt$
	\begin{equation}
		\vWDMt \equiv \dfrac{\pWDMt}{  \mWDM } \;
		\label{eq:v0_definition}
	\end{equation}
	and applying the LSSF bound on $\mWDM$. Since the average momentum today, $\pWDMt$, also depends on $\mWDM$, it is not straightforward to obtain a bound on the velocity. Thus, we need to express $\vWDMt$ in terms of the effective WDM temperature today $\TWDMt$, which is related to the WDM mass through the relic abundance $\relic$. The relation between WDM velocity and temperature is given by 
	\begin{equation}
		\vWDMt=  
		\dfrac{ \pWDMt }{ \TWDMt } \dfrac{\TWDMt}{m_{\rm WDM}} =
		\dfrac{  \pWDMrel   }{\TWDMrel} \dfrac{\TWDMt}{m_{\rm WDM}} 
		\approx 3.15 \dfrac{\TWDMt}{m_{\rm WDM}}  
		\label{eq:v0_T0}
	\end{equation}
	where  $\TWDMrel$ is the WDM temperature some time before freeze-out, with WDM being still relativistic,  $\TWDMt =  a_{\rm rel} \TWDMrel$, where $a_{\rm rel}$ is the scale factor at $T = \TWDMrel$, and $\pWDMrel  \approx 3.15 \ T^{\rm WDM}_{\rm rel}$, with WDM assumed to be a fermion.
	In order to express $\TWDMt$ in terms of the WDM mass we note that, at $T=\TWDMrel$ its number density is given by
	\begin{equation}
		n_{\rm WDM, \; rel} = \dfrac{3}{2 \pi^2} \zeta(3) \ \TWDMrel^3 
		\label{eq:nWDM_rel}
	\end{equation}
	which can be expressed in terms of its relic abundance as 
	\begin{equation}
		n_{{\rm WDM}, \; 0} = \dfrac{\rho_{c} \Omega_{\rm DM} }{m_{\rm WDM}}  \;.
		\label{eq:nWDM_0}
	\end{equation}
	Since the number of WDM particles is conserved after freeze-out, its number density today is given by $n_{\rm WDM, 0}= a_{\rm rel}^3 n_{\rm WDM, \; rel} $, with $\TWDMt = a_{\rm rel} \TWDMrel $. Therefore 
	\begin{equation}
		\TWDMt = 
		\left( \dfrac{ 2\pi^2  }{3 \zeta(3)}  \dfrac{\rho_{c} \Omega_{\rm DM} }{m_{\rm WDM}}\right)^{1/3} \;,
		\label{eq:TWDM_0}
	\end{equation}
	and the WDM velocity today can be expressed as
	\begin{equation}
		\vWDMt \approx 3.15 \left( \dfrac{ 2\pi^2  }{3 \zeta(3)}  \dfrac{\rho_{c} \Omega_{\rm DM} }{m_{\rm WDM}^4}\right)^{1/3} \approx 1.2 \times 10^{-7} \left(  \dfrac{\rm keV}{m_{\rm WDM}} \right)^{4/3} \;.
		\label{eq:v0}
	\end{equation}
	By combining this expression with the bound $m_{\rm WDM} > 6.3~\keV$ we find  an upper limit on the DM velocity such that successful LSSF can still take place. This limit has to be respected by other DM scenarios as well.  Therefore, the constraint on the velocity of the DM candidate in  this work can be written as 
	\begin{equation}
		\vev{v_{\chi ,0}} = \dfrac{m_{\chi}}{ \pDMt } < 1.03 \times 10^{-8} \;,
		\label{eq:SF-bound_v0}
	\end{equation}
	or, in terms of the DM particle mass,
	\begin{equation}
		m_{\chi}> \dfrac{\vev{p_{\chi, 0}} }{1.03 \times 10^{-8} } \;. 
		\label{eq:SF-bound}
	\end{equation}
	
	\subsection{Evolution of $\pDM$}\label{sec:pDM_evol}
	Assuming  $\pDM$ is known for a certain cosmological scenario, we can use \eqs{eq:SF-bound} to constrain the model. In order to do so, we  may approximate the evolution of the average energy $\EDM$, as shown in Appendix~\ref{app:BE_Emean}. As explained in Section~\ref{sec:DM_production}, the $S\to \chi \chi$ decay channel dominates the DM production until $\mSV < 2 \mDM$, when the process $SS\to \chi\chi $ becomes dominant. Therefore, since the LSSF bound is still expected to apply for light DM, we are only concerned with the evolution of the DM momentum  produced via the $S$ decay channel. The corresponding equation of $\pDM$ in this case takes the form
	\begin{equation}
		\dfrac{d\pDM}{dt} = -H \pDM \lrb{\dfrac{\EDM}{\pDM}}^2   \lrsb{ \lrBiggb{1  -\lrb{ \dfrac{m_{\chi}}{\EDM}}^2 } + 
			\lrb{1- \dfrac{1}{2} \dfrac{\vev{E_S}}{\EDM} } \dfrac{2\GammaDM}{H}\dfrac{ \mST  \; n_{S}^{(-1)}}{ \nDM}   } \;.
		\label{eq:dpDMdt}
	\end{equation}
	Although this equation is difficult to solve analytically, we can still examine it in order to we understand how $\pDM$ evolves. 
	In order to determine the effect of entropy injection, we introduce the quantity $\uDM \equiv \dfrac{\pDM}{T}$, which is constant away from the time of entropy injection and DM production.
	The corresponding evolution equation can be obtained in a straightforward way from \eqs{eq:dpDMdt}.  
	To simplify the discussion, we take the limit $T \gg \mDM$, where we expect DM particles to be highly relativistic, and assume that $ 4 \, \rhoR = 3 \,s\,T$. Then $\uDM$ approximately obeys     
	\begin{equation}
		\dfrac{d \log \uDM}{d\log \frac{\Ti}{T}} = 
		\lrBigb{1-  \mathcal{F}_2 } + \
		\lrb{  \dfrac{\vev{E_S}}{2T} \uDM^{-1} -1} \dfrac{d \log \NDM}{d \log \frac{\Ti}{T}}  \;.
		\label{eq:dudT_approx}
	\end{equation}
	We will study the two different terms of the \rhs separately in order to follow the evolution of $\pDM$.

	\subsubsection*{First term}
	The first term, $\mathcal{T}_1=1-\mathcal{F}_2$, is related to the entropy injection. From the discussion in Section~\ref{sec:DM_production_12}, and as can be seen from \Figs{fig:F2}, we note that this term gives always a negative contribution and vanishes in the limit of entropy conservation.  This means that when the second term of \eqs{eq:dudT_approx} vanishes, $\uDM$ is either conserved, implying that DM particles free-fall with $\pDM \sim T \sim a^{-1}$, or decreases, in which case DM momentum redshifts faster than the temperature.  
	
	\subsubsection*{Second term}
	The second term of \eqs{eq:dudT_approx},
	$$
	\mathcal{T}_2 = \lrb{   \dfrac{\vev{E_S}}{2T} \uDM^{-1} -1} \dfrac{d \log \NDM}{d \log \frac{\Ti}{T}} \; ,
	$$
	vanishes when $S$ is highly relativistic~~\footnote{We assume that DM particles are being produced only by decays. This means that over a period of time $\Delta t$, $\Delta \NDM$ particles have been created, with mean energy 
		$$
		\vev{\omega_\chi} \approx \dfrac{\Delta \NDM \, \vev{E_S} /2 }{\Delta \NDM } = \dfrac{1}{2} \vev{E_S} 
		$$
		where in the limit $T \gg \mSV$, $\vev{E_S} \sim T$. Therefore, DM particles produced from relativistic $S$ have energy proportional to the temperature of the plasma. Consequently, the mean energy of all DM particles that have been  produced up to  some time with $T \gg \mSV$, should be $\EDM \sim T$, and the second term of \eqs{eq:dudT_approx} should vanish.  }
	or when DM  production is inactive. 
	Furthermore, this term appears to be positive or zero, since $\EDM$ should be less (due to redshift) or equal to $\vev{E_S}/2$, since DM particles are produced by decays of $S$. 
	This behaviour is expected to be similar for both $\mSV > 2 \mDM$ and $\mSV < 2 \mDM$ as long as $S$ remains relativistic.  
	The difference between the two cases is  close to their freeze-in. This is because $S$ is non-relativistic (with $\vev{E_S} / T \ll 1$) when $\mSV>2\mDM$, while it is relativistic  (with $\vev{E_S}  /T\approx 3$) when $\mSV>2\mDM$.
	Furthermore, in both cases the relationship between $\TFI$ and $\TDII$ is also important, since the entropy injection causes the temperature to redshift more slowly than the momentum of free-falling DM particles.  
	\begin{figure}[t!]
		\centering 
		\begin{subfigure}[c]{0.5\textwidth}
			\centering\includegraphics[width=1\textwidth ]{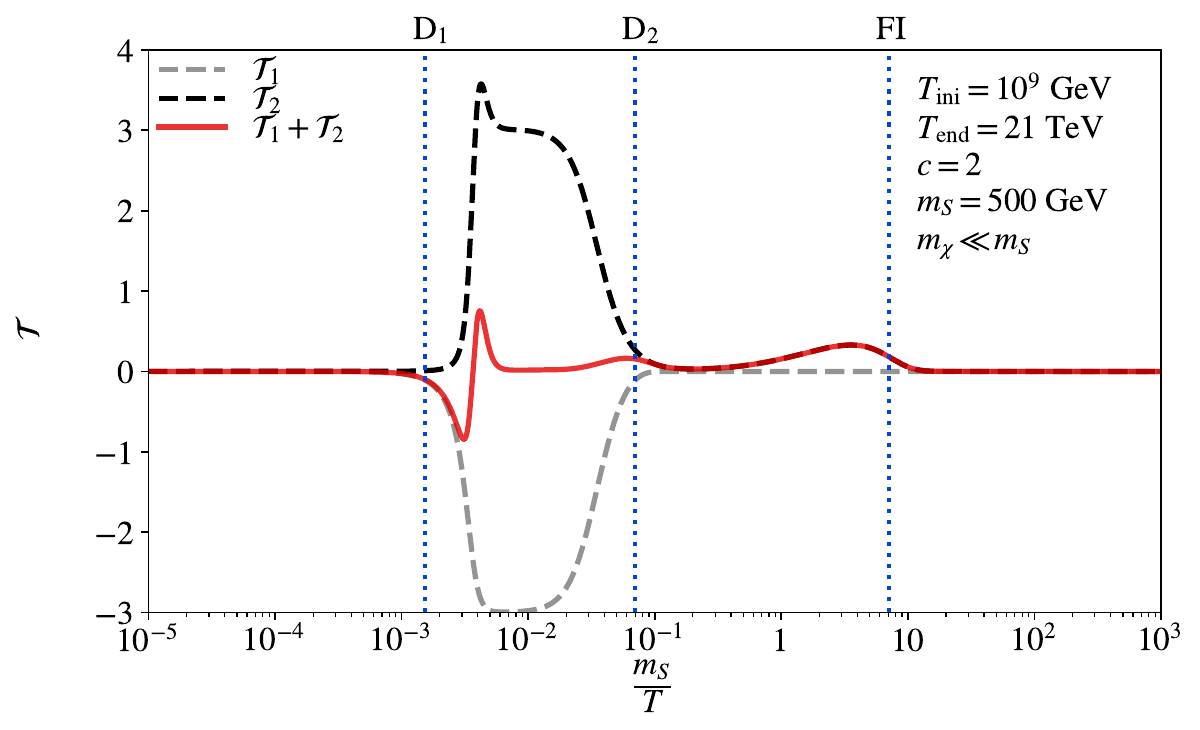}
			\caption{\label{fig:T1T2_noD}}
		\end{subfigure}%
		\begin{subfigure}[c]{0.5\textwidth}
			\centering\includegraphics[width=1\textwidth]{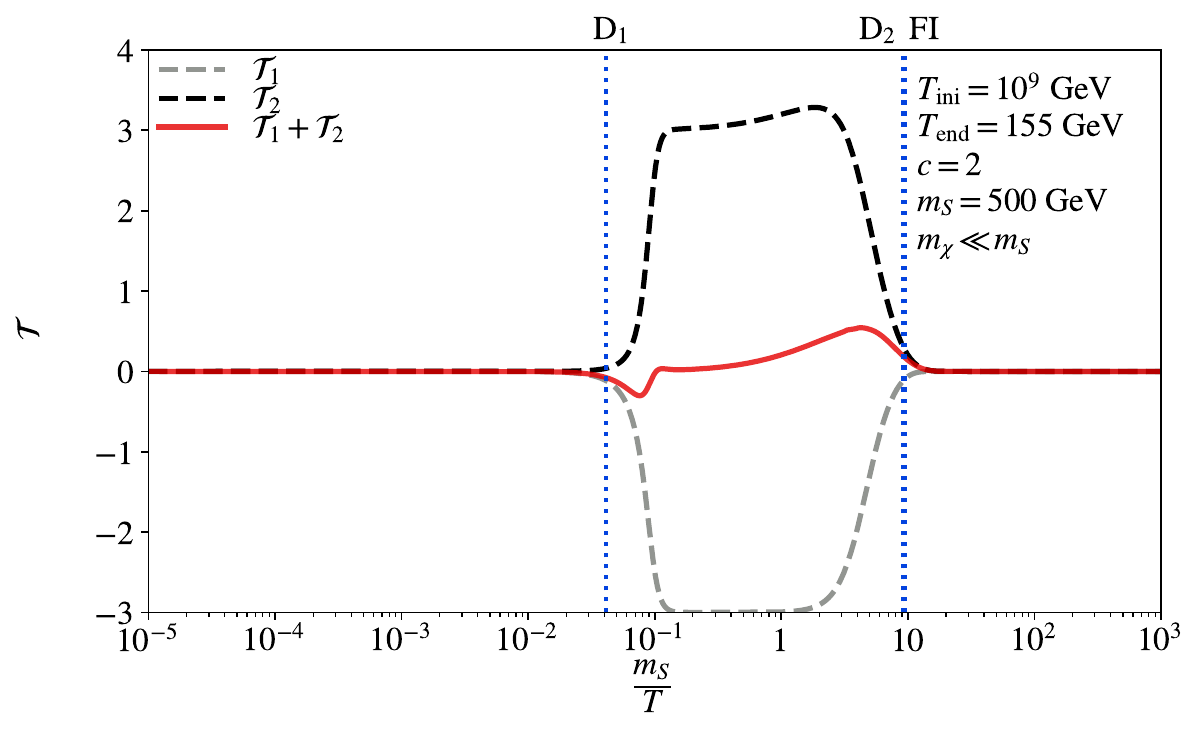}
			\caption{\label{fig:T1T2_mD}}
		\end{subfigure}
		\begin{subfigure}[c]{0.5\textwidth}
			\centering\includegraphics[width=1\textwidth]{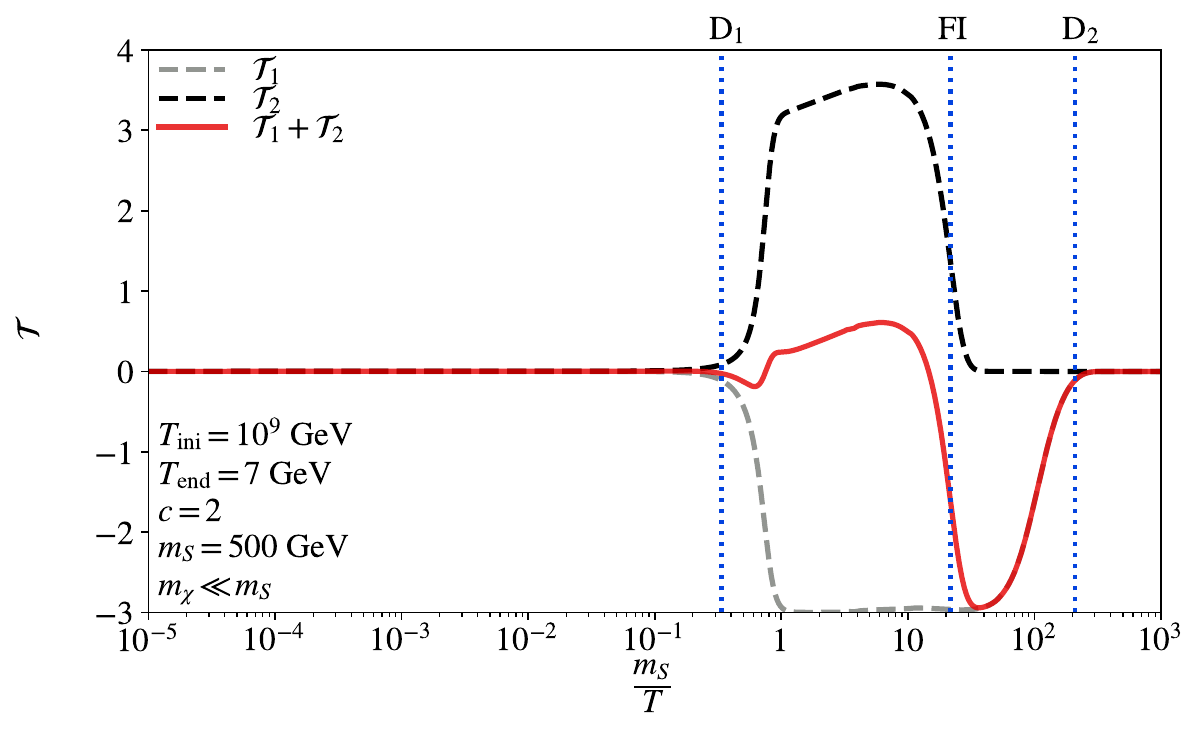}
			\caption{\label{fig:T1T2_D}}
		\end{subfigure}
		\caption{The evolution of $\mathcal{T}_{1}$ (gray line), $\mathcal{T}_{2}$ (black line), and their sum (red line) with the
			temperature for the case $\mSV \gg 2\mDM$ and the same parameter choice as in Fig.~\ref{fig:F3}. Note that after the freeze-in, $\mathcal{T}_2$ falls rapidly to zero, and if $\TFI \lesssim \TDII$, the \rhs of eq.(~\ref{eq:dudT_approx}) is dominated by $\mathcal{T}_1$ which is negative. This results in a DM momentum that redshifts faster than the temperature. The qualitative behavior is similar for $\mSV \gg 2\mDM$. }
		\label{fig:T1T2}
	\end{figure}
	
	An example for how $\mathcal{T}_{1,2}$ behave for  $\mSV \gg 2\mDM$ is shown in \Figs{fig:T1T2} for $\TFI< \TDII$ (a), $\TFI \approx \TDII$ (b), and $\TFI > \TDII$ (c) and for the other parameters set as in \Figs{fig:F3}. As expected $\mathcal{T}_1$ vanishes away from the time of entropy injection ($\TDII \lesssim T \lesssim \TDI$), and assumes a negative value between $\DI$ and $\DII$. The second term is positive as long as DM is produced, while around $\FI$ it starts to fall rapidly. These two terms seem to compete, as one is positive and the other negative with similar magnitudes, while their sum remains relatively small, and can fluctuate around zero. This can cause various minima and maxima in the evolution of $\uDM$. 
	The case of $\mSV \ll 2\mDM$ is similar. As already mentioned, their main difference is close to the time of freeze-in, since in this case $S$ is still relativistic.
	
	\subsubsection{Numerical examples}\label{sec:pDM_num}
	
\begin{figure}[t!]
		\centering 
		\begin{subfigure}[b]{0.5\textwidth}
			\centering\includegraphics[width=1\textwidth ]{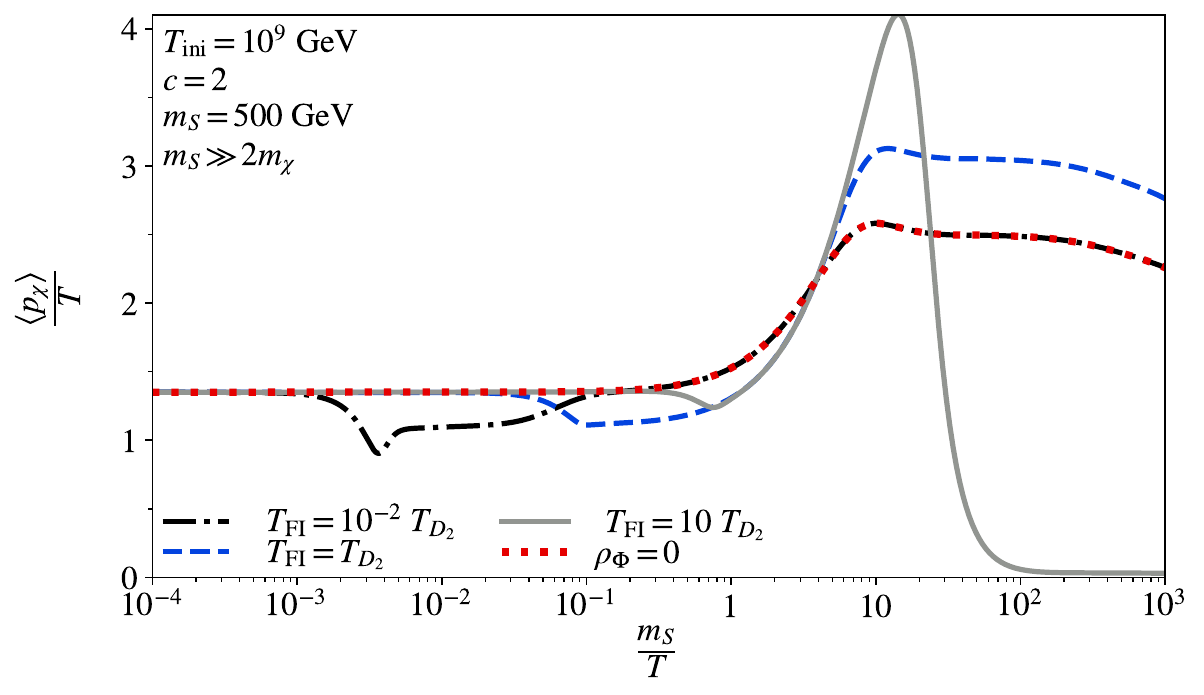}
			\caption{\label{fig:uDM_std-FI}}
		\end{subfigure}%
		\begin{subfigure}[b]{0.5\textwidth}
			\centering\includegraphics[width=1\textwidth]{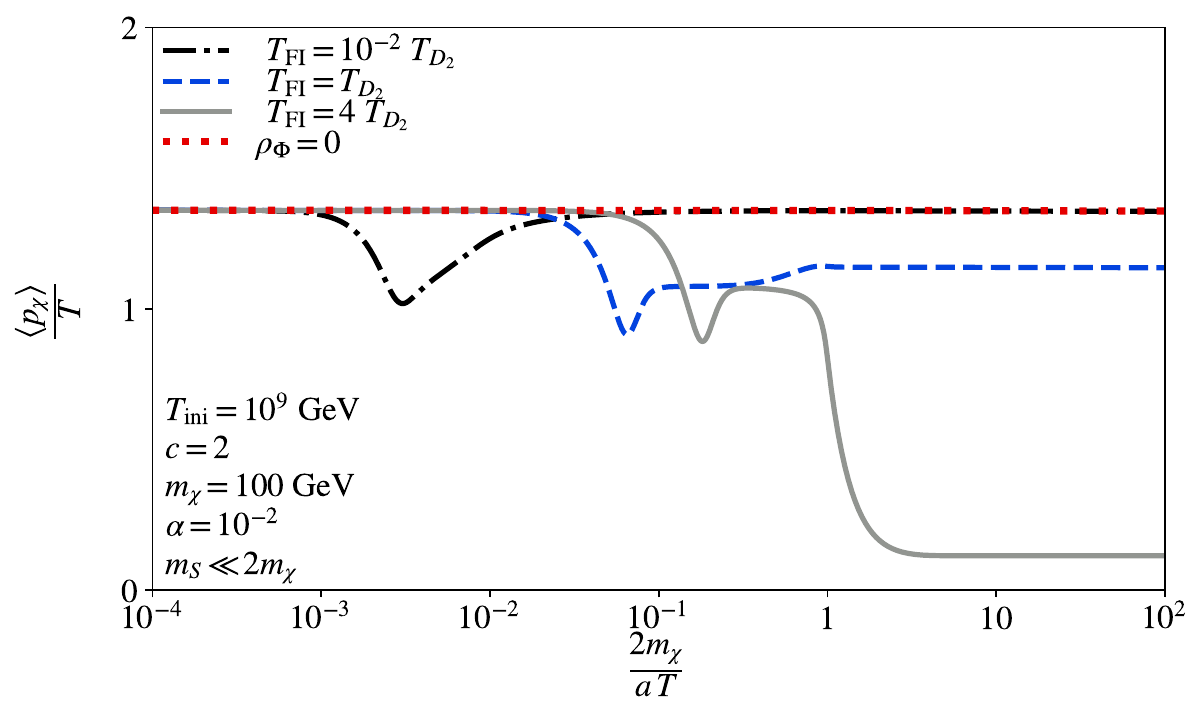}
			\caption{\label{fig:uDM_FFI}}
		\end{subfigure}
		\caption{The evolution of $\uDM$ for different production scenarios. The parameter choice for the radiation-$\Phi$ system and color coding is the same as in Fig.~\ref{fig:YDM}. The left panel shows $\uDM$ as a function of $\mSV/T$ for $\mSV = 500 ~\GeV$ and $\mDM \ll \mSV/2$, while the right panel shows the forbidden freeze-in case,  with  $\mDM= 100 ~\GeV \gg \mSV/2 $. In both panels the red line corresponds to the standard cosmological history, \ie $\rhoPhi = 0$. }
		\label{fig:uDM}
	\end{figure}

	The effects described above can also be seen in \Figs{fig:uDM}. The left panel, which corresponds to $\mSV > 2 \mDM$, shows that $\uDM$ starts as constant -- as expected since $\Ti \gg \mSV$ -- and then begins to decrease as entropy injection proceeds and $\mathcal{T}_1$ dominates; see also \Figs{fig:T1T2} for $T<\TDI$, where $\mathcal{T}_1+\mathcal{T}_2 <0$. Close to freeze-in,  where $T \approx \mSV$, $\uDM$ increases since $\vev{E_S} \sim \mSV>T$. After this point, the evolution of $\uDM$ depends on the relationship between $\TFI$ and $\TDII$. If $\DII$ occurs before freeze-in then $\uDM$ reaches its  standard cosmological value, while if  $\DII$ happens after the freeze-in then $\uDM$ decreases since $\pDM$ redshifts faster that the temperature. 
	In addition, there exists a fine-tuned case when $\TFI \approx T_{\DII}$ in which $\uDM/T$ reaches a maximum that is above the value corresponding to the standard cosmological scenario.

	The right panel of \Figs{fig:uDM} shows the dependence of $\uDM$ on $\dfrac{2\mDM}{\alpha \, T} $ for $\mSV \ll 2\mDM$. As described above, in this case $\uDM$ mostly decreases. However, there are points where $\mathcal{T}_2$ can grow, resulting in the various minima observed in the figure. 
	However, we note that after the minimization of $\uDM$, if $\DII$ takes place before the freeze-in then $\uDM$ reaches the same value as in standard cosmological  scenario, while otherwise $\pDM$ is highly redshifted. 
	Although $\uDM $ is lower than its standard cosmological history value for $\TFI \approx T_{\DII}$, it is worth noting a slight increase at $T \lesssim \TFI$ which is  a combination  of effects from slow entropy injection -- that tends to slightly decrease $\uDM$-- and from a positive $\mathcal{T}_2$.
	
	\subsubsection{The dependence of DM mean momentum on $c$}\label{sec:pDM_c}
	\begin{figure}[b!]
		\centering\includegraphics[width=0.8\textwidth ]{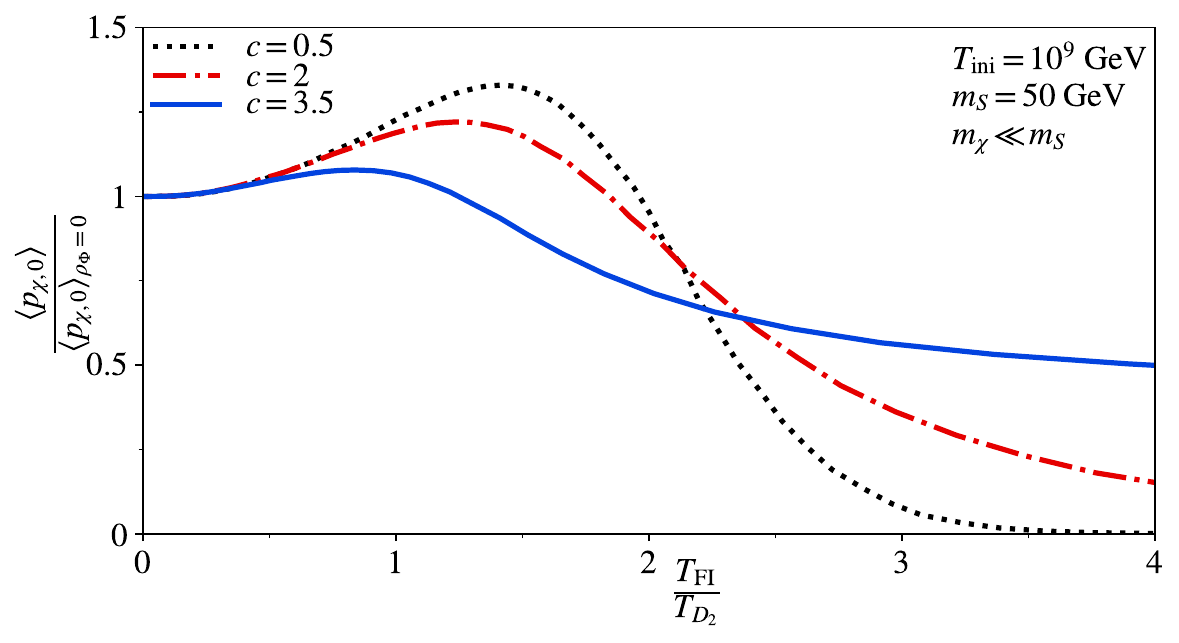}
		\caption{ The dependence of $\vev{p_{\chi , 0}}/\vev{p_{\chi , 0}}_{\rhoPhi = 0}$ on $\TFI/\TDII$, with $\TDII$ computed by varying $\TEND$ within the range $ 10^{-3} \mSV \lesssim \TEND \lesssim 10^3 \mSV$, for $\Ti=10^9~\GeV$ and $\mSV = 50 ~\GeV\gg 2 \mDM$ and for  $c= 0.5$ (black line), $c = 2$ (red line), and $c=3.5$ (blue line). 
			In each case there is a maximum exceeding one (\ie $\vev{p_{\chi , 0}}>\vev{p_{\chi , 0}}_{\rhoPhi = 0}$) for $\TFI \approx \TDII$ that can result in a stricter LSSF bound. To the right of the
			maximum there is more time for $\vev{p_{\chi , 0}}$ to redshift, and the LSSF bound is expected to be reduced -- since $\vev{p_{\chi , 0}}<\vev{p_{\chi , 0}}_{\rhoPhi = 0}$ -- provided that the thermalization bound is not violated.
		}
		\label{fig:p0T0_std-FI}
	\end{figure}
	From the inequality~(\ref{eq:SF-bound}), and as will be explicitly shown in the next Section, it is expected that the bound on the DM mass from LSSF can either increase or decrease compared to the  standard cosmological scenario. Nonetheless, the relationship between the mean DM momentum today $ \vev{p_{\chi , 0}} $ and the dilution of DM, characterized by $\TFI / \TDII$, is worth further examination.
	
	In \Figs{fig:p0T0_std-FI} we show the dependence of $\vev{p_{\chi,0}}$  (normalized to the same quantity in the absence of the fluid, \ie $\vev{p_{\chi,0}}_{\rhoPhi = 0}$) on the ratio $\TFI / \TDII$, with $\TDII$ calculated  by varying $\TEND$, for $\mSV= 50~\GeV$, $\mDM \ll \mSV/2$, $\Ti = 10^{9}~\GeV$, and some fixed values of $c$.
	First, we note that, for $\TDII \gg \TFI$ the mean DM momentum today converges to the case $\rhoPhi = 0$ for all the choices of $c$, as expected. At $\TDII \sim \TFI$  $\pDMt$ reaches a maximum. At $\TDII \ll \TFI$ the DM momentum becomes suppressed as a result of entropy injection which causes DM momentum to redshift faster than its standard cosmological history value.
	The maximimum itself, as well as the rate of suppression depends strongly on $c$ since this parameter can significantly affect the amount of entropy injection; see \eg \Figs{fig:gamma_TEND}.

	\section{Benchmark points}\label{sec:benchmarks}
	\setcounter{equation}{0}
	The DM production discussed in the previous Sections can give the observed DM relic abundance $\relic \approx 0.12$~\cite{Aghanim:2018eyx} in different regions of the parameter space. In this Section we will focus on some representative cases that exhibit different features, which will be helpful in understanding effects we will encounter in the analysis that follows. In particular, we will be interested in the parameters of the dark sector ($\mDM$, $\yDM$, and  $\alpha$) that result in the observed relic abundance while keeping $\chi$ out of equilibrium as well as respecting the LSSF bound (discussed in Section~\ref{sec:DM_momentum}). We will also examine how  different cosmological histories affect these parameters.   
	
	\subsubsection*{Thermalization of $\chi$}
	The main assumption behind the freeze-in mechanism is that  DM particles never reach equilibrium with the plasma. In most cases this is automatically ensured by the smallness of relevant couplings that is required to obtain $\relic \approx 0.12$. However, in the case of entropy injection a larger coupling may be needed in order to compensate for the dilution of the DM number density.  
	On the other hand, such a larger coupling may lead to the increase the DM number density at early times, to the point of reaching its equilibrium value $\nDM^{\rm eq}$. In this case back-reactions $\chi \chi \to SS$ and  $\chi \chi \to S$ can become frequent enough, leading to a possible DM thermalization with the plasma. 
	In order to  make sure that this does not happen, we introduce the ratio
	\begin{equation}
		R = \dfrac{\nDM}{\nDM^{\rm eq}}\;.
		\label{eq:R_definition}
	\end{equation} 
	We will derive an upper bound on its maximum value $\Rmax$ before the freeze-in as well as before $\pDM < \mDM$. This way we will ensure that $R$ remains small enough up to the point where the DM production becomes inefficient, as explained in ref.~\cite{DeRomeri:2020wng}.

	It is worth noting that, in the temperature range $T>\TFI$ and $\pDM > \mDM$  DM is expected to be mostly relativistic, \ie $\nDM^{\rm eq} \sim T^3$. Thus, we expect $R$ to satisfy an equation similar to \eqs{eq:dYdT_with_dNdT}, \ie $R \sim \YDM$. As a consequence, its maximum can occur both close as well as away from the freeze-in temperature, depending on the relation between $\TFI$ and $T_{\rm D_{1,2}}$; see \Figs{fig:F3,fig:YDM}, and the relevant discussion. This, in turn, means that $\Rmax$ has to be determined numerically, as in principle it depends on the behavior of $\mathcal{F}_{1,2,3}$.  
	
	\subsection*{Benchmark scenarios} 
	As we have already mentioned, we are interested in studying the effect of the considered NSC scenario on the allowed parameter space. In this Section we will study some representative points for which $\relic \approx 0.12$ but for which various bounds also are affected by the different NSC scenarios. 
	Since DM production in the light DM regime ($\mSV \gg \mDM$) is significantly different from the heavy DM one ($\mSV \ll\mDM$) we discuss them separately. 
	
	First we observe that in the light-DM case the pair annihilation channel is always subdominant.~\footnote{We have checked that in the entire parameter space that we consider this is indeed the case.} Therefore, this channel can be safely neglected both in the DM production as well as in the evolution of $\pDM$.
	Moreover, we note that, since we assume a fairly heavy $S$, the LSSF constraint is relevant only in this regime, since we expect it to apply for DM mass around the $\keV$ scale.

	In the heavy-DM regime the forbidden freeze-in can lead to relatively large Yukawa couplings due to its inefficiency, as argued in~\cite{Darme:2019wpd}. When combined with the effect of dilution, this can result in a further increase of $\yDM$ compared to the standard cosmological history. 
	However, this regime can also be dominated by the $2 \to 2$ channel, which can limit $\yDM$. Thus, the heavy-DM regime provides an opportunity to study how the production is affected by the appearance of two competing channels operating at different temperatures.
	Furthermore, this regime corresponds to deeply non-relativistic DM at low temperatures, regardless of NSC, and the bound~(\ref{eq:SF-bound}) is always satisfied. 
	
	Finally, in order to study in more detail the effect of the cosmological parameters $\Ti$, $\TEND$, and $c$ we also show how they can affect the parameter space and the relevant constraints in the entire DM mass range.
	
	\subsection{Light DM regime ($\mSV \gg \mDM$)}\label{sec:std-FI}
	
	In this case $\mSV$ is the only relevant energy scale around the freeze-in, resulting in $\nDM$ and $\pDM$ that are independent of $\mDM$ and $\alpha$ for $T<\TFI$.
	%
	The relic abundance then depends linearly on the DM mass as
	\begin{equation}
		\relic \approx 2.8 \times 10^8 \ Y_{\chi,0} \ \dfrac{\mDM}{\GeV} \;,
		\label{eq:relic}
	\end{equation}
	with $Y_{\chi,0}$ denoting the DM yield today. Furthermore, in this case $Y_{\chi,0} \sim \yDM^2$ since the dominant production channel is the decay of $S$. Thus, the Yukawa coupling depends on the DM mass as $\yDM^2 = F_{\rm cosm}(\mSV) \mDM$ -- we shall call  this curve the ``Planck line" -- with the factor   $F_{\rm cosm}(\mSV)$ depending on $\mSV$ as well as the parameters $\Ti$, $c$, and $\TEND$ that 
	determine the cosmological history.
	That means that for a given set of $\mSV$, $\mDM$, $\Ti$, $c$ and $\TEND$ we can find one value of $\yDM^2$ that reproduces the 
	observed relic abundance.

	\begin{figure}[t!]
		\centering 
		\begin{subfigure}[b]{0.5\textwidth}
			\centering\includegraphics[width=1\textwidth ]{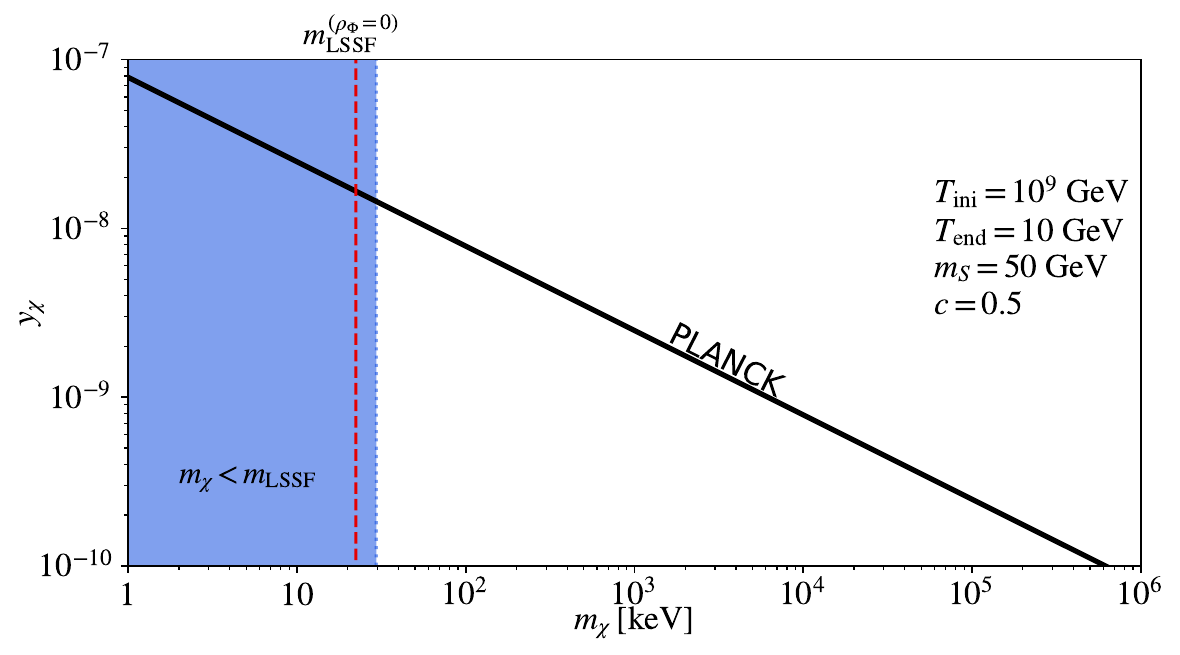}
			\caption{\label{fig:Planck_std-FI_High-LS}}
		\end{subfigure}%
		\begin{subfigure}[b]{0.5\textwidth}
			\centering\includegraphics[width=1\textwidth]{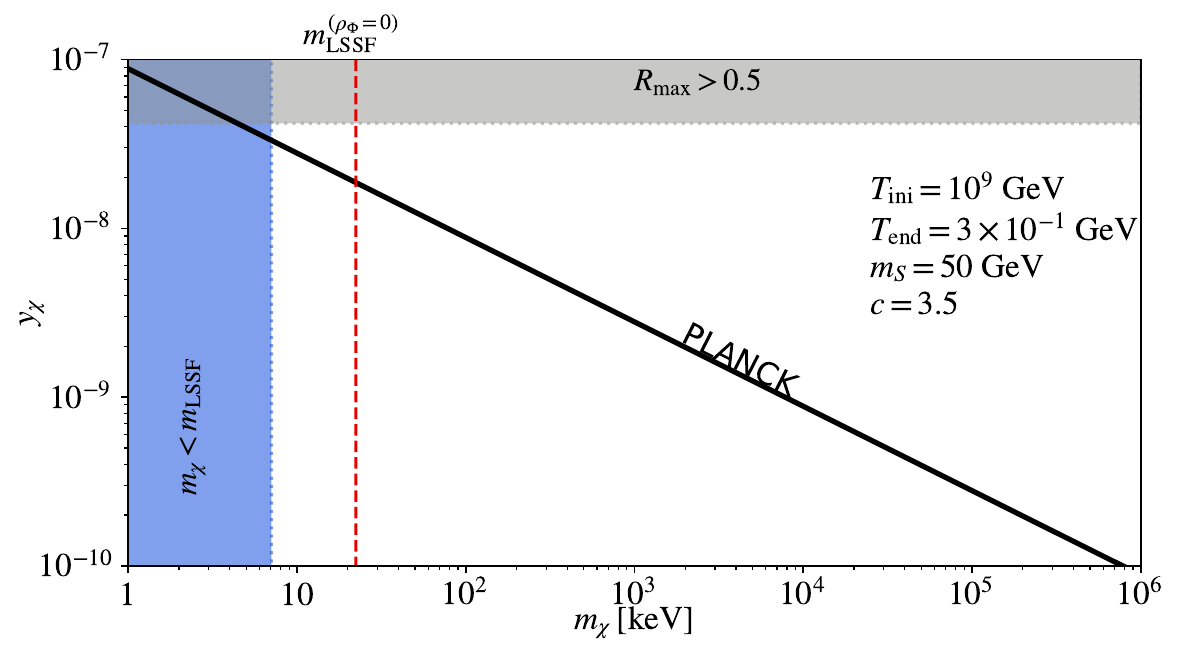}
			\caption{\label{fig:Planck_std-FI_Low-LS}}
		\end{subfigure}
		\caption{ The dependence of $\yDM$ on $\mDM$ (the Planck line) for $\Ti = 10^9~\GeV$, $\mSV = 50~\GeV \gg 2\mDM$ 
			for $\TEND = 10~\GeV$ and $c=0.5$ (a) and for $\TEND = 0.3~\GeV$ and $c=3.5$ (b). 
			In the blue shaded region the LSSF bound is violated while gray area indicates the cases where possible thermalization takes place.
			The vertical red line shows the lower LSSF bound $m_{\rm LSSF}^{(\rhoPhi = 0)} \approx 22~\keV$ in the standard cosmological scenario. 		
		}
		\label{fig:Planck_std-FI}
	\end{figure}
	In \Figs{fig:Planck_std-FI} we show some examples of the Planck line for several NSC scenarios. 
	In \Figs{fig:Planck_std-FI_High-LS} the black line of the Yukawa coupling obeying $\yDM \sim \mDM^{-1/2}$ is drawn for the choice of parameters  $\mSV = 50 ~\GeV$, $\Ti = 10^{9}~\GeV$, $\TEND = 10 ~\GeV$ and $ c=0.5$. The blue shaded region shows where the bound~\eqs{eq:SF-bound} is violated. The vertical red line shows the lower LSSF bound in the standard cosmological scenario ($m_{\rm LSSF}^{(\rhoPhi = 0)} \approx 22~\keV$).  
	Note that, in this particular case the thermalization bound $\Rmax<0.5$ is not violated, at least for $\mDM>1~\keV$.
	Interestingly, although this case corresponds to sightly diluted DM with $\TFI \approx 1.3 \ \TDII$, the LSSF constraint becomes enhanced, by around $35\%$, compared to its standard cosmological history value. This is a result of the enhancement 
	observed in \Figs{fig:uDM_std-FI} for the case $\TFI = \TDII$. 
	
	A second example is shown in \Figs{fig:Planck_std-FI_Low-LS} where the Planck (black) line corresponds to  $\mSV = 50 ~\GeV$, $\Ti = 10^{9} ~\GeV$, $\TEND = 3\times 10^{-1} ~\GeV$, and $ c=3.5$. The gray area shows where the thermalization bound is violated and the blue shaded region shows the DM mass range for which the LSSF constraint is not obeyed, with the vertical red line corresponding to the lower LSSF bound in the standard cosmological scenario.
	Note that in panel (a) the LSSF bound is stronger while in panel (b) it is weaker compared to the standard cosmological scenario.   %
	Although in both cases $\yDM$ turns out to be similar, in case (b) the LSSF constraint is weaker, by around $60\%$, compared to its standard cosmology value. Note that the lower bound on DM mass $\mDM \approx 7 ~ \keV$ almost coincides with the boundary of the thermalization constraint. This is because the DM number density redshifts rapidly, making such a low LSSF bound possible only with small amount of entropy injection -- hence the choice of $c=3.5$ -- in order to avoid thermalization.
	
	\subsection{Heavy DM regime ($m_S\ll\mDM$)}\label{sec:FFI}

	The second case we discuss corresponds to the DM production via kinematically forbidden (in the vacuum) decays  $S \to \chi \chi$. As it was pointed out in~\cite{Darme:2019wpd}, in a radiation-dominated Universe the pair annihilation production channel $SS \to \chi \chi$ is suppressed compared to thermally induced $S$ decays. 
	However, as we argued in Section~\ref{sec:DM_production}, since the decays stop at higher temperatures than the pair annihilation, the DM population produced by the latter process can dominate the DM relic abundance by being less, or not at all,  diluted.

	In contrast to the light DM regime, the dependence of the relic abundance on $\mDM$ is now more complicated. However, for both channels during radiation domination for $\mSV \ll   \mDM$ the relic abundance becomes almost independent of $\mDM$; see \eg refs.~\cite{Lebedev:2019ton,Darme:2019wpd}. 
	This is so because the Hubble rate scales as $T^2$ while, for $T \gg \mSV,\ \mDM $, the DM production rate is proportional to $T^4$, resulting in 
	\begin{equation}
		\dfrac{d\YDM}{dT} \sim - \dfrac{T^4}{H  s   T} \sim - T^{-2} \;.
		\label{eq:dYdT-std}
	\end{equation}
	This, in turn, means that $Y_{\chi,0} \sim \dfrac{1}{\TFI}$, with $\TFI$ (in both channels) being proportional to $\mDM$, since $\mSV \ll  \mDM$. 
	Therefore,  we expect $\yDM$ to depend on the DM mass only through the freeze-in temperature which, for  given values of the cosmological parameters $\Ti$, $\TEND$, and $c$, determines if DM is diluted or not. The Yukawa coupling should then be approximately independent of $\mDM$ as long as DM is produced during radiation domination, \ie  $\TFI> T_{\EI}$ or $\TFI < \TDII$.

	\begin{figure}[t!]
		\centering 
		\begin{subfigure}[b]{0.5\textwidth}
			\centering\includegraphics[width=1\textwidth]{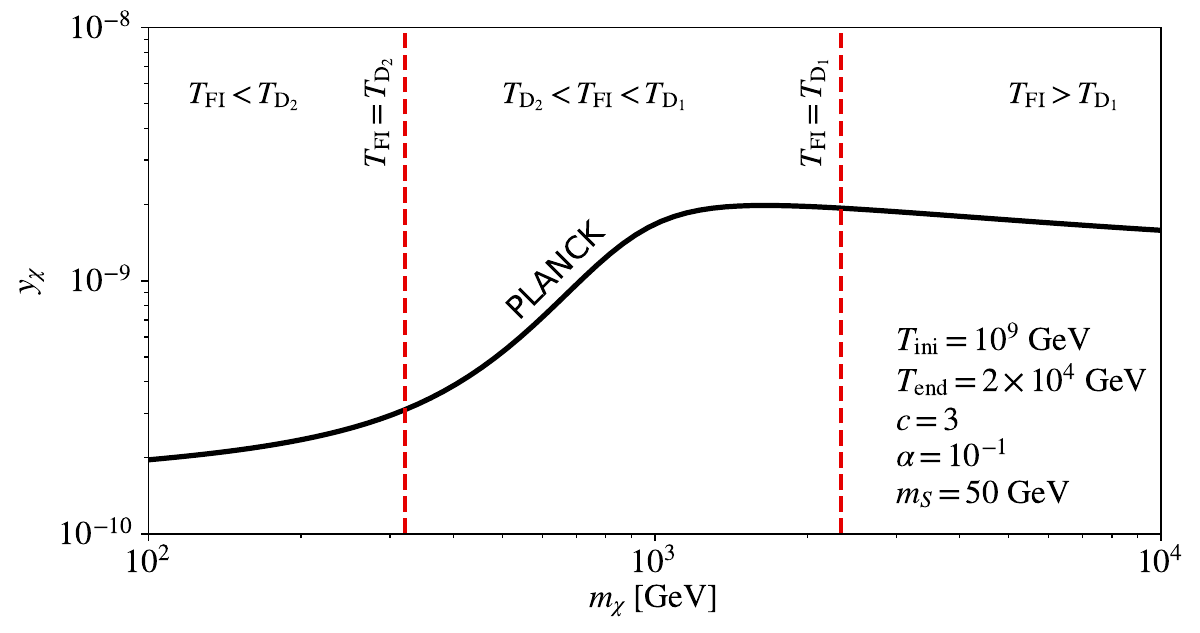}
			\caption{\label{fig:Planck_FFI_12}}
		\end{subfigure}%
		\begin{subfigure}[b]{0.5\textwidth}
			\centering\includegraphics[width=1\textwidth]{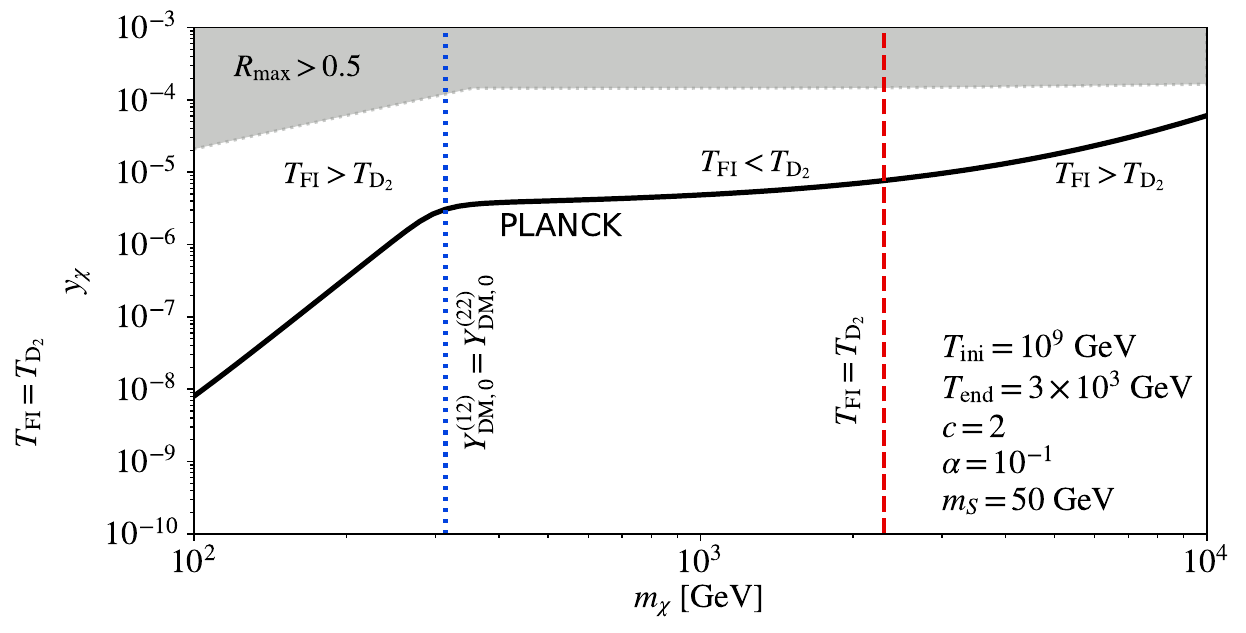}
			\caption{\label{fig:Planck_FFI_12_vs_22}}
		\end{subfigure}
		\caption{ The Planck lines for $\mSV = 50~\GeV \ll 2\mDM$, $\Ti = 10^9~\GeV$ for  $c=3$, $\TEND = 2 \times 10^4 ~\GeV$, $\alpha = 10^{-1}$ (a), and for  $c=2$, $\TEND = 2 \times 10^3 ~\GeV$, $\alpha = 10^{-1}$ (b). 
			The left panel shows that, away from entropy injection, $\yDM$ is approximately independent of $\mDM$. The right panel shows that increasing $\yDM$ (due to dilution) can result in the pair annihilation channel becoming dominant by reducing the freeze-in temperature. The result is a diluted DM population produced via $S \to \chi \chi$ and a non-diluted one from $S S \to \chi \chi$.   	The blue line indicates where both production channels contribute to the relic abundance in equal amounts.
			Note that if there is no change in the dominant channel, freeze-in temperature increases with the DM mass, since in both channels $\TFI \sim \mDM$ as discussed in the text.
		}
		\label{fig:PLANCK_FFI}
	\end{figure}

	This effect is shown in \Figs{fig:Planck_FFI_12} where we present the Planck line for $\Ti = 10^9~\GeV$, $\TEND = 2\times 10^4~\GeV$, $\mSV = 50~\GeV$, $c=3$ and $\alpha=10^{-1}$ along with the points where $\TFI$ coincides with $\TDII$ and $\TDI$ (red lines). As  expected, for $\TFI < \TDII$ $\yDM$ becomes approximately independent of $\mDM$, which also corresponds to the same value as in standard cosmological scenario (\ie DM is not diluted). As $\mDM$ increases, the freeze-in temperature also increases, and for $3 \times 10^2 ~\GeV \lesssim \mDM \lesssim 2 \times 10^3~\GeV $ the freeze-in temperature is between  $\TDI$ and $\TDII$. In this region the Yukawa coupling increases in order to compensate for the dilution due to entropy injection, until $\TFI>\TDI$. After this point, most of the DM is produced during radiation domination and $\yDM$ becomes approximately constant but still increased due to entropy injection that occurs entirely after freeze-in has ended. 
	
	In \Figs{fig:Planck_FFI_12_vs_22} we show the Planck line for $\Ti = 10^9~\GeV$, $\TEND = 3\times 10^3~\GeV$, $\mSV = 50~\GeV$, $c=2$ and $\alpha=10^{-1}$. The red line shows the DM mass where $\TFI = \TDII$ while the blue line indicates where both production channels contribute equally to the relic abundance.
	In this figure all effects discussed above can be clearly seen. In particular, for low $\mDM$ the  dominant channel is the decay of $S$ with $\TFI > \TDII$. Larger DM mass requires larger $\yDM$, because  DM becomes more diluted (since $\TFI \sim 2\mDM / \alpha$). Eventually, the Yukawa coupling becomes large enough and the pair annihilation channel starts to dominate at $\mDM \approx 300~\GeV$. At this point, the freeze-in temperature drops since for this channel $\TFI \approx \mDM$. Consequently, most DM is not diluted anymore and $\yDM$ becomes independent of $\mDM$ until DM becomes heavy enough to push the freeze-in to temperatures higher than $\DII$, $\mDM \approx 2 \times 10^3~\GeV$. For higher DM mass $\TFI$ increases causing DM to become more diluted, which results in larger values of the Yukawa coupling.

	\subsubsection*{Pair annihilation production and $\alpha$}
	\begin{figure}[b!]
		\centering	\includegraphics[width=.8\textwidth]{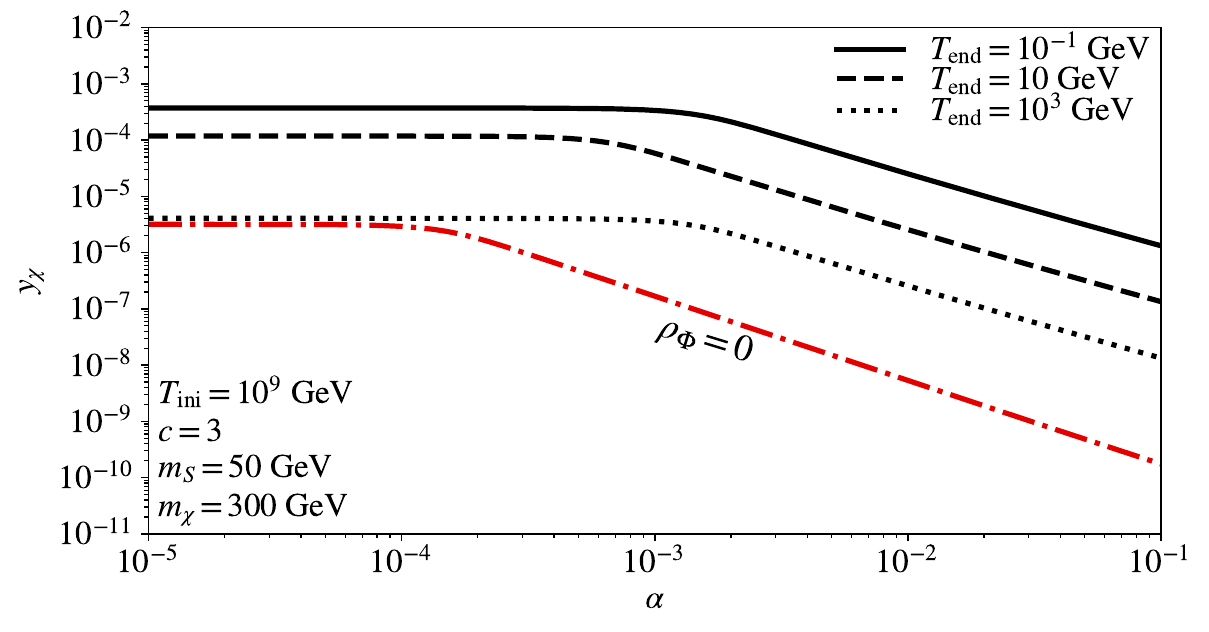}
		\caption{ The relationship between $\yDM$ and $\alpha$ that results in $\relic = 0.12$ for $\Ti = 10^9~\GeV$,  $c=3$,  $\mSV = 50~\GeV$ and $\mDM = 300~\GeV$.  The black lines corresponds to $\TEND=0.1~\GeV$ (solid), $\TEND=10~\GeV$ (dashed), and $\TEND=10^3~\GeV$ (dotted), while the red one (dashed-dotted) to the standard cosmological scenario.
			Note that taking smaller $\alpha$  requires assuming larger $\yDM$ in order to reproduce  the observed relic abundance, up to some point where they become independent. }
		\label{fig:Planck-a}
	\end{figure}
	At high temperatures $\GammaDM \sim \mST \approx \alpha  T $. This implies that the contribution of the kinematically forbidden decay channel to the relic abundance is proportional to some power of $\alpha$.~\footnote{In a radiation-dominated Universe this has been shown in ref.~\cite{Darme:2019wpd}.}
	Therefore, this channel becomes less efficient for smaller values of $\alpha$ and,  in order to maintain the correct relic abundance  the Yukawa coupling has to be bigger. 
	On the other hand, the thermalization constraint prohibits $\yDM$ from becoming too large. However, since the pair annihilation process can become dominant for a large enough coupling, we expect $\yDM$ to become independent of $\alpha$ before this constraint is reached.~\footnote{This is because $S S \to \chi \chi$ does not depend strongly on $\alpha$ since the production via this channel occurs at temperatures such that $\mST \ll \vev{E_S}$.}
	We illustrate the relationship between $\yDM$ and $\alpha$ that preserves the observed DM relic abundance in \Figs{fig:Planck-a} for $\Ti = 10^9~\GeV$,  $c=3$, $\mSV = 50~\GeV$, and $\mDM = 300~\GeV$. The black lines correspond to different values of $\TEND$, while the red one to the standard cosmological scenario. 
	As we can see, since entropy injection dilutes DM (if the freeze-in happens before $\DII$) a lower $\TEND$, which implies increased entropy injection (see \eg  \Figs{fig:gamma_TEND}), results in larger $\yDM$.
	As expected, in all cases $\yDM$ increases as $\alpha$ becomes smaller, up to some point where it becomes constant. This is the point where the pair annihilation channel starts to dominate the DM relic abundance.
	Note that, in the NSC scenarios this happens at a larger $\alpha$. This is a result  of larger $\yDM$ combined with  greater dilution of the DM population produced from the forbidden decays which terminate earlier than the pair annihilation.
	Finally, we note that  this transition leads to the lowering of the freeze-in temperature. For the line that corresponds to $\TEND = 10^3 ~\GeV$ this results in $\TFI<\TDII$ which brings $\yDM$ close to its standard cosmological value. 
	
	\subsection{Impact of the cosmological history }\label{sec:CosmoParameters}
	Since entropy injection is sensitive to the parameters that determine the cosmological scenario, we expect $\Ti$, $\TEND$ and $c$ to strongly affect the Planck lines when DM is diluted. 
	Qualitatively, values of these parameters that result in  $\gamma>1$, see, \eg \Figs{fig:gamma}, lead to the increase of $\yDM$ (given that $\TFI \gtrsim \TDII$) in order to compensate for the dilution of the DM number density.  

	\begin{figure}[t!]
		\centering 
		\begin{subfigure}[b]{0.5\textwidth}
			\centering\includegraphics[width=1\textwidth]{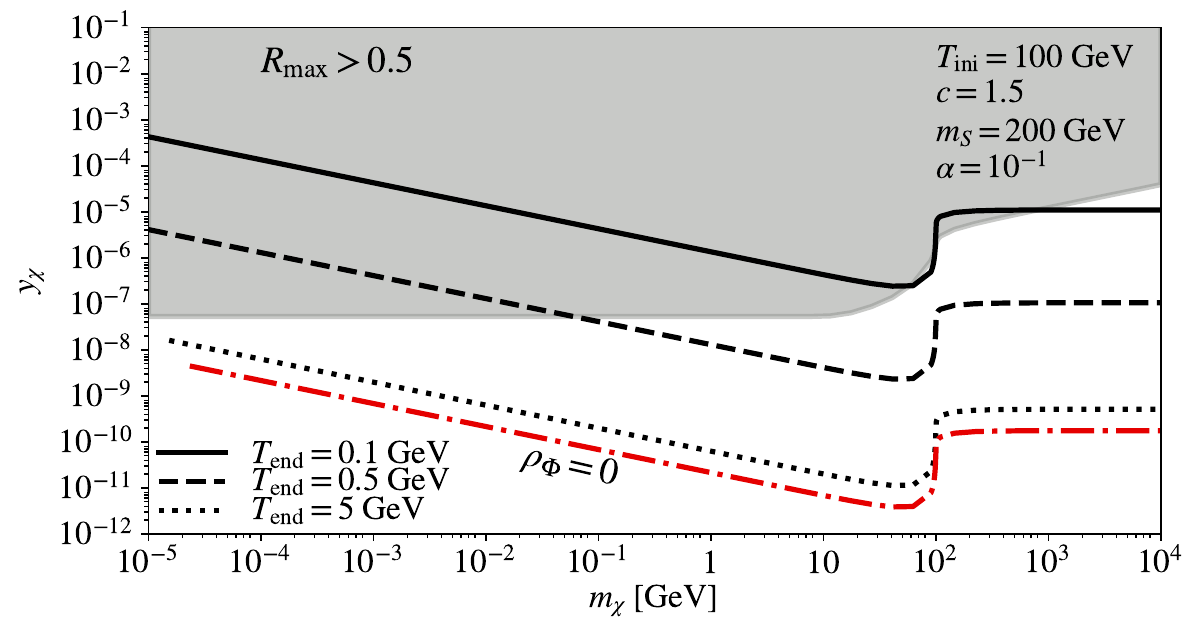}
			\caption{\label{fig:PlanckLines_low-Ti}}
		\end{subfigure}%
		\begin{subfigure}[b]{0.5\textwidth}
			\centering\includegraphics[width=1\textwidth]{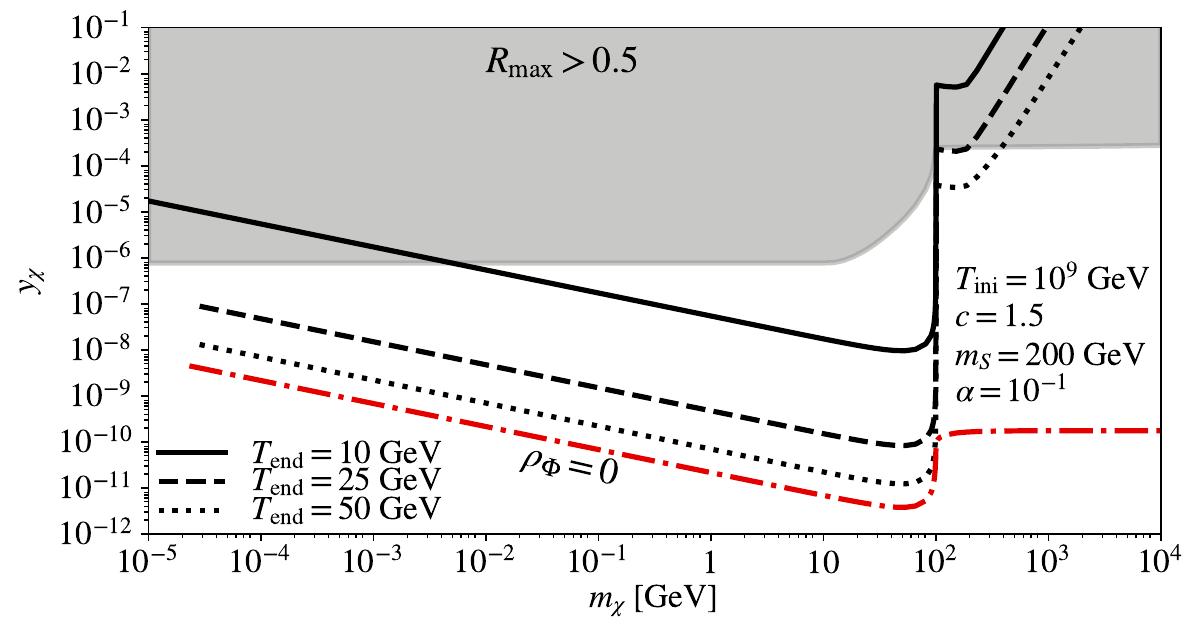}
			\caption{\label{fig:PlanckLines_high-Ti}}
		\end{subfigure}
		\caption{The Planck lines for  $c=1.5$, $\mSV = 200~\GeV$, $\alpha=10^{-1}$ and for $\Ti=100~\GeV$ (a) and $\Ti=10^9~\GeV$ (b). In the left panel the black lines correspond to $\TEND=0.1~\GeV$ (solid), $0.5~\GeV$ (dashed), and $=5~\GeV$ (dotted), while in the right panel they correspond to $\TEND=10~\GeV$ (solid), $25~\GeV$ (dashed) and $50~\GeV$ (dotted). In both panels the red line corresponds to the standard cosmological values of $\yDM$ and the gray area delineates the coupling where thermalization is possible.
			Note that the case $\rho_\Phi = 0$ results in the lowest  limit on $\yDM$ and that decreasing $\TEND$ pushes the Planck lines for NSCs scenarios towards it.  
		}
		\label{fig:PlanckLines-Ti}
	\end{figure}
	
	In \Figs{fig:PlanckLines-Ti} we show the Planck lines for $c=1.5$, $\mSV = 200~\GeV$, $\alpha=10^{-1}$ and for different values of $\TEND$ (black lines) for
	$\Ti=100~\GeV$ (a) and $\Ti=10^9~\GeV$ (b). The red line corresponds to the Planck line in the standard cosmological scenario while the gray area is the area where thermalization can occur.  
	As we have already mentioned, the Plank line that corresponds to the standard cosmological scenario  gives the lowest $\yDM$. As we have already seen, taking larger $\TEND$ leads to reducing the Yukawa coupling since the amount of entropy injection is reduced; see, \eg \Figs{fig:gamma_TEND}.

	Since in \Figs{fig:PlanckLines_low-Ti}  $\Ti$ is quite low, all DM is produced before the entropy injection begins, \ie at $\TFI>\TDI$. This gives Planck lines parallel to the case of $\rhoPhi = 0$. We note that, only in the diluted DM cases DM particle mass can be smaller than $\mDM \sim 22~\keV$ due to the more relaxed LSSF bound (similar to \Figs{fig:Planck_std-FI_Low-LS}). At lower mass, however, the thermalization bound is violated which becomes relevant for the cases  $\TEND=0.1~\GeV$ and $0.5~\GeV$.
	Furthermore, due to taking $ \TFI> \TDI$, $\Rmax$ is independent of $\TEND$ since DM production ends before the decays of $\Phi$ become relevant. For  $\mDM \ll \mSV/2 $ it is also independent of $\mDM$ since $\TFI \gg \mDM$.
	On the other hand, for $\mDM \gg \mSV/2$ the thermalization bound  increases with $\mDM$ since now $\chi$s are mostly relativistic when they are produced, with $R \sim \YDM$ -- since it obeys an equation similar to \eqs{eq:dYdT-std} -- which results in $\Rmax \sim \mDM^{-1}$.
	
	In \Figs{fig:PlanckLines_high-Ti} we observe a similar behavior for the Planck lines. However, for $\mDM \gg \mSV/2$ the pair annihilation channel dominates and $\yDM$ increases; compare \Figs{fig:Planck_FFI_12_vs_22}. The thermalization bound for $\mDM \ll \mSV/2$ depends weakly on $\TEND$ and it is only relevant for $\TEND=10~\GeV$, which is the only case that we show. 
	
	For $\mDM \gg \mSV/2$ the thermalization condition becomes independent of $\TEND$ since $\TFI \gg \TDII$. Interestingly, it is also independent of the DM particle mass. In order to understand this, we note again that $R$ obeys an equation similar to \eqs{eq:dYdT_with_dNdT}, \ie $\chi$ is relativistic, where the dilution term can become dominant before freeze-in, resulting in a maximum that does not depend on $\TFI$, \ie independent of $\mDM$. We should point out, however, that we expect that there exist cases where $\Rmax$ shows a non-trivial dependence on $\mDM$ because of the maxima that can occur, depending on the relationship between $\TFI$ and  $T_{\rm D_{1,2}}$, compare \Figs{fig:YDM}.

	\section{Parameter space}\label{sec:ParameterSpace}
	\setcounter{equation}{0}
	In this Section we perform a numerical analysis of the parameter space of the NSC scenario considered in this article. Our aim is to identify all regions where the correct relic abundance can be obtained, while taking into account the  thermalization and LSSF constraints discussed in the previous Sections.  In order to ensure that  $\Phi$  has decayed away before the process of nucleosynthesis begins, we will impose  $\TDII > 10~\MeV$. 
	In addition, results from Planck collaboration~\cite{Aghanim:2018eyx} constrain the scalar spectral index of a $\Lambda$CDM$+r$ model to $n_s = 0.9670 \pm 0.0074$. In order to apply this bound in our case we would normally need to specify a model of inflation that determines $n_s$. However, we can parameterize $n_s$ in terms of the so-called number of e-foldings $N_{k_*}$,~\footnote{It is defined as the number of e-foldings between the time when relevant perturbations, with $k_* \sim 0.05 {\rm Mpc}^{-1} $, left the horizon and the and the end of inflation.} in order to obtain a rough model independent bound in the NSCs under study. The number of e-foldings takes the form~\cite{Liddle:2003as,Rehagen:2015zma}
	\begin{equation}
		N_{k_*} = 57.6 + \dfrac{1}{4} \log r - \Delta N_{\rm reh} - \Delta N_{\Phi} \;, 
		\label{eq:Nkstar}
	\end{equation}
	with an experimental value for the tensor-to-scalar ratio $r \approx 0.064$~\cite{Akrami:2018odb} and $ \Delta N_{\rm reh}$ denoting the contribution from reheating, \ie the decay of the inflaton. The contribution due to the period of dominance of $\Phi$ is encoded in $\Delta N_{\Phi}$ which takes the form~\cite{Allahverdi:2018iod,Allahverdi:2019jsc}   
	\begin{equation}
		\Delta N_{\Phi} = \dfrac{4-c}{2 \ c} \log{ \dfrac{H_{\EI}}{H_{\EII}} } \;.
		\label{eq:DNPhi}
	\end{equation}
	Following refs.~\cite{Allahverdi:2018iod,Allahverdi:2019jsc}, we note that in general we expect $\Delta N_{\rm reh} \geq 0$~\cite{Podolsky:2005bw}. In addition, for some quite general~\cite{Roest:2013fha} classes of  single field inflation models -- see, \eg see~\cite{Starobinsky:1980te,Linde:1983gd,Goncharov:1983mw,Belinsky:1985zd} -- $N_{k_*} \approx 45$. Therefore, we obtain the approximate constraint 
	\begin{equation}
		\log{ \dfrac{H_{\EI}}{H_{\EII}} }  \lesssim \dfrac{24 \ c}{4-c} \;.
		\label{eq:DNPhi_bound_gen}
	\end{equation}
	It is clear that this bound goes away as $\Phi$ approaches a radiation-like behavior (\ie $c \to 4$), while it puts a strong constraint on cases with $c  \ll  1$. 
	
	\subsubsection*{DM related parameters}
\begin{figure}[t!]
		\centering 
		\begin{subfigure}[b]{0.5\textwidth}
			\centering\includegraphics[width=1\textwidth]{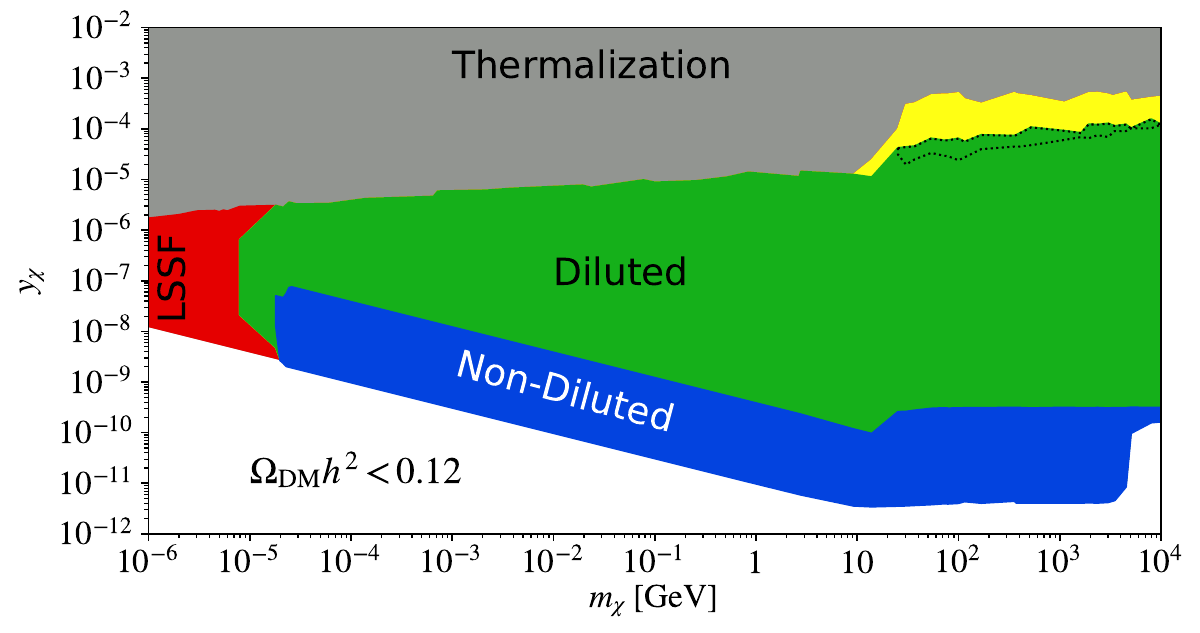}
			\caption{\label{fig:ParameterSpace_high-Ti_high-a}}
		\end{subfigure}%
		\begin{subfigure}[b]{0.5\textwidth}
			\centering\includegraphics[width=1\textwidth]{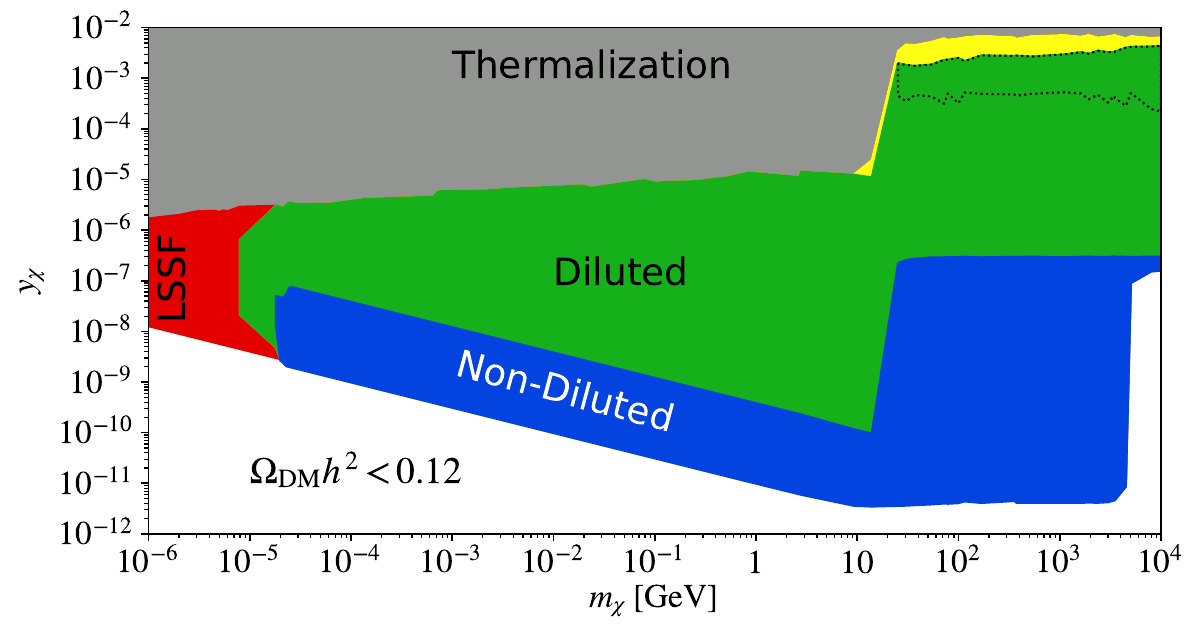}
			\caption{\label{fig:ParameterSpace_high-Ti_low-a}}
		\end{subfigure}
		\centering 
		\begin{subfigure}[b]{0.5\textwidth}
			\centering\includegraphics[width=1\textwidth]{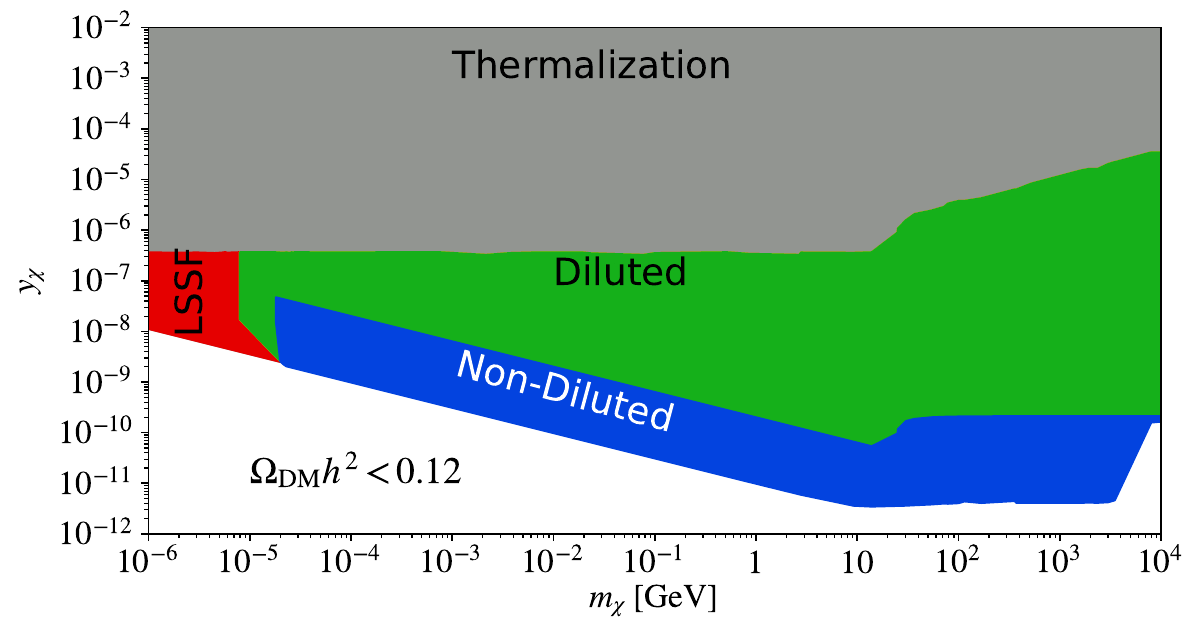}
			\caption{\label{fig:ParameterSpace_low-Ti_high-a}}
		\end{subfigure}%
		\begin{subfigure}[b]{0.5\textwidth}
			\centering\includegraphics[width=1\textwidth]{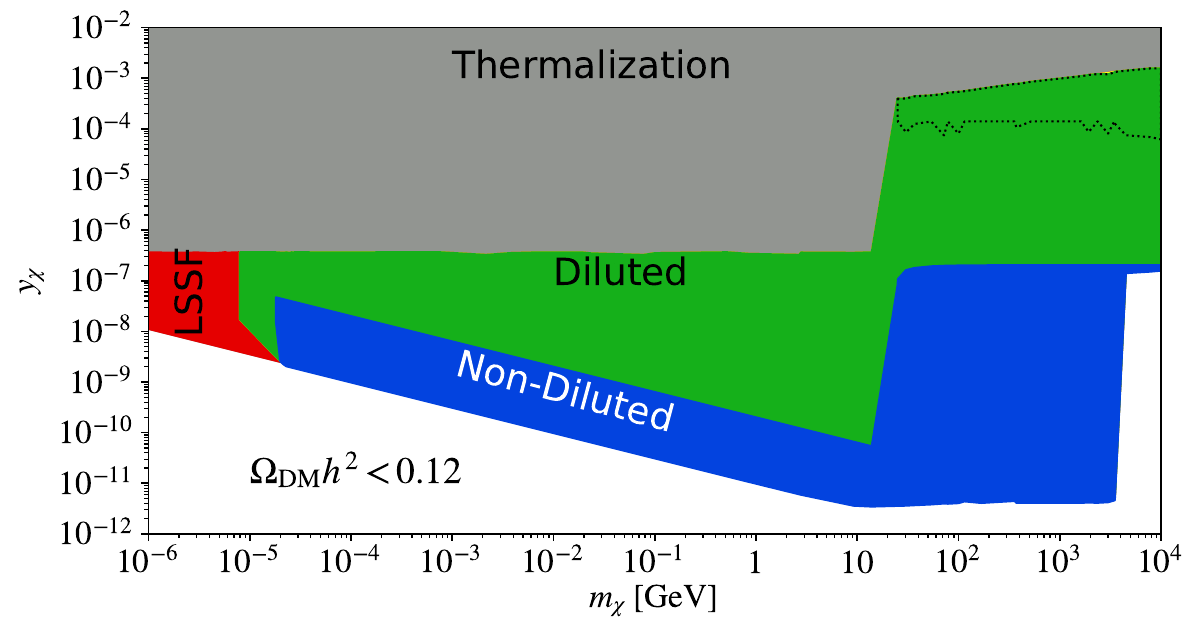}
			\caption{\label{fig:ParameterSpace_low-Ti_low-a}}
		\end{subfigure}
		\caption{The parameter space in the plane $(\mDM, \yDM)$ for: (a)  $\Ti = 10^9~\GeV$ and $\alpha =10^{-1}$; (b) $\Ti = 10^9~\GeV$ and $\alpha =10^{-3}$; (c) $\Ti = 10^2~\GeV$ and $\alpha =10^{-1}$; and (d) $\Ti = 10^2~\GeV$ and $\alpha =10^{-3}$. In all the panels we assume $10^{-2} ~\GeV \leq \TEND \leq \Ti$, $0.5 \leq c \leq 4$, $50 ~\GeV \leq \mSV\leq 10^4 ~\GeV$.
			In the gray regions the thermalization bound is violated. In the red regions the LSSF bound is not satisfied and in the yellow ones the bound~(\ref{eq:DNPhi_bound_gen}) is violated. 
			The areas delineated by the black dotted-lines show where the production via $SS\to \chi \chi$ dominates the DM relic abundance. 
			The green and blue regions are allowed by all the constraints, including $\relic = 0.12$, and correspond to the diluted and non-diluted DM cases, respectively. 
			The diluted DM region extends the allowed parameter space by both increasing $\yDM$ without leading to thermalization, as well as by relaxing the LSSF bound. In the white region DM relic density is lower than the observed one. 
		}
		\label{fig:ParameterSpace}
	\end{figure}

	We perform a numerical scan over $10 ~\MeV \leq \TEND \leq \Ti$, $0.5 \leq c \leq 4$, $50 ~\GeV \leq \mSV\leq 10^4 ~\GeV$ for four cases: (a)  $\Ti = 10^9~\GeV$ and $\alpha =10^{-1}$; (b) $\Ti = 10^9~\GeV$ and $\alpha =10^{-3}$; (c) $\Ti = 10^2~\GeV$ $\alpha =10^{-1}$, and (d) $\Ti = 10^2~\GeV$ and $\alpha =10^{-3}$. Our results are shown in \Figs{fig:ParameterSpace} in the plane $(\mDM, \yDM)$.
	In all panels the parameter regions where DM thermalization may take place ($\Rmax > 0.5$) are marked in gray. The red region is excluded by the LSSF constraint~\eqs{eq:SF-bound}. In yellow regions $\relic = 0.12$ while avoiding thermalization and the LSSF constraint but  \eqs{eq:DNPhi_bound_gen} is violated.~\footnote{We should point out that this  dependents on the model of inflation, \ie in general  the bound can be different.} 
	The allowed regions of the parameter space where all constraints are satisfied are marked green and blue, corresponding to diluted ($\TFI<\TDII$) and non-diluted ($\TFI>\TDII$) DM cases, respectively. In the region delineated by the black dotted-line the pair annihilation channel  dominates. In other words, this part of the parameter space is exclusively accessible due to contribution from this channel. 
	Finally, in the white region $\chi$ cannot constitute $100\%$ of the DM relic density of the Universe. 

	It should be noted that in all the panels of \Figs{fig:ParameterSpace} the diluted and non-diluted regions partly overlap, with the latter extending slightly below the former. The reason for this is that the non-diluted DM case corresponds to the standard cosmological scenario which results in the lowest possible values for the coupling, as was discussed in the previous Sections and can also be seen in \Figs{fig:YDM,fig:Planck-a,fig:PlanckLines-Ti}. Also, we should note that only the diluted DM case is  affected by the thermalization bound, because the observed relic abundance for non-diluted DM is obtained for suppressed Yukawa couplings, at least for the mass ranges that we focus on.
	Furthermore, entropy injection can lead to a relaxed LSSF constraint, as discussed also in Section~\ref{sec:std-FI}, when the thermalization bound is not reached. This results in a lower bound around  $\mDM \approx 7~\keV$. 

	In the high $\Ti$ case shown in \Figs{fig:ParameterSpace_high-Ti_high-a,fig:ParameterSpace_high-Ti_low-a} the two choices of $\alpha$ result in identical allowed regions for  $\mDM < 25~\GeV$. For heavier DM these regions depend strongly on $\alpha$ which plays an important role in the kinematically forbidden decays. 
	We note that reducing $\alpha$  by two orders of magnitude shift the maximum allowed coupling from $\yDM \approx  10^{-4}$ to $\yDM \approx 2 \times 10^{-3}$. This happens because the efficiency of the production via forbidden decays depends on $\alpha$, as discussed in Section~\ref{sec:FFI}.
	We also point out that lower values of $\alpha$ yield approximately the same maximum $\yDM$ since the pair annihilation channel now takes over, as shown in \Figs{fig:ParameterSpace_high-Ti_low-a}. 
	Moreover, we observe that,  for $\mDM \gg 25~\GeV$ the maximum value of $\yDM$ changes slowly.  This is caused by the fact that $\mDM$ is independent of $\Rmax$ for $\mDM \gg \mSV /2$, as it is also shown in \Figs{fig:PlanckLines_high-Ti}.
	The lower limit on $\yDM$ corresponds mostly to the standard cosmological scenario. For the region accessible solely by the forbidden decays ($\mDM > 5~\TeV$), the minimum Yukawa coupling increases more strongly for lower values of $\alpha$. 
	We also observe that the constraint~\eqs{eq:DNPhi_bound_gen} only slightly affects the parameter space in the region where $\yDM$ is around its maximum  since this corresponding to a longer period of $\Phi$ dominance.

	The low $\Ti$ case, presented in \Figs{fig:ParameterSpace_low-Ti_high-a,fig:ParameterSpace_low-Ti_low-a}, exhibits basically similar features, however, with the allowed values of $\yDM$ generally being now lower due to the effect of the thermalization constraint. This is also shown in \Figs{fig:PlanckLines-Ti}. 
	To understand this, we consider the case $\mSV \gg 2 \mDM$ with $\TDI \gg \TFI$, \ie the case of the highest possible dilution. In this case $\Rmax$ is reached before entropy injection starts, \ie $\nDM^{\rm eq} \sim T^3 \sim  a^{-3}$. On the other hand, if $\Ti=10^9~\GeV$, there are values of $\TEND$ and $c$ for which, although the amount of dilution is the same, $\TFI<\TDI$, and $\Rmax$ is obtained in an era when the temperature decreases more slowly due to the energy injection, \ie $\nDM^{\rm eq}$ decreases more slowly than $a^{-3}$. Therefore, at high $\Ti$ there is a region where, given the same amount of dilution, $\Rmax$ is lower compared to the low $\Ti$ case. 
	For $\mSV \ll 2\mDM$ and low $\Ti$, the freeze-in temperature $\TFI \approx 2\mDM/\alpha$ is always higher than $\Ti$ and $\Rmax$ is inversely proportional to the DM mass, which relaxes the bound for higher masses, as explained in the discussion around \Figs{fig:ParameterSpace_high-Ti_low-a}.  We also note that the constraint from the number of e-foldings leaves this case unaffected since the dominance of $\Phi$ is now limited due to the low value of $\Ti$.
	
	\subsubsection*{NSC related parameters}

	\begin{figure}[t!]
		\centering 
		\begin{subfigure}[b]{0.5\textwidth}
			\centering\includegraphics[width=1\textwidth]{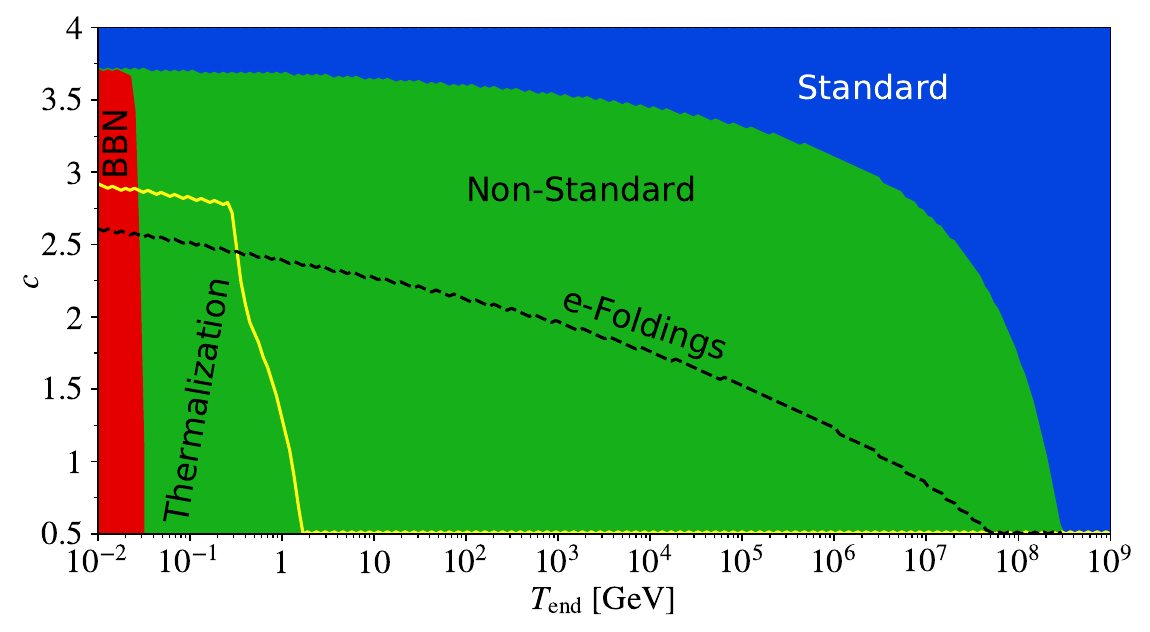}
			\caption{\label{fig:TEND_vs_c_high-Ti}}
		\end{subfigure}%
		\begin{subfigure}[b]{0.5\textwidth}
			\centering\includegraphics[width=1\textwidth]{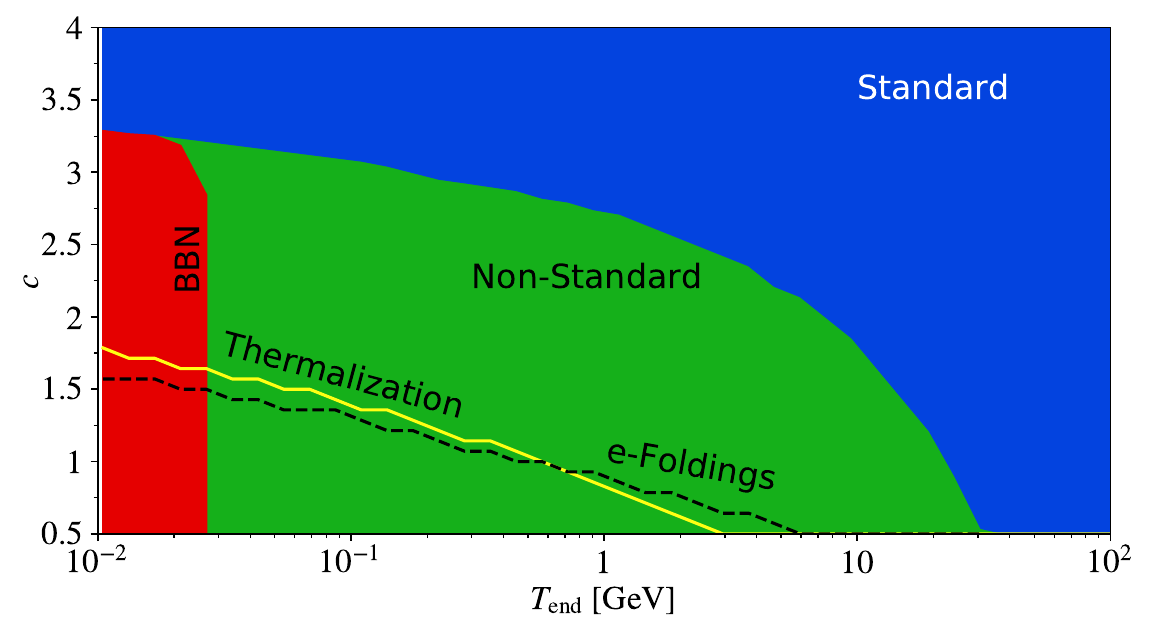}
			\caption{\label{fig:TEND_vs_c_low-Ti}}
		\end{subfigure}
		\caption{The parameter space in the $(\TEND, c)$ plane for $\alpha=10^{-1}$ and for (a) $\Ti=10^9~\GeV$ and (b) $\Ti=10^2~\GeV$. We scan over the other parameters in the following ranges: $50 ~\GeV \leq \mSV\leq 10^4 ~\GeV$, $10^{-6} ~\GeV \leq \mDM \leq 10^4 ~\GeV$, while $\yDM$ is chosen such that $\relic =0.12$. 
			The red region is excluded by the BBN constraint ($\TDII>10~\MeV$). In the region below the yellow curve  thermalization of DM is possible, while under the black line the bound on the number on e-foldings is violated. The green region corresponds to the NSCs scenario with significant entropy injection, while in the blue one it is almost identical to the standard cosmological scenario.
		}
		\label{fig:TEND_vs_c}
	\end{figure}

	The parameters $\TEND$ and $c$  that characterize the NSC scenario under study are shown in \Figs{fig:TEND_vs_c} for (a) $\Ti = 10^9~\GeV$  and (b) $\Ti = 10^2~\GeV$. In order to relate these parameters with the dark sector ones, we set  $\alpha= 10^{-1}$ and scan over the rest of the parameters as in \Figs{fig:ParameterSpace}. 
	The blue and green regions show the allowed parameter space in the standard and NSCs scenarios, respectively.~\footnote{Since  in the presence of $\Phi$ entropy injection is always non-zero, we  define the standard cosmological scenario as the case where the energy injection rate does not reach the level of $10\%$ of the dilution of $\rhoR$.} In the red region the condition $\TDII > 10~\MeV$ is violated, while the dark sector parameters only exclude the region below the yellow curve which violates the thermalization bound. 
	The cosmological constraint~(\ref{eq:DNPhi_bound_gen}) is violated below the black lines. At high $\Ti$ this constraint is quite severe and disfavors a large region of the $(\TEND, c)$ plane. At low $\Ti$ the ratio $H_{\EI} / H_{\EII}$ is small, since $\Phi$ becomes dominant only briefly compared to the high-$\Ti$ case, and the constraint~(\ref{eq:DNPhi_bound_gen}) has a limited effect on the parameter space. We also note that in this case the yellow and black lines due to the number of e-foldings and thermalization, respectively, lie close to each other, which is why \eqs{eq:DNPhi_bound_gen} does not play much role in the low $\Ti$ examples shown in \Figs{fig:ParameterSpace_low-Ti_high-a,fig:ParameterSpace_low-Ti_low-a}  where the  yellow region is almost non-existent.  
	Note that small values of $c$ and $\TEND$ are disfavored by the thermalization constraint, while $\TEND \lesssim 30~\MeV$ is not allowed as long as the decays of $\Phi$ affect, even slightly, the plasma. 	 

	As $\TEND$ increases, even for very small $c < 1$, there is a region where $\Phi$ decays very early, or  never dominates, and the standard cosmological scenario is not affected. Furthermore, as $c$ increases  $\Phi$ tends to behave more and more as a radiation component and at some point ($c \gtrsim 3$)  even at extremely low $\TEND$ the radiation component becomes unaffected.  
	Finally, clearly  the constraint $\TDII > 10 ~\MeV$  in both cases is approximately translated to $\TEND \gtrsim 30~\MeV$ as long as the NSCs scenario is realized, \ie below the blue region. This constraint overlaps with the thermalization bound for moderate values of $c$ which extends up to $\TEND \approx 2$ or  $3~\GeV$.  This depends on $\alpha$ but it should be within the same order of magnitude even for smaller values of the parameter.
	
	\section{Summary}\label{sec:sum}
	\setcounter{equation}{0} 
	In this article we studied in detail the impact of a decaying fluid on the process of DM production via freeze-in, assuming that the DM particle  is a Majorana fermion produced solely in decays and pair annihilations of a scalar, via its Yukawa interactions, that remains in thermal equilibrium with plasma. Assuming that, after an initial radiation-dominated expansion, the fluid dominates the energy density of the Universe until it decays away, increasing the entropy of the plasma, we examined the evolution of the Universe and identified the points where the  behavior of the $\Phi$-radiation system qualitatively changes the evolution relative to the standard radiation-dominated case. Next, we studied in detail the DM production as well as the evolution of its mean momentum. We showed that the entropy injection to the plasma can greatly affect both quantities since the temperature of the plasma now deviates from its standard cosmological scaling.  As a result we showed that, compared to the standard cosmological scenario, DM population becomes always more diluted today. It also  exhibits a lower mean momentum due to faster redshift. This leads to a parameter space where the DM coupling with the scalar becomes larger -- in order to compensate for the dilution -- and to a relaxed LSSF bound on the DM mass, due to the rapid redshift of its momentum.
	
	We also showed that entropy injection can dilute the DM population produced from decays, in the case where the decays are allowed only due to plasma effects, resulting in a dominant contribution from $SS \to \chi \chi$,  which  is usually a subdominant production channel.
	
	Next, we demonstrated that, in most cases entropy injection causes also the DM momentum to redshift faster compared to the plasma temperature. However, we also pointed out the presence of a finely tuned case (for $\mSV > 2\mDM$) where DM production stops close to $\DII$ with the mean DM momentum being slightly enhanced.
	
	Following this, we studied a few benchmark points that represent different regions of the parameter space. First we examined the case of light DM ($\mDM < \mSV/2$), where the dominant production channel is the decay of $S$. Furthermore, this is the only case where the LSSF bound  is relevant, as it applies for DM mass in the range of a few $\keV$.   We showed how the Planck lines behave and pointed-out that there are cases where the LSSF bound can be both slightly enhanced and relaxed.  However, it became apparent that the LSSF constraint cannot decrease by an arbitrary amount, as this happens only when DM population is diluted, which requires taking larger $\yDM$ in which case the thermalization constraint is almost always violated.
	We also examined the heavy DM scenario ($\mDM>\mSV / 2$). In this case we suggested that the LSSF bound is always satisfied and we examined how the DM dilution affects the Planck lines. We argued that, in the limit $\mDM \gg \mSV$ the Yukawa coupling should be  independent of $\mDM$ regardless of the dominant production channel and showed that deviation from this occurs if the freeze-in ends between $\DI$ and $\DII$ (\Figs{fig:Planck_FFI_12}), \ie when the amount of dilution changes with the DM mass which determines $\TFI$ in the heavy DM regime.   Moreover, we showed how the pair annihilation channel dominates when the DM population produced via decay becomes diluted. We also argued that  $\yDM$ increases as the thermal contribution to the mass of $S$ becomes smaller and showed that eventually the pair annihilation becomes dominant and limits $\yDM$ to some maximum value. 
	We also examined the effect of the parameters of the NSC scenario on the different Planck lines.  We showed once again that $\yDM$ increases with the amount of entropy injection.
	
	In a numerical scan over the parameter space we distinguished two main cases with relatively high and low $\Ti$. In both cases we examined two values of the parameter $\alpha$ that affect the Yukawa coupling for heavy DM as well as the region where the pair annihilation channel can dominate.
	We found that both parameters relevant to DM, \ie $\mDM$ and $\yDM$, are affected by the decaying fluid.  Our  results showed that, the Yukawa coupling is allowed to be much larger in the diluted DM population case, compared to the non-diluted case. Moreover, we showed that a region of very light DM of $\mDM \approx 7~\keV$ opens up, which is a result of the redishift of the DM momentum that corresponds to slightly diluted DM. 
	Moreover, we examined the regions of the allowed  $(\TEND,c)$ plane, taking into account a cosmological bound on the number of e-foldings. We found that a wide range of NSC parameters are allowed as long as $\TEND \gtrsim 30 ~\MeV$, while small values of $c$, depending on $\TEND$, are disfavored as they tend to make the Universe to expand faster than observations allow.
	
	In summary, we showed that the presence of a decaying fluid can strongly affect the DM interaction strength with the plasma that is allowed by the relic density, as well as its mean momentum. This opens up new regions of the parameter space of a model which are inaccessible assuming a standard cosmological history.  In particular, DM particle mass is allowed to values below the standard LSSF bound.
	Finally, our findings may have observational implications for detection prospects of DM produced via freeze-in since the allowed couplings can now be larger.

	\section*{Acknowledgments}
	DK and LR are supported in part by the National Science Centre, Poland, research grant No.  2015/18/A/ST2/00748.   LR is also supported  by  the  project AstroCeNT:  Particle  Astrophysics  Science  and  Technology Centre,  carried  out  within  the  International  Research  Agendas  programme  of  the Foundation for Polish Science financed by the European Union under the European Regional Development Fund. PA is thankful  to AstroCeNT for their hospitality and acknowledges support from the Polish National Agency for Academic Exchange through their Ulam Programme Scholarship and FONDECYT project 1161150.
	
	\setcounter{section}{0}
	\section*{Appendix}
	\appendix
	
	\renewcommand{\theequation}{\Alph{section}.\arabic{equation}}
	\setcounter{equation}{0}  
	
	\section{Approximate evolution of the energy densities}\label{app:approx}
	\setcounter{equation}{0}

	\subsubsection*{Initial Radiation Domination}
	We assume at some high temperature (\eg after inflation)  the energy density of the Universe  is dominated by radiation. Since this happens at high temperature, the decay rate of $\Phi$ is negligible compared to the expansion rate of the Universe ($\GammaPhi \ll H$), otherwise $\Phi$ would decay away quickly and a period of $\Phi$ dominance would be impossible. The energy densities evolve as  
	\begin{align}
		\rho_{R} &= \rho_{R,\ini}  \lrb{ \dfrac{\ai}{a} }^4 \nonumber \\
		\rho_{\Phi} &= \rho_{\Phi,\ini}  \lrb{ \dfrac{\ai}{a} }^c\;,
		\label{eq:energy_densities_R-domination}
	\end{align}
	with $\rho_{\Phi,\ini} \ll \rho_{R,\ini}$ the initial values of the energy densities at some $a=\ai$.
	As the Universe expands,  and since we have assumed $c<4$, the two energy densities become equal at $a=a_{\EI}$ given by 
	\begin{eqnarray}
		a_{\EI} &= \ai \lrb{ \dfrac{\rho_{R,\ini} }{\rho_{\Phi,\ini} } }^{\frac{1}{c-4}}  \; ,
		\label{eq:aE1_approx}
	\end{eqnarray}
	from which we obtain
	\begin{align} 
		\rho_{\EI} &= \rho_{R,\ini} \lrb{ \dfrac{\rho_{R,\ini} }{  \rho_{\Phi,\ini} } }^{\frac{4}{4-c}}   =   \rho_{\Phi,\ini} \lrb{ \dfrac{\rho_{R,\ini} }{ \rho_{\Phi,\ini}} }^{\frac{c}{4-c}} \;. \label{eq:E1}
	\end{align}

	\subsubsection*{Fluid Domination}
	After $\EI$, we consider the Universe to be dominated by $\Phi$, \ie $H \approx H_{\EI} \sqrt{ \lrb{ \dfrac{\rho_{\Phi}}{ \rho_{\EI} } }  }$, and the evolution  of the energy densities is given by 
	\begin{align}
		\dfrac{d \log \rhoPhi}{d \log \lrb{ \frac{a_{\EI}}{a}}} &= c + \dfrac{\GammaPhi}{H_{\EI}} \sqrt{ \dfrac{\rho_{\EI}}{\rhoPhi} }  \nonumber \\ 
		\dfrac{d \log \rhoR}{d \log \lrb{\frac{a_{\EI}}{a}} } &= 4 - \dfrac{\GammaPhi}{H_{\EI}} \sqrt{ \dfrac{\rho_{\EI}}{\rhoPhi} } \dfrac{\rhoPhi }{\rhoR} \;. 
		\label{eq:BEs_Phi-domination}
	\end{align}
	Before solving this system, we note here that for $a$ close to $a_{\EI}$ there should be a period where the decays do not affect radiation and both components continue to evolve as 
	\begin{subequations}
		\begin{align}
			\rhoR &= \rho_{\EI} \lrb{\dfrac{a_{\EI}}{a}}^4 \label{eq:rhoR_Phi-domination} \\
			\rhoPhi &= \rho_{\EI} \lrb{\dfrac{a_{\EI}}{a}}^c \label{eq:rhoPhi_Phi-domination} \;.
		\end{align}\label{eq:aE1r<a<aD1}
	\end{subequations}
	This period ends when $\dfrac{\GammaPhi}{H}  \dfrac{\rhoPhi }{\rhoR} = \dfrac{4}{10}$  at  $a=a_{\DI}$, which is
	\begin{equation}
		a_{\DI} = a_{\EI} \lrb{\dfrac{4 }{10} \dfrac{ H_{\EI}}{ \GammaPhi} }^{\frac{2}{8-c} }  \;,
		\label{eq:aD1_approx}
	\end{equation}
	where we observe that $a_{\DI}>a_{\EI}$ (\ie energy injection is significant since  $\Phi$ dominates) needs $\GammaPhi \lesssim \dfrac{4}{10} \ H_{\EI}$.

	During the era of $\Phi$ domination the system of equations~(\ref{eq:BEs_Phi-domination}) can be solved exactly by~\footnote{We have to assume here $c \neq 0$ in order to simplify the solution.}
	\begin{align} 
		\rhoR = &\rho_{\EI} \lrb{\dfrac{a_{\EI}}{a}}^{4}  \lrBiggcb{ 1  - \dfrac{ 2 \GammaPhi }{(8-c) H_{\EI}} \lrb{1 - \lrb{\dfrac{a_{\EI}}{a}}^{\frac{1}{2}(c-8)} } + \\
			&\dfrac{1}{c} \lrb{ \dfrac{\GammaPhi }{2 H_{\EI} } }^2 \lrsb{1 - \lrb{\dfrac{a_{\EI}}{a}}^{ -4 }  - \dfrac{8}{8-c} \lrb{1-  \lrb{\dfrac{a_{\EI}}{a}}^{\frac{1}{2}(c-8) }  }  }   }  \nonumber \\[0.5cm]
		\rhoPhi = &\rho_{\EI}  \lrb{\dfrac{a_{\EI}}{a}}^{c} \lrsb{  1- \dfrac{\GammaPhi}{c \; H_{\EI}}   
			\lrb{ \lrb{\dfrac{a_{\EI}}{a}}^{-c/2} -1 } }^2   \;.\label{eq:energy_densities_Phi-domination}
	\end{align}
	The form of these energy densities can give us a few important results. First, we get the following approximate behavior away from $a=a_{\EI}$ (assuming $\GammaPhi \ll H_{\EI}$)  
	\begin{align}
		\rhoR & \approx \rho_{\EI}   \lrsb{ \lrb{\dfrac{a_{\EI}}{a}}^{4 }  + \dfrac{2 \GammaPhi }{(8-c) H_{\EI}} \lrb{\dfrac{a_{\EI}}{a}}^{ c/2 } -
			\dfrac{1}{c} \lrb{ \dfrac{\GammaPhi}{2 H_{\EI} } }^2     } 
		\label{eq:Approx-energy_densities_Phi-domination} \\
		\rhoPhi & \approx \rho_{\EI} \lrsb{  \lrb{\dfrac{a_{\EI}}{a}}^{ c} - \dfrac{2\GammaPhi}{c \; H_{\EI}}  \lrb{\dfrac{a_{\EI}}{a}}^{c/2} + \lrb{\dfrac{\GammaPhi}{c \; H_{\EI}}}^2  }  \;,
		\nonumber
	\end{align}
	where as expected the first term in both $\rhoPhi$ and $\rhoR$ is due to the expansion of the Universe, the second term is due to the decay of $\Phi$ (positive for radiation and negative for fluid). The last term in both $\rhoPhi$ and $\rhoR$ has the opposite sign than one would expect, since we expect the decays of $\Phi$ to increase the energy of radiation, while decreasing  the energy of $\Phi$.  However, these terms emerge because, as $\Phi$ decays away, the expansion of the Universe slows down, and as a result the  rate at which $\GammaPhi/H$ increase slows down. This effect is also the reason that $\rhoPhi$ does not decay exponentially. We should point out that these terms cannot take over the evolution of the energy densities since we have neglected the contribution of radiation in the Hubble parameter, which should dominate before these terms become significant.
	
	Another result we can obtain from \eqs{eq:Approx-energy_densities_Phi-domination} is the second point of equality ($\EII$),  \ie $\rhoPhi = \rhoR$, assuming it happens at $a=a_{\EII} \gg a_{\EI}$. Keeping only the highest orders of $\lrb{\frac{a}{a_{\EI}}}$, we obtain~\footnote{ We should point out that there are two solutions for $a_{\EII}$ but the other solution  corresponds to a nonphysical increase of $\rhoPhi$ due to the third terms of \eqs{eq:Approx-energy_densities_Phi-domination}. }
	\begin{eqnarray}
		a_{\EII} \approx a_{\EI}  \lrBiggcb{  \dfrac{2c \lrb{ 16 - c \sqrt{12-c} }  }{(8-c)(c+4)}     \dfrac{H_{\EI}}{\GammaPhi} }^{\frac{2}{c}} \;,
		\label{eq:aE2_approx}
	\end{eqnarray}
	with $\rho_{\EII}$ being
	\begin{eqnarray}
		&\rho_{\EII} \approx \rho_{\EI}   \lrsb{ \dfrac{4 \lrb{ 4 +  \sqrt{12-c} } -c  }{ 4(8-c)^2 }   } \lrb{ \dfrac{\GammaPhi}{H_{\EI}} }^2 \;.
		\label{eq:Quantities_E_2}
	\end{eqnarray}
	
	\subsubsection*{Final Radiation Domination}
	The expansion of the Universe after $\EII$ is dominated by radiation, with $H \approx H_{\EII} \sqrt{\dfrac{\rhoR}{ \rho_{\EII} } } = H_{\EII} \lrb{\dfrac{T}{ T_{\EII} }}^2 $. For some time the decays of $\Phi$ are still important, and   we cannot assume that $\rhoR \sim a^{-4}$. During this period it is convenient to solve for the temperature of the plasma instead of $\rhoR$. The system of equations is
	\begin{align}
		\dfrac{d \log \rhoPhi}{d\log \lrb{\frac{a_{\EII}}{a}}} &= c + \dfrac{\GammaPhi}{ H_{\EII} } \lrb{\dfrac{T}{ T_{\EII} }}^{-2}   \nonumber \\ 
		\dfrac{d \log T}{d\log \lrb{\frac{a_{\EII} }{a}}} &= 1 - \dfrac{1}{4} \dfrac{\GammaPhi}{ H_{\EII} } \dfrac{\rhoPhi}{  \rho_{\EII} } \lrb{\dfrac{T}{ T_{\EII}}}^{-6} \;. 
		\label{eq:BEs_R-domination}
	\end{align}
	Expressing $T = T_{\EII} \lrb{\frac{a_{\EII}}{a}} \  f(a)$, the above equations take the form
	\begin{subequations}
		
		\begin{align}
			\dfrac{d \log \rhoPhi}{d\log \lrb{\frac{ a_{\EII} }{a}}} &= c + \dfrac{\GammaPhi}{ H_{\EII} } \lrb{\frac{a_{\EII}}{a}}^{-2} f^{-2}(a)  
			\label{eq:drhoPhidu_R-domination-f} \\ 
			\dfrac{d f^6(a)}{d\log \lrb{\frac{a_{\EII}}{a}}} &= - \dfrac{3}{2} \dfrac{\GammaPhi}{ H_{\EII} } \dfrac{\rhoPhi}{  \rho_{\EII} } \lrb{\frac{a_{\EII}}{a}}^{-6} \label{eq:df6dz} \;, 
		\end{align}\label{eq:BEs_R-domination-f}
	\end{subequations}
	The formal solution for the energy density of $\Phi$ is 
	\begin{equation}
		\rhoPhi = \rho_{\EII}  \lrb{\dfrac{a_{\EII}}{a}}^{c} e^{\dfrac{\GammaPhi}{H_{\EII}} \int_{0}^{z} dz^{\prime} e^{-2z} f^{-2}(z) } \; ,
		\label{eq:eq:rhoPhiFinal_R-domination-formal}
	\end{equation}
	where we have introduced $z= \log \lrb{\dfrac{a_{\EII}}{a}}$. Assuming that $\Phi$ decays fast, we only need the solution of $|z| \ll 1$. Thus, using $f(z) \approx 1$, we obtain
	\begin{equation}
		\rhoPhi \approx \rho_{\EII} e^{c   z} e^{\frac{\GammaPhi}{ 2 H_{\EII}} \lrb{ 1 - e^{-2z} } }   \; .
		\label{eq:rhoPhiFinal_R-domination}
	\end{equation}
	Expanding $1-e^{-2z} \approx 2z$, \eqs{eq:df6dz} becomes
	\begin{equation}
		\dfrac{d f^6(z)}{dz} \approx - \dfrac{3}{2} \dfrac{\GammaPhi}{ H_{\EII} } e^{   \lrb{c-6 + \frac{\GammaPhi}{ H_{\EII} }  }z} \;, 
		\label{eq:df6dz_approx}
	\end{equation}
	with a solution
	\begin{equation}
		f^6(z) = 1 - \dfrac{3}{2} \dfrac{\GammaPhi}{ \GammaPhi + H_{\EII} (c-6) } \lrb{ e^{ \lrb{ \frac{\GammaPhi}{ H_{\EII} } +c-6 }z} -1}\;. 
		\label{eq:f6_approx}
	\end{equation}
	Finally, the temperature becomes
	\begin{equation}
		T \approx T_{\EII} \ e^z \lrsb{ 1 - \dfrac{3}{2} \dfrac{\GammaPhi}{ \GammaPhi + H_{\EII} (c-6) } \lrb{ e^{ \lrb{ \frac{\GammaPhi}{ H_{\EII} } +c-6 }z} -1} }^{1/6} \;.
		\label{eq:eq:TFinal_R-domination}
	\end{equation}
	Therefore, the energy density of the plasma (for $a \approx a_{\EII}$) take the approximate form
	\begin{align}
		\rhoR &\approx  \rho_{\EII} \; \lrb{\frac{a_{\EII}}{a}}^{4} \lrsb{ 1 -  \dfrac{\GammaPhi}{ \GammaPhi + H_{\EII}(c-6) } 
			\lrb{ \lrb{\frac{a_{\EII}}{a}}^{ \lrb{ \frac{\GammaPhi}{ H_{\EII} } +c-6 }} -1} } \;. \label{eq:rhoRFinal_R-domination}
	\end{align}
	Since we have assumed that $\Phi$ decays quickly, there is a value $a_{\DII} \approx a_{\EII}$  where the decays of $\Phi$ stop affecting $\rhoR$. Although this process happens gradually, with no hard cutoff, we can define it numerically (as also mentioned in Section~\ref{sec:RhoPhi})  as the point where the energy injection rate to the plasma drops below $10 \%$ of its dilution rate, \ie $\dfrac{\GammaPhi}{H} \dfrac{\rhoPhi}{\rhoR} \Big|_{a = a_{\DII}} = \dfrac{4}{10}  $. 
	Even using the approximations \refs{eq:rhoPhiFinal_R-domination, eq:rhoRFinal_R-domination}, $\DII$ needs to be calculated numerically. However, a rough estimate is obtained by expanding $\dfrac{\GammaPhi}{H} \dfrac{\rhoPhi}{\rhoR}$ around $a = a_{\EII}$. Keeping terms up to 
	$\mathcal{O}( \log \frac{a_{\EII}}{a} )$ the scale factor at $\DII$ is
	\begin{equation}
		a_{\DII} \approx a_{\EII} \exp\lrb{ \dfrac{2}{5}\dfrac{  H_{\EII} }{ \GammaPhi } \dfrac{2 H_{\EII} -5\GammaPhi}{5\GammaPhi +2(c-6)H_{\EII}  } } \;.
		\label{eq:aD2_approx}
	\end{equation}

	During the final era, $a>a_{\DII}$,  $\rhoR$ can be considered  free with
	\begin{equation}
		\rhoR = \rho_{R, \DII }  \lrb{\frac{a_{\DII}}{a}}^4\;, 
		\label{eq:rhoRFree_R-domination} 
	\end{equation}
	while the evolution of $\rhoPhi$ is given by
	\begin{equation}
		\dfrac{d \log \rhoPhi}{d\log \lrb{\frac{a_{\DII}}{a} }} = c + \dfrac{\GammaPhi}{ H_{\DII} } \lrb{\frac{a_{\DII}}{a}}^{-2} \; .
		\label{eq:drhoPhidu_free}
	\end{equation}
	Thus, $\Phi$ continues to decay with an energy density
	\begin{equation}
		\rhoPhi = \rho_{ \Phi , \DII }  \lrb{ \frac{a_{\DII} } {a} }^c \; e^{- \frac{1}{2}\frac{\GammaPhi}{H_{\DII}} \lrsb{ \lrb{ \frac{a_{\DII} }{a} }^{-2} -1  }  }    \;.
		\label{eq:rhoPhiFree_R-domination} 
	\end{equation}

	Finally, it is worth pointing out that  the increase of the plasma entropy  between $\DI$ and $\DII$ (\ie approximately as the energy transfer is active) is estimated by
	\begin{equation}
		\gamma \equiv \dfrac{S_{\DII} }{S_{\DI} } =\dfrac{s_{\DII} \ a_{\DII}^3}{s_{\DI} \ a_{\DI}^3} \approx \lrb{ \dfrac{\rho_{R,\DII}}{\rho_{R,\DI}}  }^{3/4} \lrb{ \dfrac{a_{\DII}}{a_{\DI}} }^3 \;,
		\label{eq:gamma_approx}
	\end{equation}
	where $a_{\DI,\DII}$ are given in \eqs{eq:aD1_approx,eq:aD2_approx}, with the corresponding energy densities found from \eqs{eq:rhoR_Phi-domination,eq:rhoRFinal_R-domination}. However, \eqs{eq:gamma_approx} should be considered only a rough estimate,  and numerical determination of should be preferred.

	\subsubsection*{Comparison against the numerical solution}
\begin{figure}[t!]
		\centering	\includegraphics[width=.8\textwidth]{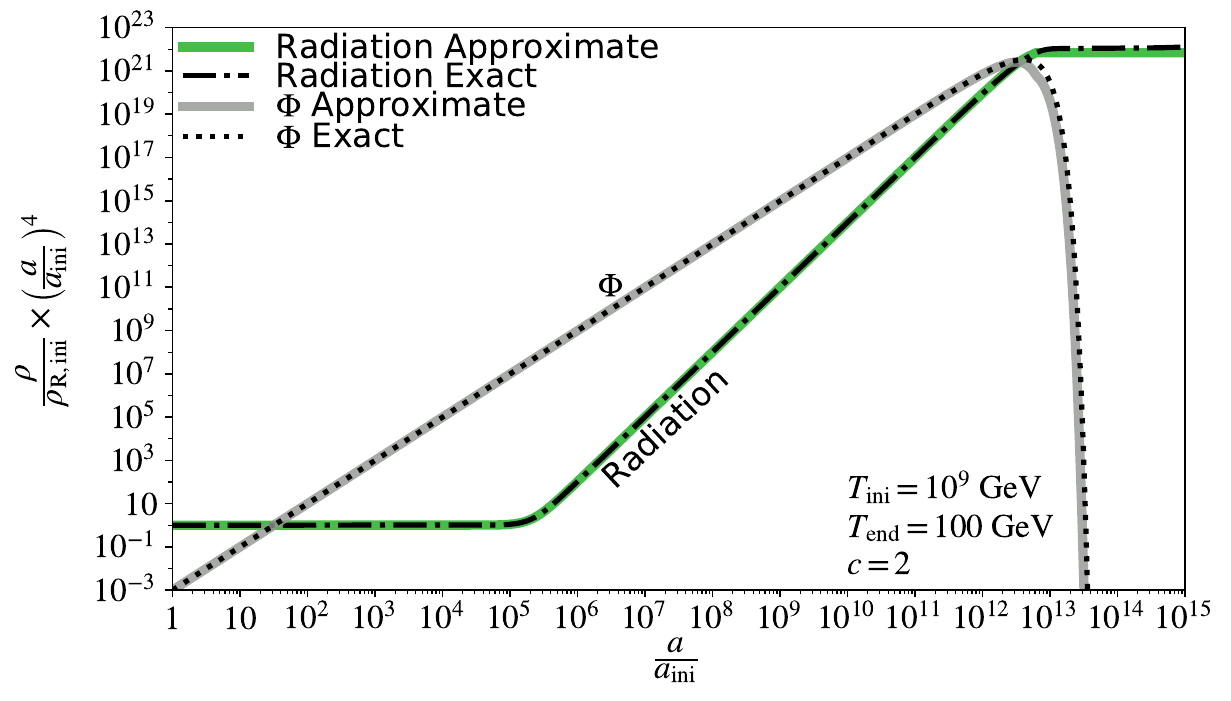}
		\caption{The evolution of the comoving energy densities of radiation and $\Phi$ for the same parameters as in Fig.~\ref{fig:rhoR_rhoPhi}. The black lines correspond to the exact solution, the other two correspond to the approximate form of $\rhoR$ (green) and $\rhoPhi$ (gray).}
		\label{fig:rho-approx}
	\end{figure}

	The approximate solutions derived in this Section are based on a number of assumptions, and we should not expect complete agreement with the numerical results. However, for choices of the parameters values compatible with these assumptions, the approximate forms of the energy densities should be accurate. In \Figs{fig:rho-approx} we show the approximations along with the numerical solutions of \eqs{eq:BE_R,eq:BE_Phi} for the same parameter choice as in \Figs{fig:rhoR_rhoPhi}, where it is apparent that the approximations capture the behavior of the energy densities with reasonable accuracy. We note that in general the approximate and numerical results agree within an order of magnitude. 

	\section{Derivation the BE for $\EDM$ }\label{app:BE_Emean}
	\setcounter{equation}{0}
	To find how  $\EDM$ scales with time, we start BE for the phase-space distribution of $S$ (assuming  production via reactions ${\rm plasma} \to {\rm DM} $)
	\begin{align}
		\lrb{ \partial_t - H p_{\chi} \partial_{p_{\chi}} }f_{\chi}(p_{\chi}) = \sum_{m,n} I^{\rm (col)}_{\rm mn} \;
		\label{eq:BE_f}
	\end{align}
	with $I^{\rm (col)}_{\rm mn}$ the collision term for $m \to n$ processes. Integrating this equation by $\dint  \dfrac{d^3 p_{\chi}}{(2\pi)^3} \times E_{\chi}$,
	the \lhs becomes

	\begin{align*}
		&\dfrac{d\rho_{\chi}}{dt} - H \dint  \dfrac{d^3 p_{\chi}}{(2\pi)^3 }  E_{\chi} p_{\chi} \partial_{p_{\chi}} f_{\chi}(p_{\chi}) =\dfrac{d\rho_{\chi}}{dt} + 4H   \rho_{\chi} -H \Bvev{\dfrac{m_{\chi}^2}{E_{\chi}} }  n_{\chi} \; .
	\end{align*}
	
	In general, the mean value of a function ($G$), can be approximated as  
	\begin{align*}
		& \bvev{G(x)} \approx G( \vev{x})  + \dfrac{1}{2} \lrb{\bvev{x^2} -\vev{x}^2 } \dfrac{d^2G}{dx^2}\Bigg|_{x=\vev{x}} \;.
	\end{align*}
	Assuming small variance, \ie keeping only the first term, the BE for $\rho_{\chi}$ can be approximated as
	\begin{align*}
		\dfrac{d\rho_{\chi}}{dt} + 4H   \rho_{\chi} -H \Bvev{\dfrac{m_{\chi}^2}{E_{\chi}} }  n_{\chi}  \approx
		\dfrac{d\rho_{\chi}}{dt} + 4H   \rho_{\chi} -H \dfrac{m_{\chi}^2}{\EDM}  n_{\chi} \; .
	\end{align*}
	Note that this approximation is exact in the highly-relativistic and deep non-relativistic limits. In the case under study, we expect the freeze-in to happen at high temperatures, where this BE is accurate. 
	
	\subsection{Production via decays}
	Focusing on DM production via decays, $S \to \chi \chi$, the \rhs ~of \eqs{eq:BE_f} is
	\begin{align*}
		\sum_{m,n} I_{mn}^{\rm (col)} \to I_{1\to 2} =&
		\dint  \dfrac{d^3 p_S}{(2\pi)^3 \, 2E_S  } f_{s}(p_S)  
		\dint \dfrac{d^3 q_{\chi_1}}{(2\pi)^3 \, 2E_{\chi_1}  } \dfrac{d^3 q_{\chi_2}}{(2\pi)^3  \, 2E_{\chi_2}  }  \  E_{\chi_2} \ \Big| \mathcal{M} \Big|^2 (2 \pi)^4 \delta^{(4)}\lrb{ p_S - (q_{\chi_1} + q_{\chi_2})   }  \;,
	\end{align*}
	Since we assume that DM particles are identical,~\footnote{If they are not identical, we can still do this as long as the DM particles have the same mass} we may make the substitution 
	\begin{align*}
		E_{\chi_{1}} \to \dfrac{E_{\chi_1}+ E_{\chi_2}}{2}= \dfrac{E_S}{2}.
	\end{align*}
	Assuming that the phase-space distribution os $S$ is sharp, the collision integral may be approximated as
	\begin{align*}
		I_{1\to 2} \approx
		\dfrac{\vev{E_S}}{2}   2\GammaDM \,  \mST \, n_{s}^{(-1)}  \;,
	\end{align*}
	with $\vev{E_S}$ the mean energy of $S$. This  leads to the BE for the evolution of $\rho_{\chi}$ 
	\begin{align}
		\dfrac{d\rho_{\chi}}{dt} + 4H   \rho_{\chi} -H \dfrac{m_{\chi}^2}{\EDM}  n_{\chi} = \dfrac{ \vev{E_S}}{2} 2\GammaDM \, \mST \, n_{s}^{(-1)}    \, .
		\label{eq_BE_rho}
	\end{align} 
	From the definition os the mean energy, $\rho_\chi = \nDM \, \EDM$, this equation takes the form
	\begin{align}
		\dfrac{d\EDM }{dt} = - \EDM  H \lrsb{ \lrBiggb{1  -\lrb{ \dfrac{m_{\chi}}{\EDM}}^2 } + 
			\lrb{1- \dfrac{1}{2} \dfrac{\vev{E_S}}{\EDM} } \dfrac{2\GammaDM}{H}\dfrac{ \mST \; n_{s}^{(-1)}}{ \nDM}   }  \, ,
		\label{eq:BE_E_t}
	\end{align}

	\subsubsection*{Initial condition}
	Initially we assume $n_{\chi}= \rho_{\chi} =0$, \ie the mean energy cannot be defined. However, from energy conservation,  we can impose the initial condition 
	$$ \lrsb{ \dfrac{1}{2}\vev{E_S}-\EDM }_{a=a_R}=0 \;,$$
	with $a_R$ corresponding to some point just after the DM production started (\eg after the end of inflation). However, this initial condition can be taken at $a=\ai$ if between $a_R$ and $\ai$ both $S$ and $\chi$ are relativistic. To show this we note that at high temperatures $\dot{ \vev{E_S} } = - H \vev{E_S}$. Expressing $\EDM$ as $\EDM =  \dfrac{1}{2}  \vev{E_S} - f(t)$,  \eqs{eq:BE_E_t} (for $\mDM \ll \EDM$) takes the form 
	$$
	\dfrac{d f(t)}{dt} = - \lrsb{ \dfrac{2\GammaDM \, \mST \, n_{s}^{(-1)}}{ \nDM} - H} \, f(t) \;.
	$$
	Since with $f(t_R) = 0$, this equation is solved by $f(t) = 0$. Therefore $2\EDM =  \vev{E_S}$ holds at any point as long as both $\chi$ and $S$ are relativistic. 
	
	\subsection{ Generalization}
	In general, DM can be produced by different channels. Following the same reasoning, we can show that if DM is produced via the annihilation of $m$ identical plasma particles ($S$) to $n$ DM particles, the BE for $\EDM$ takes the form 
	\begin{align}
		\dfrac{ d\EDM }{dt} =-H \EDM  \lrBiggcb{ \lrBiggb{1  -\lrb{ \dfrac{\mST}{\EDM}}^2 }  +  
			\sum_{m,n} \lrsb{  \lrb{1- \dfrac{m}{n} \dfrac{\vev{E_S}}{\EDM}  } \dfrac{C_{m\to n}}{H \, \nDM} }}  \, ,
		\label{eq:BE_E_mn}
	\end{align}
	where $C_{m\to n}$ are the collision terms for $m \to n $ processes, defined from $\dfrac{d\nDM}{dt}+3H \nDM=\displaystyle\sum_{m,n} C_{m\to n}$. That is,
	\begin{align}
		\sum_{m,n} C_{m\to n}  =
		\sum_{m,n} \dint  \lrBiggb{\prod_{i=1}^{m}\dfrac{d^3 p_i}{(2\pi)^3 \, 2E_i  } f_{i}(p_i) } 
		\dint \lrBiggb{\prod_{f=1}^{n}\dfrac{d^3 q_{\chi_f}}{(2\pi)^3 \, 2\omega_{\chi_f}  } }  \dfrac{n}{n!} \Big| \mathcal{M} \Big|^2 (2 \pi)^4 \delta^{(4)}\lrb{ \sum_{i=1}^m p_i -\sum_{f=1}^n q_{\chi_f}   } \;.
		\label{eq:BE_n_mn}  
	\end{align}
	Similarly to the previous case, the initial condition reads
	\begin{equation}
		\sum_{m,n}   \lrb{ \dfrac{m}{n}\vev{E_S}-\EDM }C_{m\to n} \Bigg|_{a=a_R}=0 \;.
		\label{eq:init_cond_general}
	\end{equation}
	Expressing the mean energy of DM as
	\begin{equation*}
		\EDM =  \dfrac{ \displaystyle\sum_{m,n}    \dfrac{m}{n}  C_{m\to n}}{ \displaystyle \sum_{m,n} C_{m\to n}} \vev{E_S}   
		- f(t) \;,
	\end{equation*}
	we observe that 
	\begin{equation}
		\dfrac{df(t)}{dt} = \dfrac{d \lrb { \dfrac{ \displaystyle\sum_{m,n}    \dfrac{m}{n}  C_{m\to n}}{\displaystyle\sum_{m,n} C_{m\to n}} }  }{dt} \vev{E_S}-
		\lrb{ \nDM^{-1} \displaystyle\sum_{m,n} C_{m \to n}    - H} f(t) \;.
		\label{eq:dfdt_general}
	\end{equation}
	Evidently, $f (t) = 0$ is not a solution of \eqs{eq:dfdt_general}. However, if some $m \to n$ channel dominates (between $a_R$ and  $\ai$), the first term of the \rhs of \eqs{eq:dfdt_general} vanishes. Consequently, $f (t) = 0$ becomes an approximate solution, and $\EDM = \dfrac{m}{n} \vev{E_S} $ is an appropriate initial condition at $a=\ai$.

	\bibliography{refs}{}
	\bibliographystyle{JHEP}                        
	
\end{document}